\documentclass[structabstract]{aa}  

\usepackage{graphicx, natbib, longtable, rotating, lscape}        
\usepackage{txfonts}
\usepackage{threeparttable}
\begin{document}
   \title{The molecular environment of the Galactic star forming region G19.61-0.23\thanks{Based on observations made with the 14~m FCRAO telescope, 
 the Spitzer satellite,
and the Atacama Pathfinder Experiment (APEX), ESO project: 181.C-0885.
APEX is a collaboration between the Max-Planck-Institut fur Radioastronomie,
the European Southern Observatory, and the Onsala Space Observatory.}}


   \author{G. Santangelo
           \inst{1,2,3,4}
           \and
           L. Testi
           \inst{1,4}
           \and
           S. Leurini
           \inst{1,6}
           \and
           C.M. Walmsley
           \inst{4}
            \and
           R. Cesaroni
           \inst{4}
           \and
           L. Bronfman
           \inst{7}
           \and
           S. Carey
           \inst{9}
           \and
           L. Gregorini
           \inst{2,3}
           \and
           K.M. Menten
           \inst{6}
           \and
           S. Molinari
           \inst{8}
           \and
           A. Noriega-Crespo
           \inst{9}
           \and
           L. Olmi
           \inst{4,5}
           \and
           F. Schuller
           \inst{6}
         }

   \institute{European Southern Observatory,
              Karl Schwarzschild str.2, D-85748 Garching bei Muenchen, Germany\\
              \email{santangelo@ira.inaf.it, ltesti@eso.org}
         \and
             INAF - Istituto di Radioastronomia, via Gobetti 101,
             40129 Bologna, Italy
         \and
             Dipartimento di Astronomia, Universit\`a di Bologna,
             via Ranzani 1, 40127 Bologna, Italy
         \and
             INAF - Osservatorio Astrofisico di Arcetri, Largo E. Fermi 5,
             I-50125 Firenze, Italy
         \and
             University of Puerto Rico, Rio Piedras Campus, Physics Department, 
             Box 23343, UPR Station, San Juan, Puerto Rico
         \and
             Max-Planck-Institut f{\"u}r Radioastronomie, Auf dem H{\"u}gel 69, 
             53121 Bonn, Germany
         \and    
             Departamento de Astronom\'ia, Universidad de Chile, 
             Casilla 36-D, Santiago, Chile
         \and    
             Istituto Nazionale di Astrofisica - Istituto Fisica spazio interplanetario, 
             Via Fosso del Cavaliere 100, 
             I-00133 Rome, Italy
         \and    
             Spitzer Science Center, California Institute of Technology, 
             Pasadena, CA 91125, USA
             }

   \date{Received February 2, 2010; accepted June 3, 2010}

  \abstract
   {Although current facilities allow the study of Galactic star formation at
high angular resolution, our current understanding of the high-mass
star-formation process is still very poor. In particular, we still need to
characterize the properties of clouds giving birth to high-mass stars
in our own Galaxy and use
them as templates for our understanding of
extragalactic star formation.
}
   {We present single-dish (sub)millimeter observations of gas and dust
in the Galactic high-mass star-forming region
G19.61-0.23, with the aim of studying the large-scale 
properties and physical
conditions of the molecular gas across the region.  The final aim is to compare
the large-scale (about 100~pc) properties with the small-scale 
(about 3~pc) properties and to consider
possible implications for extragalactic studies. 
}
   {We have mapped CO isotopologues in the $J=1-0$ transition using the FCRAO-14m
telescope
and the $J=2-1$
transition using the IRAM-30m telescope. We have also used APEX 870~$\mu$m
continuum data from the ATLASGAL survey and FCRAO supplementary observations of the
$^{13}$CO $J=1-0$ line from the BU-FCRAO Galactic Ring Survey, as well as
the Spitzer
infrared Galactic plane surveys GLIMPSE and MIPSGAL to characterize the star-formation 
activity within the molecular clouds.
}
   {We reveal a population of molecular clumps in the $^{13}$CO(1-0) emission,
for which we derived the physical parameters, including sizes and masses. 
Our analysis of the $^{13}$CO suggests that 
the virial parameter (ratio of kinetic to gravitational energy) varies over an order of magnitude 
between clumps that are unbound and some that are apparently "unstable".
This conclusion is independent of
whether they show evidence of ongoing star formation.
We find that the majority of ATLASGAL sources have MIPSGAL  counterparts with luminosities
in the range 10$^4$-$5\,10^4$~$L_{\odot}$ and are presumably forming relatively 
massive stars.
We compare our results with previous extragalactic studies of the nearby spiral
galaxies M31 and M33; and the blue compact dwarf galaxy Henize~2-10. We find that the main
giant molecular cloud surrounding G19.61-0.23 has physical properties
typical for Galactic GMCs and which are comparable to the GMCs in M31 and M33.
However, the GMC studied here shows smaller surface densities and masses
than the clouds identified in Henize~2-10 and associated with
super star cluster formation.
}
   {}


   \titlerunning{The molecular environment of the G19.61-0.23 region}
   \authorrunning{Santangelo et al.}
   \maketitle
%

\section{Introduction}\label{intro}

High-mass stars (OB spectral type, $M>8\,M_{\odot}$ and $L>10^3\,L_{\odot}$), although few in
number, play a major role in the energy budget of galaxies, through their
radiation, wind and the supernovae.  They are believed to form by accretion in dense cores
within molecular cloud complexes
(\citealt{yorke2002};
\citealt{mckee2003}; \citealt{keto2003,keto2005}) and/or coalescence 
(e.g. \citealt{bonnell2001}). The intense
radiation field emitted by a newly-formed central star heats and ionizes its
parental molecular cloud, leading to the formation of a hot
core (HC, e.g. \citealt{helmich1997}) and afterwards an HII region.
Our current understanding of their formation remains poor, especially
concerning the earliest phases of the process.  The main observational
difficulty is that high-mass stars are fewer in number than low-mass stars
and the molecular clouds that are able to form high-mass stars are
statistically more distant than those forming low-mass stars. 
Therefore,
current observational studies of high-mass star formation suffer both from the lack
of spatial resolution and, consequently,
from a lack of theoretical understanding.

The high-mass star-forming region G19.61-0.23 is an interesting target for the
study of star cluster formation given its richness in terms of young
stellar objects (YSOs), as indicated by studies at centimeter (e.g.
\citealt{garay1998}; \citealt{forster&caswell2000}), millimeter (e.g.
\citealt{furuya05}) and mid-infrared (MIR; \citealt{debuizer2003}) wavelengths.
G19.61-0.23 is  located at a distance of 12.6 kpc (see \citealt{kolpak2003}),
based on the 21 cm HI absorption spectrum toward the source.  The total
bolometric luminosity (\citealt{walsh1997})  is about $2\times10^6\,L_{\odot}$.
The region contains OH (\citealt{caswell1983}; \citealt{forster1989}), water
(\citealt{hofner1996}) and methanol (\citealt{caswell1995};
\citealt{valtts2000}; \citealt{szymczak2000}) masers, and a grouping of UC~HII
regions and extended radio continuum, indicating that it is an active region of
massive star formation. The radio continuum emission from this region comes
from five main sources, all of which are discussed in detail in
\cite{garay1998}. The region has been mapped in CS, NH$_3$ and CO
(\citealt{larionov1999}; \citealt{plume1992}; \citealt{garay1998}).  Extended
mid-infrared emission associated with the region is also detected by the MSX
satellite (\citealt{crowther 2003}).  Single-dish observations of molecular
lines with high critical densities show the presence of dense molecular gas
over a broad range in velocity (\citealt{plume1992}).   

In this paper we present a spectroscopic study of the region surrounding
G19.61-0.23 in several transitions of CO isotopologues at an angular resolution
of $\sim46^{\prime\prime}$ (about 2.8~pc at the distance of 12.6~kpc), on a
large region of about $23^{\prime}\times23^{\prime}$ (roughly 85~pc) centered
on G19.61-0.23.  Millimeter observations of carbon monoxide provide useful
information on the physical properties of dense interstellar clouds as well as
on their dynamical state.  Moreover, supplementary measurements of continuum 
emission in the sub-millimeter range with APEX, based on ATLASGAL data, and in the mid-infrared with
Spitzer, based on GLIMPSE and MIPSGAL data, are presented.  The aim is to study the global large-scale physical
properties and their relation with the small-scale characteristics.  We
analyzed the physical conditions and the velocity structure of the molecular
components across the region.  We finally consider the possible implications
for extragalactic observations of well-studied nearby starburst galaxies at the
same linear resolution.

In Sect.~\ref{observations} we describe the dataset of
observations used in this paper; in Sect.~\ref{largescale-integrated_simbad}
we present the
large-scale morphology of the emission; in Sect.~\ref{identifica} we describe
the identification of the clumps: from the molecular line emission and from the
sub-mm continuum emission; in Sect.~\ref{starformationassociation} we discuss the
association of the clumps with star-formation tracers; in
Sect.~\ref{section:parameter} and Sect.~\ref{section:stability} we derive
the physical parameters of the clumps, from the molecular line emission and
from the continuum emission, and we discuss the implications of the
results for the structure of the clumps; in Sect.~\ref{section:SED} we discuss
the spectral energy distributions of the IR sources associated with the APEX
continuum sources; in Sect.~\ref{comparison_extragal} we compare our results with
studies of nearby galaxies; Sect.~\ref{summary}
contains our summary and conclusions.

\section{Observational dataset}\label{observations}

\subsection{FCRAO $^{13}$CO(1-0), C$^{18}$O(1-0) and C$^{17}$O(1-0) data}

The data were obtained in the period from 2000 28th October to 7th November and in
2001 May, using the 14-m Five College Radio Astronomy Observatory (FCRAO)
telescope, located near New Salem (Massachusetts, USA). The observations were
performed with the SEQUOIA (Second Quabbin Observatory Imaging Array) 16 beam
array receiver. 

An area of about 23$^{\prime}\times$23$^{\prime}$ was mapped in the $^{13}$CO(1-0) line.
The system temperature was between 294~K and
680~K.  The telescope beam size is approximately 46$^{\prime\prime}$ at the
frequency of the observations, according to \cite{ladd}. 

The C$^{18}$O(1-0) and C$^{17}$O(1-0) emission lines were observed over smaller
regions: the C$^{18}$O(1-0) line in three regions of about $5.\!\!^{\prime}6
\times 5.\!\!^{\prime}6$ (with 22$^{\prime\prime}$ sampling, i.e. Nyquist), 
in a diagonal direction through the center of the
$^{13}$CO map (see Fig.~\ref{integr_tot}) from North-East to South-West; the
C$^{17}$O(1-0) line in the central region of about $5.\!\!^{\prime}2 \times
5.\!\!^{\prime}2$, on a 45$^{\prime\prime}$ sampled grid (i.e. undersampled). The observations were
performed in frequency switching mode. 
The system temperature and beam sizes at these frequencies are reported in 
Table~\ref{line_param}.

In each molecular line, the same area was covered several times to obtain
complete but independent maps of the whole region.  Observations related to the
same pointing were first added together.  The final maps are thus the weighted
average of all the data.  
The line intensities of all the spectra were converted into main beam
temperature units, using the main beam efficiency values\footnote{http://www.astro.umass.edu/~fcrao/observer/status14m.html\#\\ANTENNA} given in 
Table~\ref{line_param}.

The original velocity resolution for the
three lines was 0.2~km~s$^{-1}$, but the data were smoothed to a velocity
resolution of 1~km~s$^{-1}$, to increase the signal to noise ratio.  
The final rms
noise is 0.2~K per channel for the $^{13}$CO(1-0) line, 0.05~K per channel for
the C$^{18}$O(1-0) line and 0.02~K per channel for the C$^{17}$O(1-0) line.
However, part of the analysis was made at a velocity resolution of
0.5~km~s$^{-1}$, as explained below (see Sects.~\ref{kinematics} and \ref{identif}), 
to clearly distinguish and derive the parameters of some
narrow features in the emission.
\begin{table}
\caption{Summary of the parameters of the spectral line observations.
}
\label{line_param}      
\centering      
\begin{tabular}{c c c c c c }
\hline\hline       
Line & $\nu$ & $\Theta_{beam}$ & T$_{sys}$ & $\Delta$v & $\eta_{\rm mb}$\\
& (GHz) & (arcsec) & (K) & (km s$^{-1}$) &\\
(1)&(2)&(3)&(4)&(5)&(6)\\
\hline                    
\\
\multicolumn{6}{c}{\underline{FCRAO~14~m}}\\
 $^{13}$CO(1-0) & 110.201 & 46 & 294-680 & 0.2 & 0.49\\  
 C$^{18}$O(1-0) & 109.782 & 46 & 291-697 & 0.2 & 0.49 \\
 C$^{17}$O(1-0) & 112.359 & 46 & 290-371 & 0.2 & 0.47\\ 
\hline
\\
\multicolumn{6}{c}{\underline{IRAM~30~m}}\\
$^{13}$CO(2-1) & 220.399 & 11   & 639-1129 & 0.1 & 0.52 \\  
C$^{18}$O(2-1) & 219.569 & 11   & 659-1061 & 0.1 & 0.52\\   
C$^{17}$O(2-1) & 224.714 & 11   & 383-517 & 0.1 & 0.536\\
\hline
\end{tabular}
\begin{tablenotes}
\item[] NOTES. -- Col.(1): the observed line. Col.(2): the frequency of the transition. Col.(3): the beam size of the observations. Col.(4): the typical system temperature, T$_{sys}$, of the observations. Col.(5): the velocity resolution, $\Delta$v, of the observations. Col.(6): the beam efficiency used for every specific frequency.
\end{tablenotes}
\end{table}

\subsection{IRAM $^{13}$CO(2-1), C$^{18}$O(2-1) and C$^{17}$O(2-1)}

The data were obtained using the IRAM-30m telescope on Pico Veleta in its
On-The-Fly (OTF) observing mode.
The $^{13}$CO(2-1) and C$^{18}$O(2-1) observations are described in \cite{lopezsepulcre2009}.
The C$^{17}$O(2-1) line was observed in 2004 February, in frequency switching
mode, on the same region. The system temperature was between 383~K and 517~K.

The rms noise is 0.8~K per 1~km~s$^{-1}$ channel for the $^{13}$CO(2-1) line, 0.9~K per channel for
the C$^{18}$O(2-1) line and 0.3~K per channel for the C$^{17}$O(2-1) line.  The
telescope beam size at the frequencies of the observations is
11$^{\prime\prime}$\footnote{http://www.iram.es/IRAMES/telescope/telescopeSummary/\\
telescope\_summary.html}.  A summary of all the observational parameters is
given in Table~\ref{line_param}.\\

The data reduction was performed with the standard data analysis program
GILDAS\footnote{http://iram.fr/IRAMFR/GILDAS/}, developed at IRAM and Observatoire de Grenoble.

\subsection{APEX 870~$\mu$m continuum data}

Observations of the 870~$\mu$m continuum emission
were made using the Atacama 
Pathfinder Experiment (APEX)
12m telescope on July 2007.  The data are part of the APEX Telescope Large Area
Survey of the Galaxy (ATLASGAL, \citealt{schuller2009}), performed with the
Large APEX Bolometer Camera (LABOCA, \citealt{siringo2007}, 2009).  The rms
noise of the map is 40~mJy~beam$^{-1}$.  The beam size at this wavelength is
$18.\!\!^{\prime\prime}2$.
The survey is optimized for recovering compact sources 
and extended uniform emission on scales
larger than $\sim 2.\!\!^{\prime}5$ is filtered out (\citealt{schuller2009}).

\subsection{BU-FCRAO Galactic ring survey}

We retrieved FCRAO $^{13}$CO(1-0) observations of a slightly larger area (about
$27^{\prime} \times 27^{\prime}$) than the one sampled with our FCRAO
observations, from the Boston University-Five College Radio Astronomy
Observatory Galactic Ring Survey (hereafter BU-FCRAO
GRS\footnote{http://www.bu.edu/galacticring/new\_data.html}), a survey of the
Galactic $^{13}$CO $J=1-0$ emission (see \citealt{jackson 2006}). 

The velocity coverage of the survey is -5 to 135~km~s$^{-1}$ for Galactic
longitudes $l \leq 40^{\circ}$.  At the velocity resolution of
0.21~km~s$^{-1}$, the typical rms sensitivity is $\sigma(T^*_{\rm A}) \sim 0.2$~K,
which corresponds to the rms noise of our $^{13}$CO(1-0) observations,
for a main-beam efficiency $\eta_{\rm mb}$ of 0.48.  The intensities of
the BU-FCRAO GRS data are in agreement with our data within $\sim 10$~\%. We
used the BU-FCRAO GRS as a comparison with our data of the same line in the
regions covered by both observations and to cover the regions which have not
been sampled by our observations.

\subsection{Spitzer data}

Mid infrared observations of a large region around G19.61-0.23 were extracted
from the Spitzer Galactic plane surveys GLIMPSE (\citealt{benjamin2003}) and
MIPSGAL (\citealt{carey2005}). Data from 3.6~$\mu$m through 24~$\mu$m were
extracted directly from the Spitzer science archive, while for the 70~$\mu$m
maps we used the latest processing from the MIPSGAL team which substantially
improved the quality of the maps as compared with the standard pipeline results
(\citealt{carey2009}).

All the data were
analyzed with the KVIS tool, which is part of the KARMA suite of image
visualization tools (\citealt{gooch1996}).

\section{Morphology of the large-scale molecular line emission}\label{largescale-integrated_simbad}

\begin{figure*}
\centering
\includegraphics[width=1.8\columnwidth]{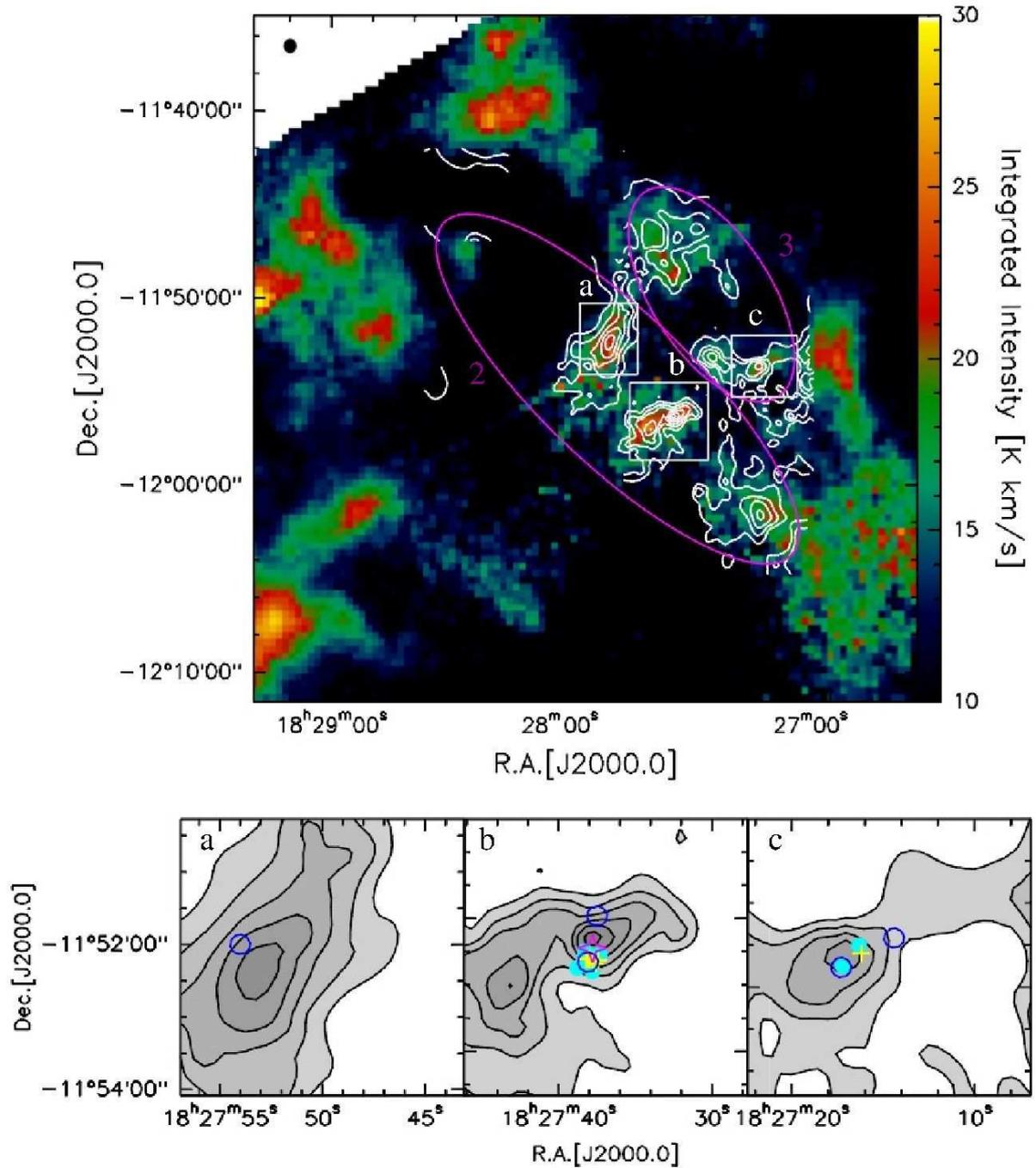}
\caption{FCRAO velocity-integrated $^{13}$CO(1-0) emission in 
white contours from our data,
overlaid on the $^{13}$CO(1-0) emission from the BU-FCRAO GRS (color). 
Both maps are integrated between 25 and
77~km~s$^{-1}$. The contour levels range from 10~$\sigma$ in steps of
3~$\sigma$ ($\sigma$= 2~K~km~s$^{-1}$). 
The magenta ellipses represent: 1) the Giant Molecular 
Cloud (GMC) surrounding G19.61-0.23, which is the
32-50 km~s$^{-1}$ molecular gas (cloud~2) discussed
in Sect.~\ref{kinematics}; and 2) the 54-63 km~s$^{-1}$ molecular gas (cloud~3)
discussed in Sect.~\ref{kinematics}.
The three white boxes, labelled as ``a'', ``b'' and ``c'', highlight the three regions 
displayed in the bottom part of the figure.
The cyan filled circles represent the HII regions, 
the yellow crosses are the water masers, the magenta open stars are the OH masers and the
blue open circles are the methanol masers (see Table~\ref{SIMBAD}).
}
\label{integr_tot}
\end{figure*}

Figure \ref{integr_tot} shows the FCRAO $^{13}$CO(1-0) integrated intensity map
overlaid over a larger area map of the $^{13}$CO(1-0) emission from the
BU-FCRAO GRS.  The two maps are integrated over the same velocity range 
(25--77~km~s$^{-1}$).  The figure, at the angular resolution of
46$^{\prime\prime}$, highlights the complex structure of the molecular gas.
The colored symbols in the figure indicate the positions of the known
sources\footnote{Retrieved from the SIMBAD Astronomical Database,
http://simbad.u-strasbg.fr/simbad/} in the molecular complex associated with
G19.61-0.23. A summary of all these
observations with associated references is given in Table~\ref{SIMBAD}, in
the Online Material Sect..

The large scale morphology and extent of the molecular gas can be seen.  The
main cloud of the complex (defined as cloud~2 in Sect.~\ref{kinematics} and
highlighted by the larger magenta ellipse in Fig.~\ref{integr_tot}) contains
G19.61-0.23, which is identified in the center of the field, around R.A.(J2000) =
18$^{\rm h}$27$^{\rm m}$38$^{\rm s}$ and  Dec.(J2000) =
-11$^\circ$56$^{\prime}$17$^{\prime\prime}$.
This main cloud is also seen in the central region of the C$^{18}$O(1-0)
velocity-integrated map (available in Fig.~\ref{integr_tot2} in the Online
material Sect.), as well as in all other tracers of star formation, such as
masers, HII regions and IR emission (see Table~\ref{SIMBAD}).  A good
correlation between the $^{13}$CO(1-0) and C$^{18}$O(1-0) emission is seen also
in the other two regions sampled in the C$^{18}$O(1-0) line, which present the
same morphology as the $^{13}$CO(1-0) emission.

\subsection{A closer view of the molecular line emission}\label{closerview}

\begin{figure*}
\centering
\includegraphics[angle=-90,scale=0.73]{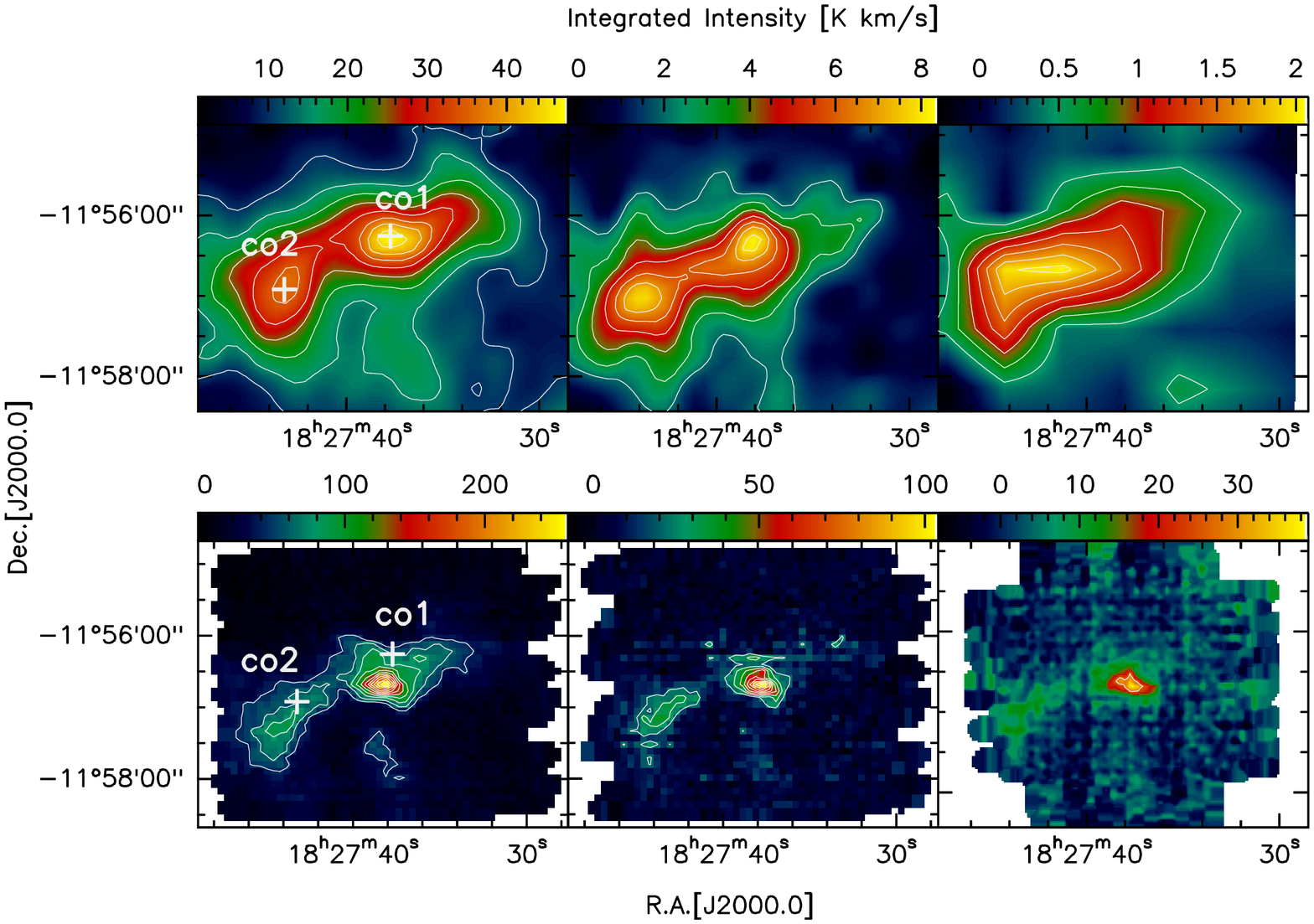}   
\caption{\emph{Top}: FCRAO emission maps, integrated between 
32 and 50 km~s$^{-1}$, of the central
$4.\!\!^{\prime}7 \times 3.\!\!^{\prime}6$ region. \emph{Left}: $^{13}$CO(1-0)
emission; the contour levels range
from 10~$\sigma$ ($\sigma = 1$~K~km~s$^{-1}$) in steps of 5~$\sigma$.  
The white crosses are the positions 
of the emission peaks of the two central clumps, 
which are labelled as ``co1'' and ``co2'' (see Table~\ref{param_nubi}).
\emph{Middle}: C$^{18}$O(1-0)
emission; the contour levels range
from 10~$\sigma$ ($\sigma = 0.2$~K~km~s$^{-1}$) in steps of 5~$\sigma$.
\emph{Right}: C$^{17}$O(1-0) emission;
the contour levels range from 10~$\sigma$ ($\sigma = 0.05$~K~km~s$^{-1}$) in
steps of 5~$\sigma$.  \emph{Bottom}: IRAM-30m emission maps, integrated between 
32 and 50 km~s$^{-1}$, of the
central 5$.\!\!^{\prime}$5$\times$5$.\!\!^{\prime}$5 region. \emph{Left}:
$^{13}$CO(2-1) emission; the contour
levels range from 10~$\sigma$ ($\sigma = 4$~K~km~s$^{-1}$) in steps of
5~$\sigma$. The white crosses are the same as before.
\emph{Middle}: C$^{18}$O(2-1) emission; the contour levels range from 10~$\sigma$ ($\sigma =
2$~K~km~s$^{-1}$) in steps of 5~$\sigma$.  \emph{Right}: C$^{17}$O(2-1)
emission; the contour levels range
from 10~$\sigma$ ($\sigma = 2.7$~K~km~s$^{-1}$) in steps of 5~$\sigma$.}
\label{integr_centr}
\end{figure*}

Observations of both transitions of all three CO isotopologues were performed
only in the central region of about $4.\!\!^{\prime}7 \times 3.\!\!^{\prime}6$
(17 pc $\times$ 13 pc), containing the main complex of the molecular emission.
Figure \ref{integr_centr} shows in the top panel the FCRAO $^{13}$CO(1-0),
C$^{18}$O(1-0) and C$^{17}$O(1-0) maps, integrated between 32 and 50
km~s$^{-1}$, in the central region covering the main cloud of the complex. The
bottom panels of Fig.~\ref{integr_centr} give the IRAM-30m $^{13}$CO(2-1),
C$^{18}$O(2-1) and C$^{17}$O(2-1) maps, integrated over the same velocity range.

In the $^{13}$CO(1-0) and C$^{18}$O(1-0) emission maps, the molecular gas is
concentrated within two main clumps (clump~co1 and clump~co2, see
Sect.~\ref{identif}).  This is not confirmed in the C$^{17}$O(1-0) emission
map, where the two peaks are not clearly distinguished.  This is maybe
affected by the low sampling of the region in the C$^{17}$O(1-0) line. The same
morphological characteristics are confirmed in the $^{13}$CO(2-1) line, where
the two clumps can be clearly distinguished, and partially in the
C$^{18}$O(2-1) line, where clump~co2 is still visible, although very faint.  In
the C$^{17}$O(2-1) emission, clump~co2 is even fainter and, due to the lower
signal to noise radio, it is hard to be distinguished.

\subsection{Kinematics of the emission and cloud definition}\label{kinematics}

Figure \ref{13cototcanali}, in the Online Material Sect., shows the
large-scale channel maps of the $^{13}$CO(1-0) emission (i.e., the emission 
in 2~km~s$^{-1}$ wide velocity intervals), which reveals the
complex spatial and velocity structure of the molecular gas.  The contours of
the emission are from 10~$\sigma$ ($\sigma$ = 0.2~K per channel) in steps of
5~$\sigma$.  The numbered crosses represent the molecular features (clumps),
which we identified in the emission (see Sect.~\ref{identif}).  

Throughout this paper we will use the
word ``\emph{cloud}''
for the four main molecular complexes described below and
identified in the $^{13}$CO(1-0) emission
and ``\emph{clump}'' for the smaller scale structures
resolved within the four main \emph{clouds} (see Sect.~\ref{identif}).

\begin{figure*}
\centering
\includegraphics[angle=-90,scale=0.73]{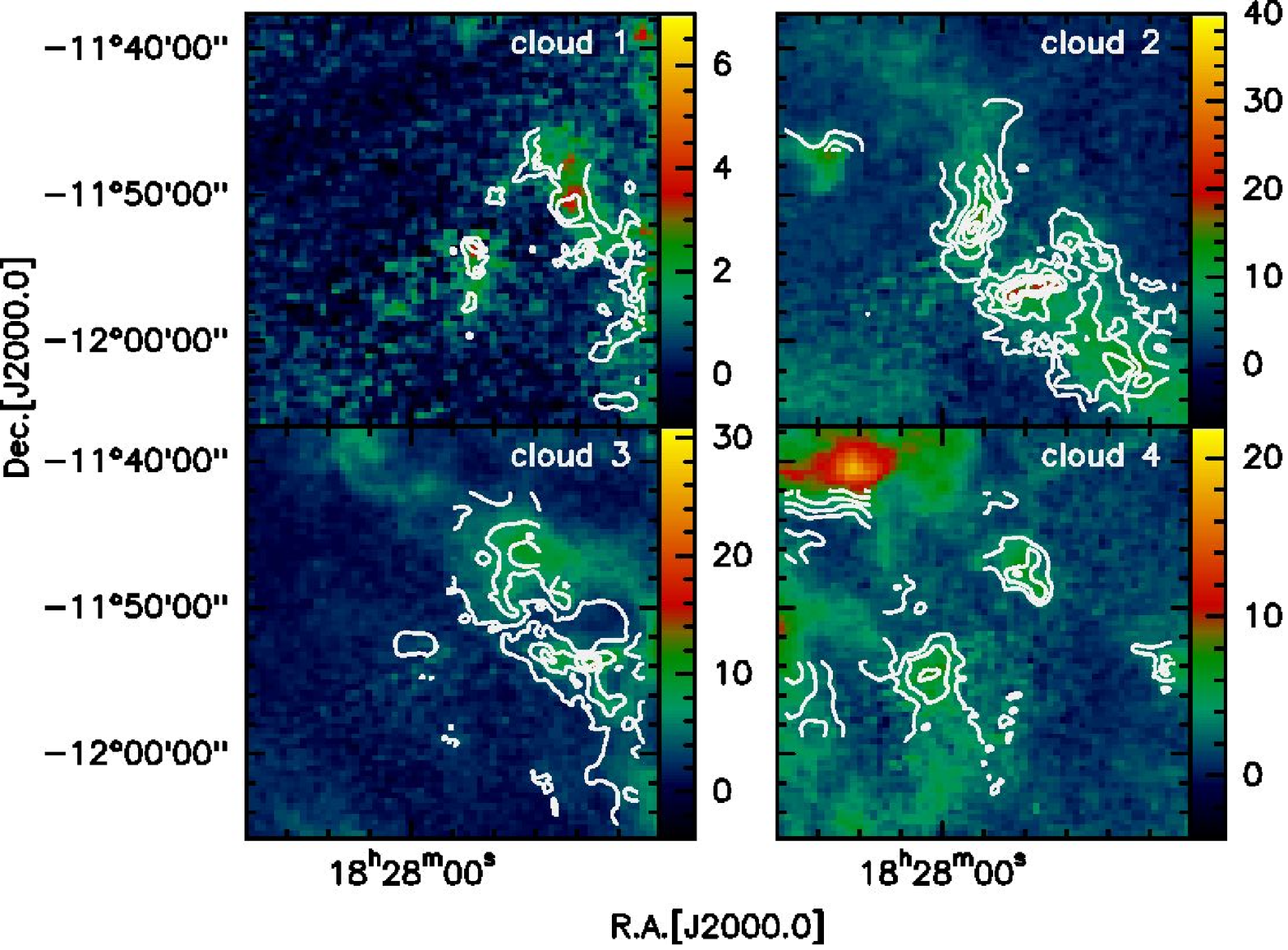}
\caption{Maps of the four molecular clouds identified in $^{13}$CO(1-0) emission, 
integrated over the velocity channels of emission of each cloud (see col.~2 of 
Table~\ref{cloud_param}). The FCRAO $^{13}$CO(1-0) emission from our data is represented 
by the white contours and is overlaid on the $^{13}$CO(1-0) emission from the BU-FCRAO GRS.}
\label{clouds_integrate}
\end{figure*}

We distinguish
four velocity intervals, corresponding to the main features (clouds) that can
be identified in the molecular gas. These are four molecular clouds, well separated 
in velocity and/or spatial distribution.
The main parameters of these four molecular clouds are given in Table~\ref{cloud_param}
and Fig.~\ref{clouds_integrate} shows the integrated emission from each of them.
In particular, we identify:
\begin{description}
\item[Cloud~1] (28--32 km~s$^{-1}$): the lowest velocity channels show no
significant emission, with the exception of one narrow feature (clump~co4, see 
Table~\ref{param_nubi}), which
is clearly visible above the 10~$\sigma$ level in the 29~km~s$^{-1}$ 
velocity channel. In order to analyze this feature, we created a $^{13}$CO(1-0)
data cube with a velocity resolution of 0.5~km~s$^{-1}$.
\item[Cloud~2] 
(32--50 km~s$^{-1}$): this is the velocity range of the main
complex of the emission, which is the GMC surrounding G19.61-0.23. It is
represented by the larger magenta ellipse in Fig.~\ref{integr_tot}, with linear
semi-axes of 49~pc and 15~pc. All the emission is spatially concentrated in
the same area, with an elliptical shape.  This suggests that
the emission is all at the same distance, i.e. at the distance of the
main central region at 12.6~kpc.
Several molecular features can
be distinguished in the emission, both in the spatial and velocity
distributions.  The identification of the single clumps is discussed 
in Sect.~\ref{identifica}; in particular, 
one of them, clump~co5 (see Table~\ref{param_nubi}) has been identified
using the BU-FCRAO GRS, because it is at the edge of our observed region.
\item[Cloud~3] (54--63 km~s$^{-1}$): these velocity channels show several
features in the emission as well, spread over 10~km~s$^{-1}$.  Also, in this
case the emission is all concentrated in the same area, in the north-west of
the sampled area, which suggests that it is all at the same distance.  The
smaller ellipse in Fig.~\ref{integr_tot} delineates the approximate outline of
this GMC.  It is worth noting that the narrow feature, seen only in the 60~km~s$^{-1}$
velocity channel (clump~co19, see 
Table~\ref{param_nubi}), has been analyzed with the
higher velocity resolution of 0.5~km~s$^{-1}$.
\item[Cloud~4] (64--73 km~s$^{-1}$): the highest velocity channels show 
four main features in the emission and two of them (clump~co14 and clump~co17)
have been identified using the BU-FCRAO GRS.
The emission at the east edge of the map has not been
included in the further analysis.
\end{description}

Both the near and far distances of each one of the four clouds have been
computed, using the rotation curve of \cite{brand1993}.  The derived distances
as well as the galactocentric distances ($D_{\rm GC}$) are given in
Table~\ref{cloud_param} for the four clouds identified in the FCRAO
$^{13}$CO(1-0) emission, with the respective velocity interval of the emission.
\begin{table}
\caption{Summary of the parameters of the four molecular clouds identified in
the FCRAO $^{13}$CO(1-0) emission.}
\label{cloud_param}      
\centering      
\begin{tabular}{c c c c c}
\hline\hline       
Cloud & $\Delta$v & $d_{\rm FAR}$ & $d_{\rm NEAR}$ & $D_{\rm GC}$ \\
& (km~s$^{-1}$) & (kpc) & (kpc) & (kpc) \\
(1)&(2)&(3)&(4)&(5)\\
\hline                    
1 & 28-32 & 13.4 &  2.6  &  6.1       \\ 
2 & 33-48 & 12.6 &  3.4  &  5.4       \\ 
3 & 54-63 & 11.7 &  4.3  &  4.7       \\ 
4 & 64-73 & 11.3 &  4.7  &  4.3       \\ 
\hline
\end{tabular}
\begin{tablenotes}
\item[] NOTES. -- Col.(1): the cloud. Col.(2): the velocity interval of the emission. Col.(3): the far distance. Col.(4): the near distance. Col.(5): the galactocentric distance.
\end{tablenotes}
\end{table}

Finally, we point out that the association of the emission, which is close in
velocity and position, as part of the same GMC is fairly arbitrary and has
consequences for the distance estimates of the different clouds.

\section{Identification of the clumps}\label{identifica}

\subsection{The $^{13}$CO emission}\label{identif}

Source identification was based on the visual inspection of all velocity channels 
of the FCRAO $^{13}$CO(1-0) map of the large-scale emission and of the BU-FCRAO GRS.
The detection threshold used for the identification of the different clumps in the $^{13}$CO(1-0)
emission was set to 10~$\sigma$ ($\sigma = 0.2$~K).
We identified 19 clumps in the FCRAO $^{13}$CO(1-0) emission, which are 
indicated in Fig.~\ref{13cototcanali} (in the Online Material Sect.),
in the channel corresponding to 
their peak emission (see also Table~\ref{param_nubi}).
The clumps co5, co10, co12, co11, co14 and co17, as shown in Fig.~\ref{13cototcanali}, are at the edges of the sampled region 
and we thus used the BU-FCRAO GRS data to derive their physical parameters for the analysis.

The angular extent of each identified clump was determined by finding the area, A$_{1/2}$, within the 
50\% intensity contour in the FCRAO $^{13}$CO(1-0) maps, integrated  
over the channels of the emission,  
and computing the equivalent radius of a circle with the same area. 
The angular diameter of each clump was derived 
by deconvolving the diameter,
assuming source and 
beam to be Gaussian: thus $\Theta_{\rm S} = \sqrt{4A_{1/2}/\pi-\Theta_{\rm beam}^2}$. 
The derived angular sizes are given in Table~\ref{param_nubi}.
All the identified clumps  
have deconvolved sizes which are larger than the beam size of the observations, indicating that they are 
well resolved.
\begin{sidewaystable*}
\caption{Physical parameters of the clumps identified in the FCRAO $^{13}$CO(1-0) emission.}\label{param_nubi}
\centering 
\smallskip 
\begin{threeparttable}
\renewcommand{\footnoterule}{} 
\begin{tabular}{l c c c r r r r r r r c l l} 
\hline\hline  
\\
Clump & \multicolumn{2}{c}{Peak Position} & Velocity Range & $\Theta_{\rm S}$\tnote{a} &   T$_{peak}$ & v$_{peak}$ & FWHM\tnote{b} & $\int$T$_{\rm B}{\rm dv}$ & $d_{\rm FAR}$ & $d_{\rm NEAR}$ & $D_{\rm GC}$ & ${\frac{^{12}{\rm C}}{^{13}{\rm C}}}^{\tnote{e}}$ & ${\frac{^{16}{\rm O}}{^{18}{\rm O}}}^{\tnote{e}}$\\ 
&     R.A.[J2000.0] & Dec.[J2000.0] & (km~s${-1}$) & (arcsec) &  (K) & (km/s) & (km/s) & (K km/s) & (kpc) & (kpc) & (kpc) & &\\
(1)&(2)&(3)&(4)&(5)&(6)&(7)&(8)&(9)&(10)&(11)&(12)&(13)&(14)\\
\hline
co1              &   18:27:37.63  &  -11:56:15.7  &  38--48     & 48.9  & 4.8 & 42.9 &  7.3         &  37.3 & 12.6 & 3.4 & 5.4 & 48.1  & 354.6 \\ 
co2              &   18:27:43.21  &  -11:56:52.7  &  38--48     & 67.4  & 5.0 & 42.8 &  5.5         & 29.8 & 12.6 & 3.4 & 5.4 & 48.1  & 354.6 \\ 
co3              &   18:27:51.73  &  -11:51:57.7  &  38--44     & 135.2 & 4.8 & 41.0 &  5.2        & 27.1 & 12.6 & 3.4 & 5.4 & 48.1  & 354.6 \\ 
co4\tnote{c}     &   18:27:45.22  &  -11:53:18.7  &  28.5--29.5 & 75.6  & 3.9 & 28.8 &  0.8  &  3.9 & 13.4 & 2.6 & 6.1 & 53.4 & 395.8 \\ 
co5\tnote{d}     &   18:28:25.15  &  -11:47:12.6  &  44--46     & 123.4 & 9.8 & 44.7 &  1.6    & 19.9  & 12.6 & 3.4 & 5.4 & 48.1  & 354.6 \\   
co6              &   18:27:17.01  &  -12:02:03.0  &  35--44     & 87.5  & 2.4 & 38.7 &  8.2         & 21.5 & 12.6 & 3.4 & 5.4 & 48.1  & 354.6 \\ 
co7              &   18:27:17.03  &  -11:53:41.4  &  56--62     & 60.9  & 4.4 & 59.0 &  4.8  & 22.8 & 11.7 & 4.3 & 4.7 & 42.9 & 313.5 \\   
co8             &   18:27:29.61  &  -11:53:12.0  &  34--39     & 51.0  & 3.0 & 37.0 &  5.2        &  16.8 & 12.6 & 3.4 & 5.4 & 48.1  & 354.6 \\ 
co9             &   18:27:22.07  &  -11:54:47.6  &  34--39     & 53.0  & 2.7 & 35.2 &  3.5         &  10.4 & 12.6 & 3.4 & 5.4 & 48.1  & 354.6 \\ 
co10\tnote{d}    &   18:27:34.05  &  -11:45:26.1  &  54--62   & 135.9 & 2.5 & 57.4 &  6.5  & 17.2 & 11.7 & 4.3 & 4.7 & 42.9 & 313.5 \\   
co11\tnote{d} &   18:27:24.62  &  -11:46:40.3  &  59--63 & 172.2 & 3.0 & 59.8 &  5.0  &  16.3 & 11.7 & 4.3 & 4.7 & 42.9 & 313.5 \\    
co12\tnote{d}    &   18:27:23.49  &  -11:49:08.2  &  56--59     & 49.6  & 2.7 & 58.2 &  5.7  & 16.9 & 11.7 & 4.3 & 4.7 & 42.9 & 313.5 \\ 
co13              &   18:27:27.60  &  -11:53:19.2  &  56--60     & 90.3  & 3.7 & 57.6 &  4.1  & 16.3 & 11.7 & 4.3 & 4.7 & 42.9 & 313.5 \\   
co14\tnote{d}  &   18:28:18.23  &  -11:39:54.2  &  63--67     & 179.5 & 5.9 & 65.4 &  4.1  & 26.3 & 11.3 & 4.7 & 4.3 & 39.9 & 289.9 \\ 
co15             &   18:27:35.66  &  -11:50:22.0  &  57--59     & 83.1  & 3.6 & 57.9 &  2.4  & 9.7  & 11.7 & 4.3 & 4.7 & 42.9 & 313.5 \\ 
co16             &   18:27:41.68  &  -11:47:39.9  &  65--67     & 142.2 & 4.4 & 65.8 &  1.7  &  9.1 & 11.3 & 4.7 & 4.3 & 39.9 & 289.9 \\ 
co17\tnote{d}   &   18:28:30.14  &  -11:41:02.6  &  69--72     & 217.2 & 4.1 & 69.1 &  3.4  & 15.5  & 11.3 & 4.7 & 4.3 & 39.9 & 289.9 \\
co18             &   18:28:02.29  &  -11:54:25.9  &  69--73     & 99.4  & 2.5 & 71.0 &  4.5  &  12.3& 11.3 & 4.7 & 4.3 & 39.9 & 289.9 \\ 
co19\tnote{c}    &   18:27:28.11  &  -12:02:32.0  &  59.5--61   & 61.5  & 2.6 & 60.2 &  1.4  &  4.2 & 11.7 & 4.3 & 4.7 & 42.9 & 313.5 \\ 
\hline
\end{tabular}
\begin{tablenotes}
\item[] NOTES. -- Col.(1): Clump. Col.(2)-(3): Peak position of the emission. Col.(4): Velocity range of the clump emission. Col.(5): Deconvolved angular diameter of the clump. Col.(6): Peak T$_{\rm mb}$ of the spectrum. Col.(7): Peak velocity of the spectrum. Col.(8): Deconvolved FWHM line-width of the spectrum. Col.(9): Integrated intensity emission from the FWHM contour of the clump. Col.(10):  Far distance of the clump. Col.(11): Near distance of the clump. Col.(12): Galactocentric distance of the clump. Col.(13): Carbon isotope ratio. Col.(14): Oxygen isotope ratio.
\item[a] Deconvolved for the beam size.
\item[b] Deconvolved for the velocity resolution.
\item[c] The parameters are derived from a high velocity resolution resample (0.5~km/s).
\item[d] The physical parameters are derived from the BU-FCRAO GRS data, because the clump is at the edge of our map.
\item[e] The carbon and oxygen isotope ratios have been obtained using: ${^{12}\rm C}/{^{13}\rm C} = 7.5\,D_{\rm GC}+7.6$ and ${^{16}\rm O}/{^{18}\rm O} = 58.8\,D_{\rm GC}+37.1$ (\citealt{wilson1994}), where $D_{\rm GC}$ is the galactocentric distance (col.~11).
\end{tablenotes}
\end{threeparttable}
\end{sidewaystable*}

As explained above, we computed both the near and far
distances for each of the four velocity intervals of the emission,
corresponding to the four main molecular clouds in the whole region. Therefore,
assuming that each of the four molecular clouds is made from material at the
same distance, which is reasonable given the localized morphology of the
emission, each clump is assumed to be at the same distance as the cloud it
belongs to (see Table~\ref{param_nubi}).
This leaves the question of the ``near-far ambiguity'', which we discuss 
in Sect.~\ref{starformationassociation}.

The spectrum of the emission of each identified clump was obtained by
integrating the $^{13}$CO(1-0) data cube in the channels of the emission of the clump, 
over the area 
enclosed by the deconvolved 50\% contour level of the
$^{13}$CO(1-0) emission.  The spectra of clump~co4 and clump~co19 have been derived
from the higher velocity resolution (0.5~km~s$^{-1}$) data-cube, as discussed
in the previous section.  The parameters of each clump were determined by
fitting a Gaussian profile to each produced spectrum.  Line profiles
showing more than one velocity component were analyzed by fitting more than one
Gaussian component, in order to remove the contribution to the emission by
other clumps along the line of sight from the emission coming from the
clump of interest.  The results of this analysis are reported in
Table~\ref{param_nubi}.

\subsection{The continuum emission}\label{continuum-identif}

\begin{figure*}
\centering
\includegraphics[angle=-90,scale=0.7]{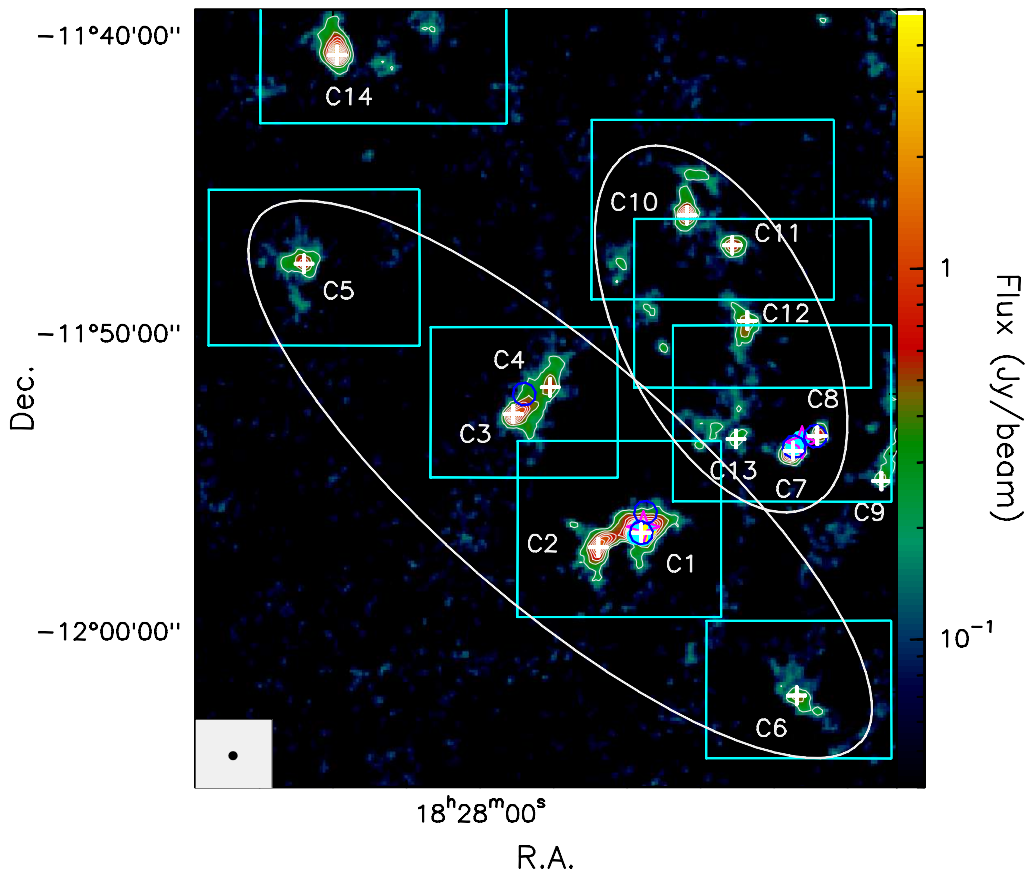}
\caption{APEX 870~$\mu$m continuum emission.  The contour levels are from
5~$\sigma$ ($\sigma = 40$~mJy~beam$^{-1}$) in steps of 5~$\sigma$. The label of
each identified source is indicated and  the positions of each emission peak is
marked with a white cross. Symbols are as in Fig.~\ref{integr_tot} 
(see Table~\ref{SIMBAD}).  The cyan boxes highlight the fields of
Fig.~\ref{CO-apex-spitzer}.  The white ellipses represent: 1) the GMC
surrounding G19.61-0.23, which is the 33--48 km~s$^{-1}$ molecular gas (cloud~2)
discussed in Sect.~\ref{kinematics}; and 2) the 54--63 km~s$^{-1}$ molecular gas
(cloud~3) discussed in Sect.~\ref{kinematics} (see also
Fig.~\ref{integr_tot}).}
\label{identif-cont}
\end{figure*}

Figure \ref{identif-cont} shows the APEX 870~$\mu$m continuum emission from the
same region we observed in the $^{13}$CO(1-0) emission line (see
Fig.~\ref{integr_tot}).  
The data are part of the ATLASGAL project (\citealt{schuller2009}).
The white ellipses represent: 1) the GMC surrounding
G19.61-0.23, which is the 33--48 km~s$^{-1}$ molecular gas (cloud~2) discussed
in Sect.~\ref{kinematics}; and 2) the 54--63 km~s$^{-1}$ molecular gas (cloud~3)
discussed in Sect.~\ref{kinematics} (see also Fig.~\ref{integr_tot}).  
We decided to use a threshold of 10~$\sigma$ to
identify the different sources in the continuum emission.  In this way, we
identified in the APEX continuum emission 14 sources, which are shown in
Fig.~\ref{identif-cont} with their respective labels.

Most of the APEX continuum sources have counterparts in one of the FCRAO
$^{13}$CO(1-0) clumps (see also Fig.~\ref{CO-apex-spitzer} in the Online
Material Sect.), with the exception of sources C8 and C9.
Source C8 is associated with significant emission in the $^{13}$CO(1-0) line,
but over a region slightly to the south of C8 (see Fig.~\ref{CO-apex-spitzer}). 
The $^{13}$CO(1-0) emission in this case corresponds to $^{13}$CO clump~co7, which 
``contains'' the APEX sources C8 and C7. We thus assume for both C8 and C7 
the distance corresponding to $^{13}$CO~8.
C9 is associated with $^{13}$CO emission at 62~km~s$^{-1}$ and hence probably to cloud~3.

\begin{table*}
\centering 
\caption{Physical parameters of the sources identified in the APEX 870~$\mu$m continuum emission.}\label{param_cont}
\smallskip 
\begin{threeparttable}
\renewcommand{\footnoterule}{} 
\begin{tabular}{l c c c r r r r r r r} 
\hline\hline  
Source & \multicolumn{2}{c}{Peak Position} &  $\Theta_{\rm C}$\tnote{a} &  F$_{peak}$ & v$_{^{13}{\rm CO}}$ & Flux & Counterpart & $d_{\rm FAR}$ & $d_{\rm NEAR}$ & $M_{\rm cont}$\tnote{b}\\ 
&     R.A.[J2000.0] & Dec.[J2000.0] &      (arcsec)  &  (Jy/beam)  & (km~s$^{-1}$) & (Jy) &&(kpc) & (kpc) & ($\times 10^2\, M_{\odot}$) \\
(1)&(2)&(3)&(4)&(5)&(6)&(7)&(8)&(9)&(10)&(11)\\
\hline
C1  &  18:27:38.05 &  -11:56:38.3 &  15.2 &  16.3  &  42.9    &  14.1     &    co1      &       12.6      &        3.4   & 203.4  \\ 
C2  &  18:27:43.88 &  -11:57:07.9 &  36.9 &   1.4  &  42.8    &   3.6	  &    co2      &       12.6      &        3.4   & 52.5   \\   
C3  &  18:27:55.43 &  -11:52:39.7 &  30.6 &   1.6  &  41.0    &   2.9	  &    co3      &       12.6      &        3.4   & 42.1   \\   
C4  &  18:27:50.42 &  -11:51:45.7 &  38.1 &   0.7  &  41.0    &   1.7	  &    co3      &       12.6      &        3.4   & 23.8   \\   
C5  &  18:28:23.88 &  -11:47:36.8 &  24.7 &   0.9  &  44.7    &   1.2	  &    co5      &       12.6      &        3.4   & 17.1   \\   
C6  &  18:27:16.91 &  -12:02:06.3 &  32.8 &   0.5  &  38.7    &   1.1	  &    co6      &       12.6      &        3.4   & 15.2   \\   
C7  &  18:27:17.36 &  -11:53:53.7 &  21.6 &   2.2  &  59.0    &   2.5	  &    co7      &       11.7      &        4.3   & 31.5   \\   
C8  &  18:27:14.08 &  -11:53:23.5 &  20.5 &   0.8  &  57.7    &   0.9	  &    -        &       11.7      &        4.3   & 11.1   \\      
C9  &  18:27:05.36 &  -11:54:54.3 &  29.0 &   0.4  &  62.5    &   0.8	  &    -        &       11.6      &        4.4   & 9.5    \\       
C10 &  18:27:31.77 &  -11:45:59.5 &  20.5 &   1.5  &  57.7    &   1.7	  &    co10     &       11.7      &        4.3   & 21.2   \\ 
C11 &  18:27:25.62 &  -11:46:59.9 &  22.7 &   0.9  &  59.8    &   1.1	  &    co11     &       11.7      &        4.3   & 13.5   \\ 
C12 &  18:27:23.58 &  -11:49:31.7 &  45.4 &   0.5  &  58.2    &   1.9	  &    co12     &       11.7      &        4.3   & 24.1   \\ 
C13 &  18:27:25.13 &  -11:53:29.9 &  25.6 &   0.4  &  57.6    &   0.6     &    co13     &       11.7      &        4.3   & 7.3    \\   
C14 &  18:28:19.32 &  -11:40:37.2 &  27.4 &   2.4  &  65.4    &   3.8     &    co14     &       11.3      &        4.7   & 7.6    \\   
\hline
\end{tabular}
\begin{tablenotes}
\item[] NOTES. -- Col.(1): Sources. Col.(2)-(3): Peak positions of the emission. 
Col.(4): Deconvolved angular diameters of the sources. Col.(5): Peak flux densities of the sources. Col.(6): Velocity of the sources from the association with the FCRAO $^{13}$CO(1-0) emission. Col.(7): Integrated emission from the FWHM contour of the sources. Col.(8): FCRAO $^{13}$CO(1-0) counterpart. Col.(9): Far distance of the source from the $^{13}$CO(1-0) emission (see Table~\ref{param_nubi}). Col.(10): Near distance of the source from the $^{13}$CO(1-0) emission (see Table~\ref{param_nubi}). Col.(11): Mass from the continuum emission of the sources, assuming $T=20$~K.
\item[a] Deconvolved.
\item[b] All the sources are assumed to be at the far kinematic distances, except C14 (see text).
\end{tablenotes}
\end{threeparttable}
\end{table*}

Moreover, given the lower resolution of the $^{13}$CO data, the continuum sources 
C3 and C4 correspond both to clump~co3 in the FCRAO $^{13}$CO(1-0) emission.
Column~8 of Table~\ref{param_cont} indicates the counterpart, if any, of each APEX continuum source, 
as identified from the comparison between the $^{13}$CO emission and the APEX continuum emission.
It is worth noting that the two maps ($^{13}$CO map and APEX continuum maps) 
have significantly different resolutions, with the APEX resolution 
being $18.\!\!^{\prime\prime}2$ at 870~$\mu$m and the FCRAO resolution being 46$^{\prime\prime}$ at the 
frequency of the $^{13}$CO(1-0) line.
Therefore it is not surprising that the sources identified in the APEX continuum emission are more compact 
than the $^{13}$CO clumps (as seen also in Fig.~\ref{CO-apex-spitzer}). 
Moreover, the 870~$\mu$m continuum emission 
probably traces 
dense cores embedded in the $^{13}$CO(1-0) clumps.

For the APEX continuum sources that have a counterpart in the FCRAO $^{13}$CO(1-0) emission, we
assume as distance the one of the corresponding $^{13}$CO(1-0) clump. The obtained distances 
are reported in col.~9-10 of Table~\ref{param_cont}
(see Sect.~\ref{identif} and Table~\ref{param_nubi}).

\section{Association with infrared emission}\label{starformationassociation}

One of our aims 
is to compare the properties of the molecular
clumps with and without star formation within them. With this in mind, we have
compared images  from the GLIMPSE (\citealt{benjamin2003}) and MIPSGAL mid
infrared surveys (\citealt{carey2005}) with both ATLASGAL maps and our FCRAO
$^{13}$CO data (supplemented by the BU-FCRAO GRS).  It is worth recalling that
the MIPSGAL 70~$\mu$m survey has a ``beam'' of $18^{\prime\prime}$,  which is 
comparable to that of ATLASGAL. Moreover, the GLIMPSE 8 $\mu $m data
traces PAH emission excited by UV from OB stars close to GMCs whereas the 24
$\mu$m MIPSGAL radiation often traces dust heated by embedded 
proto-stellar objects. Also, the
4.5~$\mu$m GLIMPSE data has been found often to trace molecular hydrogen
emission associated with outflows.  In Fig.~\ref{CO-apex-spitzer} (online
version), we superpose FCRAO $^{13}$CO and ATLASGAL maps to Spitzer images at
3.6, 8 and 24~$\mu$m.

One sees here that there are several
ATLASGAL sources associated with  strong continuum emission
in the Spitzer bands. Table~\ref{SFassociations} summarizes these associations
(within 1$^{\prime}$) as well as the information about maser emission and
HII regions close to the positions of continuum emission.
\begin{table}
\centering 
\caption{Association (within 1$^{\prime}$) of the APEX 870~$\mu$m sources with 
active star-formation tracers, such as Spitzer 3.6~$\mu$m, 8~$\mu$m and 24~$\mu$m; water, 
methanol and OH masers; HII regions; and IRAS sources.}\label{SFassociations}
\smallskip 
\begin{threeparttable}
\renewcommand{\footnoterule}{} 
\begin{tabular}{l | c c c | c c c | c | c} 
\hline\hline  
APEX & \multicolumn{3}{|c|}{Spitzer} & \multicolumn{3}{|c|}{Masers} & HII & IRAS \\   
& 3.6 & 8 & 24 & H$_2$O & OH & CH$_3$OH & &\\   
(1)&(2)&(3)&(4)&(5)&(6)&(7)&(8)&(9)\\
\hline
C1       & {\bf +} & {\bf +} & {\bf +} & {\bf +} & {\bf +} & {\bf +} & {\bf +} & {\bf +}  \\  
C2 	 & {\bf +} & {\bf +} & {\bf +} & --  & --  & --  & --  & --  \\
C3 	 & {\bf +} & {\bf +} & {\bf +} & --  & --  & {\bf +} & --  & {\bf +}  \\  
C4 	 & {\bf +} & {\bf +} & {\bf +} & --  & --  & {\bf +} & --  & --  \\
C5 	 & {\bf +} & {\bf +} & {\bf +} & --  & --  & --  & --  & --  \\
C6 	 & {\bf +} & {\bf +} & {\bf +} & --  & --  & --  & --  & {\bf +}  \\  
C7 	 & {\bf +} & {\bf +} & {\bf +} & --  & {\bf +} & {\bf +} & {\bf +} & {\bf +} \\ 
C8 	 & {\bf +} & {\bf +}  & {\bf +} & --  & --  & {\bf +} & --  & --  \\
C9 	 & {\bf +} & {\bf +} & {\bf +} & --  & --  & --  & --  & --  \\
C10	 & {\bf +} & {\bf +} & {\bf +} & --  & --  & --  & --  & {\bf +} \\ 
C11	 & {\bf +} & {\bf +} & {\bf +} & --  & --  & --  & --  & --  \\
C12	 & {\bf +} & {\bf +} & {\bf +} & --  & --  & --  & {\bf +}  & {\bf +}  \\ 
C13	 & {\bf +} & {\bf +} & {\bf +} & --  & --  & --  & --  & --  \\    
C14	 & {\bf +} & {\bf +} & {\bf +} & --  & --  & --  & --  & {\bf +}  \\ 
\hline
\end{tabular}
\begin{tablenotes}
\item[] NOTES. -- The ``{\bf +}'' symbol indicates that the association with the tracer has been observed; while the ``--'' symbol indicates that it has not been observed.
\end{tablenotes}
\end{threeparttable}
\end{table}
Not surprisingly,  there is strong mid infrared emission  from the vicinity of
clumps C1 and C2 which are associated with the HII region complex G19.61-0.23
but one also notes strong emission associated with the C7/C8 complex and with
C12.  In all of these cases, it is reasonable to assume that there is an
embedded cluster of young stars producing ultra--violet radiation responsible
for exciting the PAH and small grain emission observed at 8 and 24~$\mu$m. Less
obvious in Fig.~\ref{CO-apex-spitzer} is the fact that in many cases there are
point-like ($<$ 6 arc sec.) continuum sources at 24 $\mu$m close to the
ATLASGAL 870 micron peaks.
It is noticeable that there are 3 ATLASGAL
sources  without clear 24~$\mu$m counterparts and we presume this implies a
relatively low dust temperature (below 25~K). These are perhaps similar to the
infrared dark clouds (IRDCs) observed associated with star-forming regions
closer to the sun but lacking a strong infrared background.  We note also that
we have searched without success for extended emission in the 4.5~$\mu$m IRAC
band of the type often found associated with outflows in nearby star-forming
regions.  Finally, all the ATLASGAL sources show association with extended
Spitzer 8~$\mu$m emission, except source C14, which corresponds to the 
$^{13}$CO clump~co14, and the $^{13}$CO clump~co17 (see Table~\ref{param_nubi}). 
Both clumps correspond
to IRDCs seen against the 8~$\mu$m emission and identified by \cite{simon2006a,
simon2006b}.

The above information suggests to us that all the ATLASGAL clumps but C14 
are at the ``far'' (around 12 kpc) distance rather
than the near (around 4 kpc). This is likely to be true for the
clumps with velocity around 42 km~s$^{-1}$ (cloud 2 in the nomenclature of
Table~\ref{cloud_param}), as shown by \cite{kolpak2003} and confirmed by 
\cite{anderson2009}. We therefore use the
far distance as a working hypothesis in what follows for all clumps, with the
exception of continuum emission clump C14 and clumps~co14 and co17, seen in 
$^{13}$CO(1-0) emission, for which we used the near distance.

\begin{figure}
\centering
\includegraphics[width=\columnwidth]{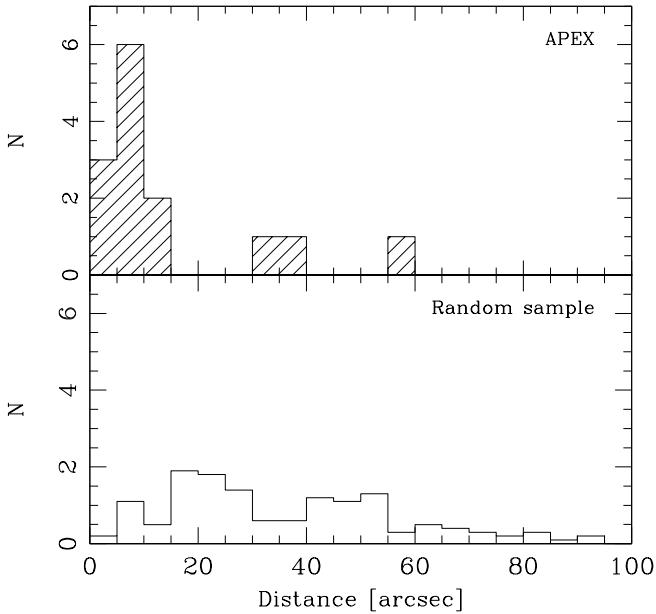}
\caption{\emph{Top}: Histogram of the distances of the APEX 870~$\mu$m
continuum sources from the closest Spitzer 24$\mu$m sources. \emph{Bottom}:
Histogram of the average distribution of the distance from the closest Spitzer
24~$\mu$m source for ten random samples of positions within the whole
$23^{\prime}\times23^{\prime}$ region (see Fig.~\ref{integr_tot}).}
\label{histo}
\end{figure}

We demonstrate the association between ATLASGAL sources and
24~$\mu$m MIPSGAL sources in Fig.~\ref{histo} where we plot a histogram
(upper panel) of the angular separations  between the 870~$\mu$m
peak and the nearest 24 $\mu$m source. 
To estimate the reliability of the associations of the infrared 
sources with the millimeter continuum cores, we have simulated randomly
located samples of 14 cores and associated them with the closest infrared
source. The bottom panel of Fig.~\ref{histo} shows the
distribution of the average separation between the infrared sources and the
random sample of millimeter cores.
One sees that the histogram for the real millimeter cores has
a strong peak for separations of less than 10$^{\prime\prime}$
(roughly half the APEX beam) which is not present for the random samples.
A statistical test on the two histograms shows that there is only a 
$\sim$0.1\%\ probability that they are drawn from the same parent
distribution. Indeed, the
simulation shows that one roughly expects 2 chance coincidence within
15$^{\prime\prime}$ whereas there are 11 APEX sources within
15$^{\prime\prime}$ of the nearest Spitzer 24~$\mu$m source.
We thus conclude that, with the exception of the three sources with separations
larger than 30$^{\prime\prime}$, the associations of the ATLASGAL
870~$\mu$m cores with their neighboring Spitzer 24~$\mu$m sources are real.

\section{Physical parameters of the clumps}\label{section:parameter}

\subsection{The $^{13}$CO(1-0) emission}\label{parameters}

\begin{figure}
\centering
\includegraphics[width=0.75\columnwidth]{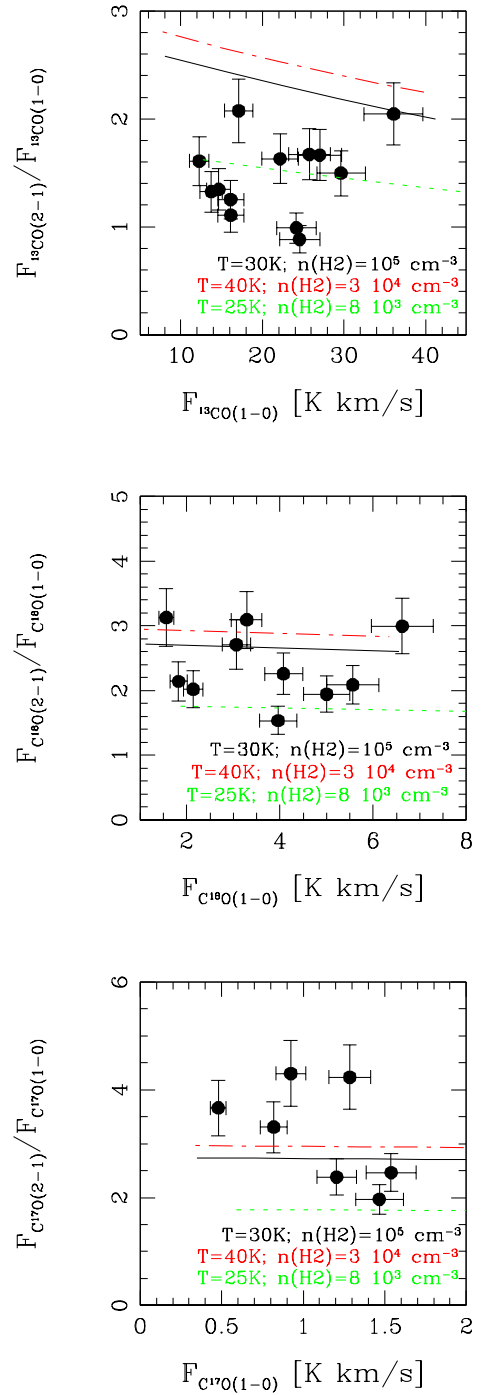}
\caption{Integrated (2-1)/(1-0) line ratios over the central 
$4.\!\!^{\prime}7 \times 3.\!\!^{\prime}6$
region as a function of the 
integrated line intensities, after smoothing the 30-m maps to the angular 
resolution of the FCRAO maps (46$^{\prime\prime}$). 
The points in the plots correspond to the values of the line ratios taken over 46$^{\prime\prime}$ beams.
The curves are the 
theoretical curves from LVG models, correspondent to: $T=30$~K and 
n(H$_2$)=$10^5$~cm$^{-3}$ (black solid curve); $T=40$~K and 
n(H$_2$)=$3\,10^4$~cm$^{-3}$ (red dot-dashed curve); and 
$T=25$~K and n(H$_2$)=$8\,10^3$~cm$^{-3}$ (green dashed curve).
\emph{Top}: ratio between IRAM $^{13}$CO(2-1) emission and FCRAO
$^{13}$CO(1-0) emission 
against the FCRAO
$^{13}$CO(1-0) emission. \emph{Middle}: ratio between C$^{18}$O(2-1)
emission and C$^{18}$O(1-0) emission 
against the C$^{18}$O(1-0) emission. 
\emph{Bottom}: ratio between C$^{17}$O(2-1)
emission and C$^{17}$O(1-0) emission 
against the C$^{17}$O(1-0) emission.}
\label{fig:opt-thin}
\end{figure}

The central $5.\!\!^{\prime}2 \times 5.\!\!^{\prime}2$ is the only
region where we have observations of the $J=1-0$ and the $J=2-1$ transitions of
the three isotopologues and therefore the only region where we can derive the
optical depth of the $^{13}$CO(1-0) line.  The optical depth, 
$\tau_{\rm ^{13}CO}$, of the $^{13}$CO(1-0) transition can be estimated from the line
ratio of the two isotopes, $^{13}$CO(1-0) and C$^{18}$O(1-0). We derive
values for the mean line optical depth of $\le 0.5$ suggesting 
that while the line peak may be moderately optically thick 
($\tau_{\rm ^{13}CO} \sim 1.4-1.6$ at most),
the integrated $^{13}$CO(1-0) emission is optically thin over the 
central region. This is also consistent with the observed lack of variation 
of the integrated (2-1)/(1-0) line ratios,
as a function of the integrated line intensities (Fig.~\ref{fig:opt-thin}).
The points in the plots correspond to the values of the line ratios taken over 46$^{\prime\prime}$ beams.

Given that the $^{13}$CO emission at the peak of the line is close to 
be optically thick, as an estimate of the excitation temperature of the 
molecular gas we assumed 20~K, the peak line brightness temperature as
measured in the clump with the highest optical depth.
In the following we have assumed this excitation temperature for all 
the clumps.

The total column density of the $^{13}$CO molecule for each identified clump
was derived using the $J=1-0$ transition, under the assumption of local
thermodynamic equilibrium (LTE) at an excitation temperature $T_{ex}$. 
The column density of this molecule,
in the optically thin limit, is given by
\begin{equation}
N_{^{13}{\rm CO}} = A  \, \frac{T_{ex}+(h\,B_{\rm ^{13}CO}/3\,k)}{e^{-h\nu_{\rm ^{13}CO}/k\,T_{ex}}}  \,\int T_{\rm B} {\rm dv} \,\,\,\,\,\,\,\,\,\,\,\,\,\,\,\,\,\,\,\,\mathrm{[cm^{-2}]} \\
\label{coldens}
\end{equation}
where:
\begin{equation}
A = 10^5\,\frac{3\,k^2}{8\,h\,\pi^3\,B_{\rm ^{13}CO}\,\mu_{\rm ^{13}CO}^2\,\nu_{\rm ^{13}CO}}
\end{equation}
(see Eq. [A1] and [A4] of \citealt{scoville1986}).
$\int T_{\rm B} {\rm dv}$ is the
integrated line brightness temperature in K~km~s$^{-1}$ of the transition with frequency
$\nu_{\rm ^{13}CO}$ (Hz), $B_{\rm ^{13}CO}$ is the rotational constant of the
molecule and $\mu_{\rm ^{13}CO}$ is $^{13}$CO's permanent electric dipole moment, which
is taken to be 0.1101 Debye.  Assuming $T_{ex} = 20$~K for every clump, 
we computed the total column density of each clump identified
in the FCRAO $^{13}$CO(1-0) emission.  The results are shown in col.~2 of
Table~\ref{mass_nubi}.

The total LTE mass of gas in the identified clumps, $M_{\rm LTE}$, can be computed from the $^{13}$CO column density 
as follows:
\begin{equation}
M_{\rm LTE} = N_{\rm ^{13}CO}\,\frac{[\rm H_2]}{[\rm ^{13}CO]}\,\mu_{\rm G}\,m_{\rm H}\,\frac{\pi\,\Theta_{\rm S}^2}{4}\,d^2
\label{LTEmass}
\end{equation}
(\citealt{scoville1986}), where [H$_2$]/[$^{13}$CO] is the abundance ratio of
molecular hydrogen to $^{13}$CO, $\mu_{\rm G} = 2.72$ is the mean molecular
weight of the gas, $m_{\rm H}$ is the mass of the hydrogen atom, $\Theta_{\rm
S}$ is the angular diameter of each clump (deconvolved FWHM, see col.~5 of
Table \ref{param_nubi}) and $d$ is the distance of the source.  Adopting an
abundance ratio [H$_2$]/[$^{12}$CO$] = 10^{4}$ (e.g. \citealt{scoville1986}),
[H$_2$]/[$^{13}$CO]=([H$_2$]/[$^{12}$CO])$\times$([$^{12}$C]/[$^{13}$C]) can be
computed for each clump, using the values of [$^{12}$C]/[$^{13}$C] presented in
col.~13 of Table~\ref{param_nubi}.  We can thus derive the gas mass of each
clump (col.~3 of Table \ref{mass_nubi}). 
\begin{table*}
\centering 
\caption{Masses and surface densities of the clumps identified in the FCRAO $^{13}$CO(1-0) emission.}\label{mass_nubi}
\smallskip 
\begin{threeparttable}
\renewcommand{\footnoterule}{}
\begin{tabular}{l r r r r r l r} 
\hline\hline  
Clump &  $N_{^{13}{\rm CO}}$ & $M_{\rm LTE}$\tnote{c}  & $\Sigma_{\rm LTE}$ & n$_{\rm LTE}$\tnote{c} & $M_{\rm VIR}$\tnote{c} & $\Sigma_{\rm VIR}$\tnote{c}  & n$_{\rm VIR}$\tnote{c} \\   
& ($\times 10^{15}$~cm$^{-2}$) &  ($\times 10^2\, M_{\odot}$) & (g~cm$^{-2}$) & ($\times 10^2$~cm$^{-3}$) & ($\times 10^2\, M_{\odot}$) &  (g~cm$^{-2}$) & ($\times 10^2$~cm$^{-3}$)\\
(1)&(2)&(3)&(4)&(5)&(6)&(7)&(8)\\
\hline
co1              &  47.5   &  34.6  &  0.10      &    37.2    &  164.9 & 0.49   &    177.1        \\
co2              &  37.9   &  52.4  &  0.08      &    21.5     &  128.9 & 0.20   &    52.9           \\ 
co3              &  34.5   & 192.3  &  0.08      &    9.8      &  230.9 & 0.09   &    11.7          \\
co4\tnote{a}     &   4.9   &  10.8  &  0.01      &    2.6      &  3.2   & 0.003  &    0.8           \\
co5\tnote{b}     &  25.4   & 117.6  &  0.06      &    7.9     &  20.3  & 0.01   &    1.4           \\ 
co6              &  27.3   &  63.8  &  0.06      &    11.9     &  381.0 & 0.35   &    71.4           \\
co7              &  29.1   &  25.2  &  0.06      &    17.5    &  83.5  & 0.19   &    58.0          \\
co8             &  21.3   &  16.9  &  0.05      &    16.0      &  89.9  & 0.25   &    85.2          \\
co9             &  13.2   &  11.3  &  0.03      &    9.5     &  41.1  & 0.10   &    34.7           \\
co10\tnote{b}    &  22.0   &  94.9  &  0.04      &    5.9    &  342.7 & 0.15   &    21.4           \\ 
co11\tnote{b} &  20.8   & 144.3  &  0.04      &    4.4     &  251.6 & 0.07   &    7.7             \\
co12\tnote{b}    &  21.5   &  12.4  &  0.04      &    15.9     &  96.0  & 0.32   &    123.4          \\
co13              &  20.7   &  39.5  &  0.04      &    8.4      &  88.4  & 0.09   &    18.9          \\
co14\tnote{b}  &  33.6   &  38.0  &  0.06      &    15.9     &  71.6 & 0.11   &    30.0              \\ 
co15             &  12.3   &  19.9  &  0.02      &    5.5     &  27.6  & 0.03   &    7.5            \\
co16             &  11.6   &  47.7  &  0.02      &    2.9    &  22.4  & 0.01   &    1.4             \\
co17\tnote{b}   &  19.7   &  32.6  &  0.04      &    7.7     &  59.5 & 0.06   &    14.0            \\
co18             &  15.7   &  31.5  &  0.03      &    5.6     &  117.2 & 0.11   &    20.8            \\
co19\tnote{a}    &   5.3   &   4.7  &  0.01      &    3.2    &  7.6   & 0.02   &    5.1            \\
\hline
\end{tabular}
\begin{tablenotes}
\item[] NOTES. -- Col.(1): Clump. Col.(2)-(4): Column density, LTE mass and surface density for T$_{ex} = 20$~K. Col.(5) Density from the LTE mass. Col.(6)-(7): Virial mass and surface density. Col.(8) Density from the virial mass.
\item[a] The parameters are derived from a high velocity resolution resample (0.5~km/s).
\item[b] The values are derived from the BU-FCRAO GRS data, because the clump is at the edge of our map.
\item[c] All the clumps are assumed to be at the far kinematic distances, except clump~co14 and clump~co17.
\end{tablenotes}
\end{threeparttable}
\end{table*}

The estimated gas masses can be compared with 
the masses computed under the assumption of virial equilibrium for an 
homogeneous 
sphere, neglecting contributions from the magnetic field and surface pressure:
\begin{equation}
M_{\rm VIR}=0.509\,d(\rm kpc)\,\Theta_{\rm S}(\rm arcsec)\,\Delta{\rm v_{1/2}}^2(\rm km\,s^{-1})
\end{equation}
(e.g. MacLaren et al. 1988), where $\Delta{\rm v_{1/2}}$ is the $^{13}$CO(1-0)
line width in kilometers per second and $\Theta_{\rm S}$ is the angular
diameter of each clump in arcsec within the 50\% intensity contour of the
integrated $^{13}$CO(1-0) emission (see col.~5 of Table \ref{param_nubi}).
The derived virial masses are presented in col.~6 of Table~\ref{mass_nubi}.

We also computed the surface density, $\Sigma$=M/($\pi$R$^2$), of each clump,
using both the LTE masses and the virial masses and they are given in col.~4
and col.~7 of Table~\ref{mass_nubi}, respectively.
The surface density is a measure of pressure for virialized clouds, since 
the gas pressure needed to support a virialized cloud 
against gravity in the absence of other forces is of the order $G\,\Sigma^2$, 
independent of the shape of the cloud \citep[e.g.][]{mckee1993,bertoldi1992}.
From Table~\ref{mass_nubi}, one sees that the surface densities derived from
$^{13}$CO ($\Sigma _{\rm LTE}$), which are essentially distance independent, vary
in the range 0.01 to 0.1~g~cm$^{-2}$, which is roughly equivalent to a range of
visual extinction 2.5 to 25 magnitudes for a local ISM dust-to-gas ratio. This
is comparable to the surface density averaged over the whole cloud~2 GMC (see
Table~\ref{param-13co63TOTspec} and Sect.~\ref{comparison_extragal}) and so
these clumps are mostly at pressures of the order typically experienced in the
GMC. The masses range from $3\,10^2$ to $2 \, 10^4 \, M_{\odot}$ and thus overlap with the
range studied by \citet{williams1998} in the Rosette cloud and in G216-2.5.

For each clump we also
derived an average density, using the deconvolved half power sizes we
determined from the $^{13}$CO(1-0) emission (see col.~5 of Table~\ref{param_nubi}). The
values we obtained are reported in col.~5 and 8 of Table~\ref{mass_nubi},
respectively from the LTE masses and the virial masses, and are in the
range $10^2$-$10^4$~cm$^{-3}$.
These values can be compared with the results from LVG models, 
based on the approach of \citet[][spherically symmetric 
homogeneous model with linear dependence of $v$ upon $r$, collisional
rates from \citealt{flower1985}]{dejong1975}.
Trapping has been accounted for
using $N({\rm CO})/\Delta v$ as an independent variable where $N({\rm CO})$
can stand for any of the CO isotopologues of interest and we 
scale between them using the ratios given in Table~\ref{param_nubi}
(line-width assumed to be the observed value of 5~km~s$^{-1}$). In practice,
trapping corrections are only of importance for $^{13}$CO and we
find that the observed excitation, from the integrated (2-1)/(1-0) line 
ratios (see Fig.~\ref{fig:opt-thin}), is compatible with $T=30-40$~K and 
n(H$_2$)$=3\,10^4-10^5$~cm$^{-3}$ for the C$^{18}$O and C$^{17}$O lines and 
with $T=25-30$~K and n(H$_2$)$\lesssim 10^4$~cm$^{-3}$ 
for the $^{13}$CO line.
The expected densities from LVG models are thus slightly 
larger than the average densities reported in Table~\ref{mass_nubi}. 
However, this can be explained with a moderate amount of clumping in the region.

\subsubsection{The [$^{18}$O]/[$^{17}$O] ratio in G19.61-0.23}\label{18oSU17o}

Using the FCRAO $J=1-0$ and IRAM $J=2-1$ data of C$^{18}$O and
C$^{17}$O, we can derive the $^{18}$O/$^{17}$O isotopic ratio in the central
$5.\!\!^{\prime}2 \times 5.\!\!^{\prime}2$ region centered on G19.61-0.23.
Assuming that the two isotopologues have the same excitation temperature and
that C$^{18}$O is optically thin, we derive an average $^{18}$O/$^{17}$O
isotopic ratio of $4.2\pm0.9$ for clump~co1 and $3.9\pm0.9$ for clump~co2, from the
line integrated intensity ratio between the FCRAO $J=1-0$ transitions of C$^{18}$O and C$^{17}$O,
and of $3.0\pm0.7$ for clump~co1 and $3.0\pm0.8$ for clump~co2, from the line ratio
between the IRAM $J=2-1$ transitions of C$^{18}$O and C$^{17}$O.

The values that we find are
marginally consistent with the value discussed by
\cite{wilson1994}, $3.2\pm0.2$.  Moreover, we find a quite constant isotope ratio
over the central region, including clump~co1 and clump~co2, in both transitions.

\subsubsection{CO selective photodissociation}\label{selective_photodissociation}

\begin{figure*}
\centering
\includegraphics[width=1.95\columnwidth]{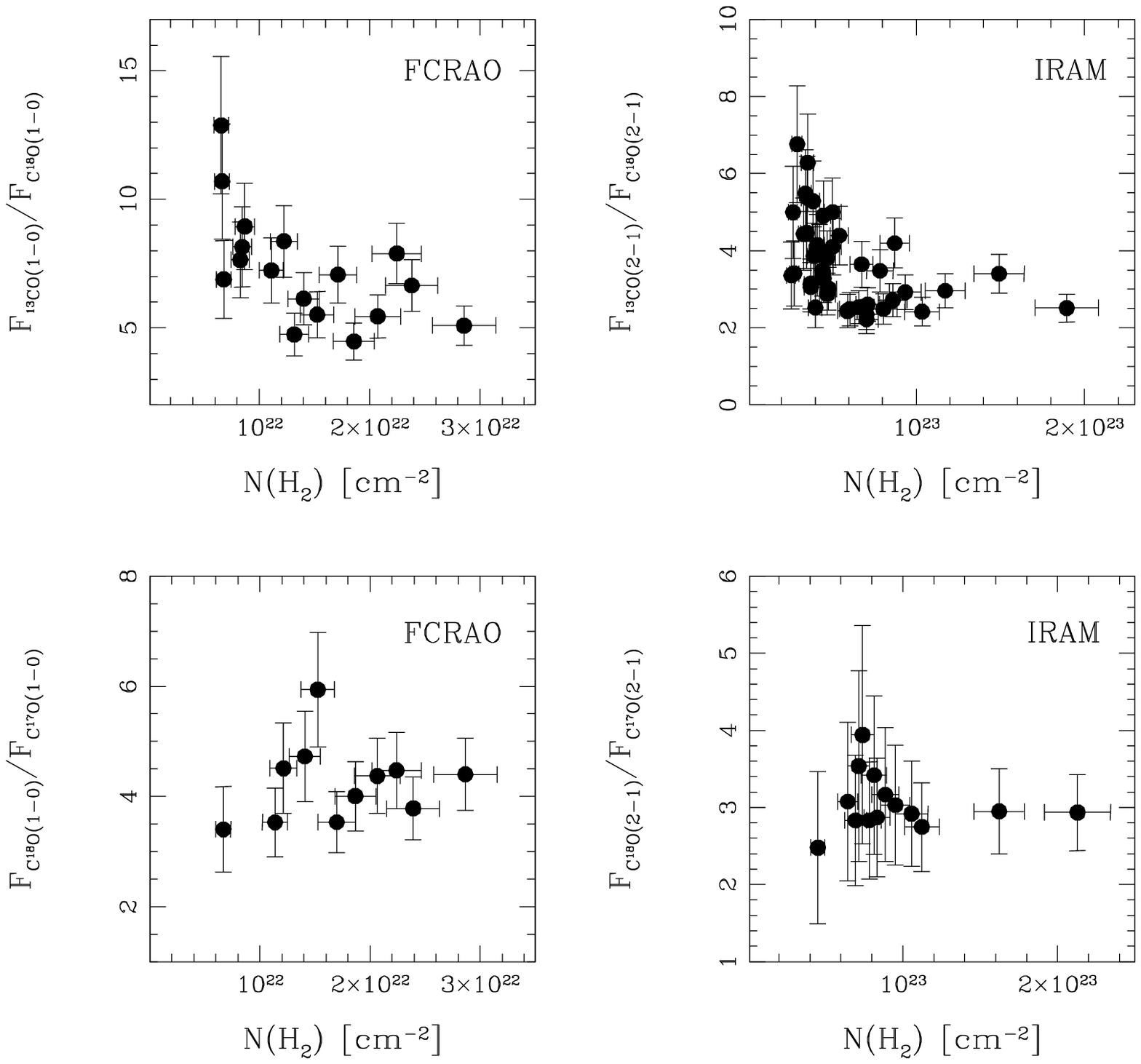}
\caption{
The points in the plots correspond to the values of the line ratios taken 
over every beam, in regions of the map with high enough S/N ratio 
(larger than 5).
\emph{Top left}: ratio between FCRAO $^{13}$CO(1-0) emission and
C$^{18}$O(1-0) against H$_2$ column density derived from the C$^{18}$O(1-0)
line intensity. \emph{Top right}: ratio between IRAM $^{13}$CO(2-1) emission
and C$^{18}$O(2-1) against H$_2$ column density derived from the C$^{18}$O(2-1)
line.  \emph{Bottom left}: ratio between FCRAO C$^{18}$O(1-0) emission and
C$^{17}$O(1-0), against H$_2$ column density derived from the C$^{18}$O(1-0)
line intensity.  \emph{Bottom right}: ratio between IRAM C$^{18}$O(2-1)
emission and C$^{17}$O(2-1) against H$_2$ column density derived from the
C$^{18}$O(2-1) line. 
}
\label{fig:photodissociation}
\end{figure*}

Figure~\ref{fig:photodissociation} shows the line ratios for the CO isotopes
for the central region of our map, as a function of the inferred H$_2$ column density. 
The points in the plots correspond to the values of the line ratios taken over every beam.
At high values of line intensity (high
column density), we do not see a variation of the line ratios. This is 
consistent with the fact that the integrated line intensities are mostly
optically thin. At low column densities, there is a tendency of an increase of
the $^{13}$CO(1-0)/C$^{18}$O(1-0) ratio, which is also seen in the ratio of
the (2-1) lines.  This trend is consistent with the expectations for selective
photodissociation at the edges of the clouds (see e.g.
\citealt{vandishoeck1988}).
The enhancement in the  $^{13}$CO/C$^{18}$O ratio which we find 
is roughly a factor 1.5 in the FCRAO $J=1-0$ data, for A$_{\rm V}\sim6$ and 
N(H$_2)\sim 10^{21}$~cm$^{-2}$. This can be compared 
with the results of \cite{vandishoeck1988} (see Table~7) 
for A$_{\rm V}=9$, $T=30$~K, n$_{\rm H}=2000$~cm$^{-3}$ and high UV field, which 
predict an enhancement of 1.7. We thus find that it is reasonable to assume that the 
trend that we find in the $^{13}$CO(1-0)/C$^{18}$O(1-0) ratio at 
low column densities is due to selective photodissociation.

\subsection{The continuum emission}\label{continuum-para}

To derive the angular extent of each identified source, we used the same method
as described in Sect.~\ref{identif} for the FCRAO $^{13}$CO(1-0) emission. 
The derived angular sizes, $\Theta_{\rm C}$, are shown in Table~\ref{param_cont} (col.~4).  
As already pointed out in Sect.~\ref{continuum-identif}, 
all the continuum sources that have a $^{13}$CO
counterpart (see column~8 of Table~\ref{param_cont}) have smaller sizes than
the respective $^{13}$CO clumps.  This 
is possibly indicating
that the continuum sources trace
the densest cores within the CO clumps.
The integrated flux densities of
the continuum sources are given in column~7 of Table~\ref{param_cont}.

Dust emission is generally optically thin in the sub-millimeter continuum.
Therefore, following \citet{hildebrand1983} and assuming constant gas-to-dust mass
ratio equal to 100, optically thin emission and isothermal conditions, the
total gas+dust mass of each source is directly proportional to the continuum
flux density integrated over the source:
\begin{equation}
 M_{\rm cont} = \frac{ F_{\nu}\, d^2}{\kappa_{\nu}\, B_{\nu}({\rm T})}
\label{gas-dust}
\end{equation}
where $F_{\nu}$ is the integrated flux, $\kappa_{\nu}$ is the dust absorption
coefficient per gram of gas, $B_{\nu}({\rm T})$ is the Planck function
calculated at the dust temperature $T$ and $d$ is the distance of each source.
Adopting a dust absorption coefficient $\kappa_{\nu}=\kappa_0\,(\nu /
\nu_0)^{\beta}$, with $\kappa_0 = 0.005$~cm$^2$~g$^{-1}$ at $\nu_0 = 230$~GHz
(e.g. \citealt{hildebrand1983}; \citealt{andre2000}) and $\beta = 2$
(\citealt{hildebrand1983}), we derive $\kappa_{\nu} = 0.0112$~cm$^2$~g$^{-1}$
at $\nu = 870$~$\mu$m.  For a dust temperature $T=20$~K, the derived masses for
each continuum source are given in Table~\ref{param_cont}, col.~11.  The
masses range from 700 to $2 \, 10^4 \, M_{\odot}$ and are similar to results for 
the $^{13}$CO, though the continuum sources are more compact.

\section{Stability of the clumps}\label{section:stability}

\begin{figure}
\centering
\includegraphics[width=\columnwidth]{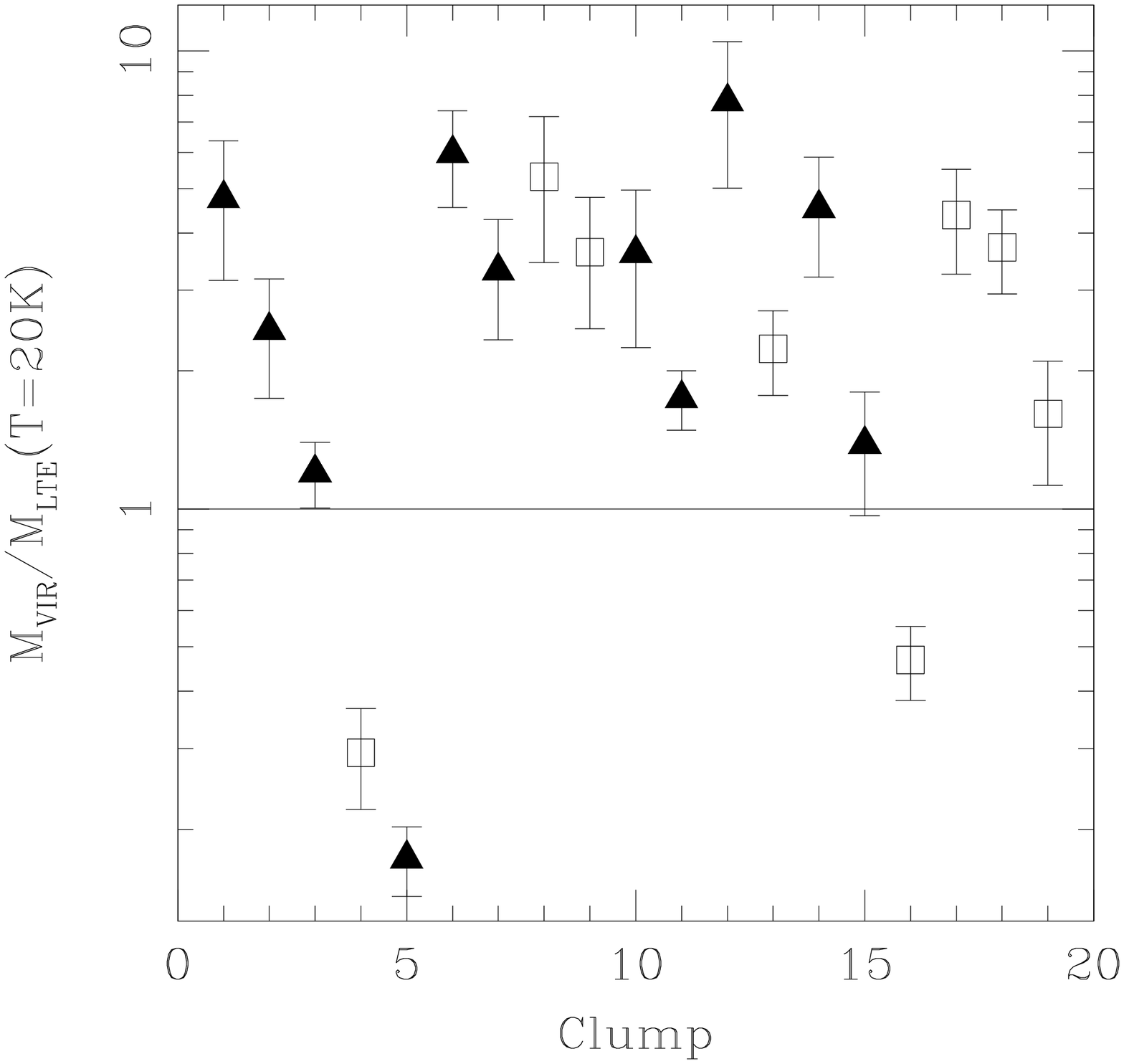}
\caption{Plot of the ratio between the mass derived assuming virial
equilibrium ($M_{\rm VIR}$) and the LTE mass for the each clump ($M_{\rm LTE}$).
The numbers on the x-axis correspond to the CO clumps, according to col.~1 of 
Table~\ref{param_nubi}.
Filled triangles indicate clumps associated with active star-formation tracers
(see Table~\ref{SFassociations}) and empty squares represent clumps which are
not associated with star-formation tracers.  The straight line corresponds to
$M_{\rm VIR}=M_{\rm LTE}$.}
\label{stability-plot}
\end{figure}

Figure~\ref{stability-plot} shows the ratio between the virial mass and the
$^{13}$CO LTE mass for each of the 19 clumps identified in the FCRAO
$^{13}$CO(1-0) emission, assuming the far distance for all clumps but
clump~co14.
The filled triangles indicate the clumps associated with active star-formation
tracers (see Table~\ref{SFassociations}) and the empty squares represent the clumps
without star-formation tracers.  The ratio is
larger than 1 for all the sources but 3 (clumps~co4, co5 and co16), which
indicates that the clumps range from gravitational unbound to 
unstable in few cases, with an average value for the ratio of about 3.
Most of the clumps thus seem to be 
transient molecular structures, as highlighted from the $^{13}$CO(1-0) 
analysis.
This result does not seem to depend on whether the clumps are 
associated with star formation (see however \citealt{williams1998}; \citealt{williams1995}).

We can compare this result with the study of \cite{fontani2002}, who observed the molecular 
clumps associated with 12 ultracompact (UC) HII regions in CH$_3$C$_2$H and found that the 
clumps are unstable against gravitational collapse (see also \citealt{cesaroni1991} 
and \citealt{hofner2000} for the C$^{34}$S and the C$^{17}$O lines). 
However, G19.61-0.23, also studied by \cite{fontani2002}, is one of only two sources with 
$M_{\rm VIR}/M_{\rm gas}$ ratio very close to one.
Therefore, in order to investigate the discrepancy between \cite{fontani2002} and us,
we repeated the analysis using the C$^{17}$O(2-1) from our IRAM-30m observations, 
for the two central clumps~co1 and co2. 
We find for both clumps a much smaller $M_{\rm VIR}/M_{\rm LTE}$ ratio, which
is $<1$ for clump~co2. We thus find evidence that the $M_{\rm VIR}/M_{\rm LTE}$ ratio changes 
with the tracer: the gas traced by the $^{13}$CO seems to be more virialized than the gas 
traced by the C$^{17}$O. 
Since C$^{17}$O (as well as C$^{34}$S and CH$_3$C$_2$H) traces higher density gas 
than $^{13}$CO, our result seems to indicate that the clumps are globally in equilibrium 
but locally unstable.
New observations of all the identified clumps in high density tracers are needed to further 
investigate this point.

It is worth to further discuss clump~co5, which corresponds to continuum 
emission source C5, an interesting case 
because of its small virial mass compared with its LTE mass. 
From Fig.~\ref{stability-plot}, the clump seems not to be in virial equilibrium against
gravitational collapse. Moreover,
it appears also to be associated with active star formation (see
Fig.~\ref{CO-apex-spitzer} and Table~\ref{SFassociations}), which excludes the
possibility that we overestimated the temperature in computing the LTE mass of
the clump, from Eq.(\ref{LTEmass}).  A possible explanation for the small ratio
$M_{\rm VIR}/M_{\rm LTE}$ of clump~co5 is that the magnetic field might be
playing an important role in stabilizing the clump.

It is also interesting to compare the continuum masses (column~10 of
Table~\ref{param_cont}) of all the continuum sources that have a $^{13}$CO
counterpart with the LTE masses of the respective $^{13}$CO clumps (column~4 of
Table~\ref{mass_nubi}).  All the continuum sources with a $^{13}$CO counterpart
have $M_{\rm LTE} \succsim M_{\rm cont}$, which is probably due to the fact that the 
$^{13}$CO clumps have larger sizes than the respective continuum counterparts.
However, there are three exceptions: the continuum source
C1, which is associated with $^{13}$CO clump~co1; the continuum source C7, which is
associated with $^{13}$CO clump~co7; and the continuum source C12, which is associated with
$^{13}$CO clump~co12. A possible explanation might be that we underestimated the
temperature for these three sources, therefore we overestimated the continuum
mass and we underestimated the LTE mass. This explanation is consistent with
the association of the three sources with active star-formation tracers, in
particular with Spitzer emission at 3.6~$\mu$m, 8~$\mu$m and 24~$\mu$m.
Therefore we obtain the same result on the gravitational stability of the clumps
also from the continuum masses.

\section{SED and luminosities}\label{section:SED}

In Sect.~\ref{starformationassociation} we demonstrated that associations
between the ATLASGAL sources and the 24~$\mu$m MIPSGAL emission exist.
In the following we assume that these associations indeed 
are real and combine the Spitzer and
APEX results in order to derive crude spectral energy distributions from the
24 and 70~$\mu$m MIPSGAL data and ATLASGAL 870~$\mu$m observations.

\begin{figure}
\centering
\includegraphics[width=\columnwidth]{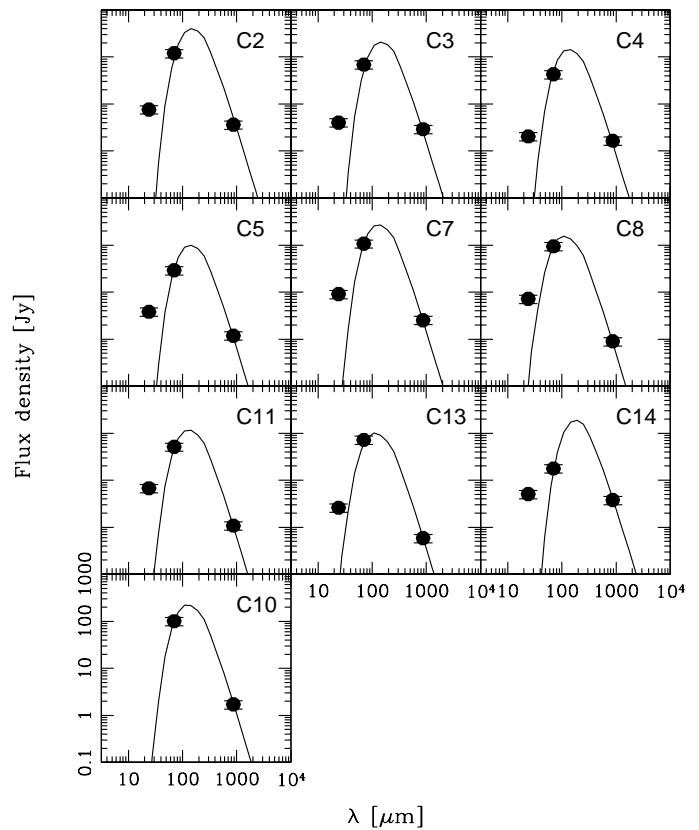}
\caption{Spectral energy distributions (SEDs) of the APEX 870~$\mu$m continuum
sources that are associated with Spitzer 24~$\mu$m emission (see 
Sect.~\ref{starformationassociation}).
The SEDs, shown as solid lines, are derived from 
fitting the 870 and 70~$\mu$m points with a modified black body of given 
density, dust temperature, and radius (see text for the details).
The continuum sources C6, C9 and C12 are not presented in the
plot because they do not seem to be directly associated with Spitzer 24~$\mu$m
emission and the continuum source C1 has been excluded 
as well because its emission at 24~$\mu$m and 70~$\mu$m is saturated.}
\label{sed}
\end{figure}

We have integrated the MIPSGAL 24 and 70~$\mu$m images over the area of
ATLASGAL sub-mm continuum emission in order to derive the spectral energy distributions shown
in Fig.~\ref{sed}. 
The Spitzer counterparts of ATLASGAL sources are mostly close to point-like
(6$^{\prime\prime}$ at 24~$\mu$m and 18$^{\prime\prime}$ at 70~$\mu$m), thus
more compact than the ATLASGAL sources, as one would expect for dust heated  by
a central object.  We clearly  do not have sufficient frequency coverage to get
a precise estimate of the luminosity but we have derived crude estimates 
by fitting the 870 and 70~$\mu$m points with a modified black
body of given density, dust temperature, and radius.  This is not
unique and always underestimates the 24~$\mu$m flux.  However, it gives a
reasonable first guess to the dust temperature appropriate for the
ATLASGAL data and hence for the determination of the mass associated with the
870~$\mu$m emission. Results for the 10 sources fitted in this way are given in
Table~\ref{apex-ir-luminosity} and are shown in Fig.~\ref{sed}. 
\begin{table*}      
\centering
\begin{threeparttable}
\caption{Properties of the ATLASGAL sources with a counterpart in the mid-infrared emission, derived by fitting the 870 and 70~$\mu$m points with a modified black body.}
\label{apex-ir-luminosity}      
\begin{tabular}{c c r c c | c c c c }
\hline\hline       
Source &  Radius & $d$\tnote{a} & F$_{24\mu m}$ & F$_{70\mu m}$ &   T$_d$ &  L & M & $\Sigma$ \\ 
& (pc) & (kpc) & (Jy) & (Jy) &   (K) & ($\times 10^4\, L_{\odot}$) & ($\times 10^2\, M_{\odot}$) & (g~cm$^{-2}$)\\ 
(1)&(2)&(3)&(4)&(5)&(6)&(7)&(8)&(9)\\
\hline                    
  C2  &  0.7    &  12.6   &  7.6$\pm$1.5 &  119.8$\pm$24.0    &      25  &  5   &  60  &  0.81  \\  
  C3  &  0.7    &  12.6   &  4.0$\pm$0.8 &   68.8$\pm$13.8    &      25  &  3   &  30  &  0.41  \\   
  C4  &  0.7    &  12.6   &  2.1$\pm$0.4 &   42.8$\pm$8.6     &      27  &  2   &  10  &  0.14  \\   
  C5  &  0.7    &  12.6   &  3.8$\pm$0.8 &   29.0$\pm$5.8     &      26  &  1   &  10  &  0.14  \\   
  C7  &  0.7    &  11.7   &  9.1$\pm$1.8 &  107.3$\pm$21.5    &      28  &  3   &  20  &  0.27  \\  
  C8  &  0.7    &  11.7   &  7.2$\pm$1.4 &   94.2$\pm$18.8    &      33  &  2   &  6   &  0.08  \\   
  C10 &  0.7    &  11.7   &    --        &  100.6$\pm$20.1    &      29  &  3   &  10  &  0.14  \\  
  C11 &  0.7    &  11.7   &  6.8$\pm$1.4 &   51.4$\pm$10.3    &      28  &  2   &  9   &  0.12  \\ 
  C13 &  0.6    &  11.7   &  2.6$\pm$0.5 &   72.0$\pm$14.4    &      32  &  1   &  4   &  0.07  \\ 
  C14 &  0.5    &   4.7   &  5.1$\pm$1.0 &   17.8$\pm$3.6     &      20  &  0.3 &  9   &  0.24  \\  
\hline
\end{tabular}
\begin{tablenotes}
\item[] NOTES. -- Col.(1): ATLASGAL continuum source. Col.(2): radius of the APEX continuum source. Col.(3): distance of the source. Col(4): flux density at 24~$\mu$m of the IR source associated with the ATLASGAL 870~$\mu$m source. Col.(5): flux density at 70~$\mu$m of the IR source associated with the ATLASGAL 870~$\mu$m source. Col.(6)-(8): output of the modified black body fit. In particular: the dust temperature; the luminosity; the mass; and the surface density.
\item[a] All the sources are assumed to be at the far kinematic distances, except C14.
\end{tablenotes}
\end{threeparttable}
\end{table*}
For the grey body fits we derive dust
temperatures of 20--33~K and luminosities of 3000--50000~$L_{\odot}$, with
corresponding gas masses of 10$^2$--10$^4$~$M_{\odot}$ and typical surface 
densities of 0.07--0.8~g~cm$^{-2}$. 
So the results
are in the range of values expected for young high mass (proto-)stars 
(e.g. \citealt{molinari2008}).

\begin{figure*}
\centering
\includegraphics[width=1.5\columnwidth]{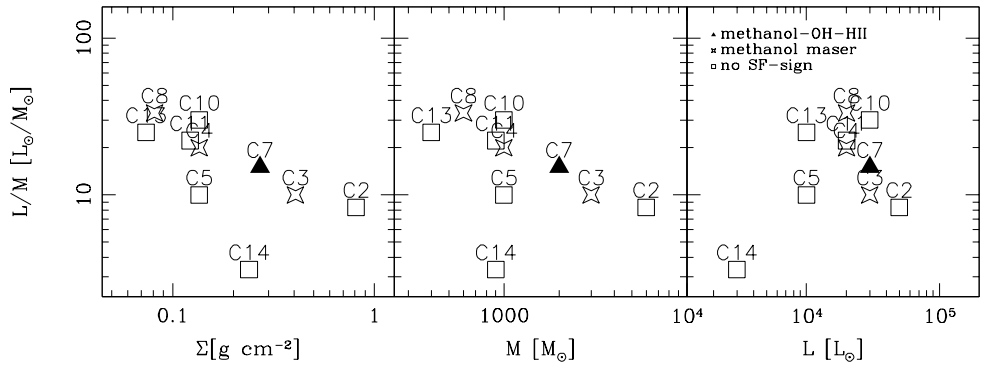} 
\caption{Luminosity to mass ratio, $L/M$, for the 870~$\mu$m continuum sources associated 
with the Spitzer emission at 24~$\mu$m (see Table~\ref{apex-ir-luminosity}), as a function of (1) the surface density, $\Sigma$ 
(\emph{left panel}); (2) the mass (\emph{middle panel}); and the luminosity (\emph{right panel}). We note that source C14 is the only source 
at the ``near'' kinematic distance (see Table~\ref{apex-ir-luminosity}).
}
\label{sigma-LsuM}
\end{figure*}

One useful application of these results is to consider the inferred 
luminosity to gas mass ratio, $L/M$, for the
ATLASGAL sources. This is a distance independent parameter and hence
useful for comparison with extragalactic star-formation indicators
(see \citealt{plume1992,plume1997}).
Since bolometric luminosity is a rough measure
of star formation rate (for a given initial mass function, IMF), 
$L/M$ is a measure of the gas
exhaustion timescale of the clumps, which is typically much larger than the
free fall time (see \citealt{krumholz2005}); hence, an indicator of the 
clump evolutionary timescale. 
One might expect $L/M$ to
correlate with the surface density $\Sigma $ (also a distance
independent quantity roughly speaking and a measure of pressure for
virialized clumps) and we thus show in Fig.~\ref{sigma-LsuM} a plot of
$L/M$ against $\Sigma $ \ for our sample.  
One expects $L/M$ to increase with time as molecular gas is converted
into stars as star formation proceeds. One might
also expect star formation rates to increase in regions of high pressure or
surface density. 
It is interesting in this context that \citet{sridharan2002}
find evidence of an increase of $L/M$ going from regions without 
3.6~cm radio emission (at the mJy level with the VLA) and clumps
associated with UC~HII regions. It would be useful to test this
result for more homogeneous samples similar to the present one.
We note, however, that \citet{faundez2004} find such result marginal and offer 
an alternative interpretation of $L/M$ as an indicator of the luminosity 
of the most massive star embedded within the clump. However, these results 
are based on much less sensitive radio continuum measurements, using a  
highly inhomogeneous sample of objects
with distances varying over an order of magnitude.
It is clear that larger samples 
with both high angular resolution far-infrared measurements and
sensitive radio observations are needed to make further progress in
this area.

A surprising feature of our results (see Fig.~\ref{sigma-LsuM}) 
is that low column density sources (below 0.5~g~cm$^{-2}$) appear 
to have higher $L/M$ than high column density sources, which is hard 
to understand theoretically. However, this finding is marginal given the
poor statistics.

These estimates for $L/M$ can be compared with the results of \cite{wu2005}
who used HCN as a dense gas tracer in order to estimate the star forming
efficiency both in Galactic cores and extragalactic star-forming
regions. From a survey of 31 Galactic cores, one can derive a median
$L/M$ ratio of 63 L$_{\odot}$/M$_{\odot}$ for regions with luminosities above
$3\, 10^4\, L_{\odot}$; the ratio for extragalactic star-forming regions
is similar.  Below $3\, 10^4\, L_{\odot}$, the ratio drops rapidly. Our
median value of $L/M$ from the black body fit is roughly a factor of 4
lower but our sample is small and straddles the $3\, 10^4\, L_{\odot}$
limiting value for the \cite{wu2005} relation.  However, given both
the dispersion in the \cite{wu2005} result and the uncertainties in our
determinations of both L and M, we conclude that one needs both more
complete SEDs and more reliable mass determinations to make further progress.
The former will be undertaken by future surveys with HERSCHEL (such as
HI-GAL; \citealt{molinari2010})
but the latter will require careful calibration and intercomparison of
the various approaches to determining clump masses.

\section{Comparison with extragalactic observations}\label{comparison_extragal}

G19.61-0.23 is one of the most luminous ($\sim$2$\times10^6\,L_{\odot}$, 
see Sect.~\ref{intro}) known Galactic massive star-forming regions and therefore
one of the best templates to study locally the 
starburst phenomenon.
Several high resolution studies have been carried out in this region as well as 
in other well known Galactic high mass star-forming regions. 
However, the aim of this work is to analyze the 
large-scale properties and physical conditions of the molecular gas in the region 
and the implications for extragalactic studies of nearby starburst galaxies.

Of particular interest are the youngest extragalactic embedded super star clusters 
(SSCs), which have been seen in merging systems (e.g. \citealt{whitmore2002})
and in some dwarf irregular galaxies (e.g. \citealt{elmegreen2002}; \citealt{johnson2003})
and have been identified by the free-free radio emission from their associated
HII regions (e.g. \citealt{johnson2003}).
One can see that the molecular clouds associated 
with the SSCs
show lower surface densities than their end product (the star clusters). 
Since extragalactic studies are mostly done with single-dish telescopes and in 
low J transitions of CO,
this result may be due to: (1) the limited linear resolution
reached with current facilities; and/or (2) the molecular tracers used in the 
extragalactic studies, which are not tracing the real active sites of cluster formation, 
but rather lower density associated material.

In order to analyze the effect of low angular resolution in extragalactic studies, 
we compared the results from our $^{13}$CO low resolution observations of the Galactic region 
G19.61-0.23 with the physical parameters that have been found for some nearby extragalactic 
starbursts, in particular with CO(2-1) high angular resolution observations of Henize~2-10, 
a dwarf irregular starburst galaxy at the distance 
of 9~Mpc ($H_0$=75 km~s$^{-1}$~Mpc$^{-1}$, \citealt{vacca1992}; \citealt{kobulnicky1999}).
We observed Henize~2-10 (\citealt{santangelo2009}) at high angular resolution 
(1$.\!\!^{\prime\prime}9\times1.\!\!^{\prime\prime}3$,
which corresponds to a linear resolution of $80\times60$~pc at the distance of Henize~2-10), 
revealing a rich population of molecular clouds with estimated masses and surface densities 
systematically larger than those measured in our Galaxy and we found possible evidence that the super 
star clusters are associated with very massive and dense molecular cores.
We also compared our results with studies of M31 and M33 from  
\cite{sheth2008} and \cite{rosolowsky2003}. They present CO(1-0) observations of the two 
spiral galaxies made with the BIMA array at a linear resolution of $\sim 20$~pc.

\subsection{The total $^{13}$CO(1-0) emission}

The total region of G19.61-0.23 that we sampled in $^{13}$CO(1-0) with FCRAO 
is about $27^{\prime}\times 27^{\prime}$ large (using the data from the BU-FCRAO GRS to cover the 
regions which were not sampled by our observations), 
which corresponds approximately to a size of 98~pc at the distance of the main complex of G19.61-0.23 (12.6~kpc).
This value corresponds approximately to the size of each cloud identified in the SMA CO(2-1) 
emission of Henize~2-10 (see Table~\ref{param-13co63TOTspec} and Table~3 of \citealt{santangelo2009}).
We can thus compare the properties of the total FCRAO $^{13}$CO(1-0) emission 
from the main cloud of the complex, containing G19.61-0.23 (cloud~2; see Sect.~\ref{kinematics} and Table~\ref{cloud_param}),
with the physical parameters of the 14 clouds resolved in the SMA CO(2-1) emission from Henize~2-10.

With this aim, 
we analyzed the total FCRAO $^{13}$CO(1-0) emission from cloud~2 
as if we were observing a single molecular cloud 
at the same linear resolution with which we observed Henize~2-10 ($\sim 80\times60$~pc).
In particular, we integrated all the $^{13}$CO(1-0) emission from the BU-FCRAO GRS data of cloud~2, 
between 32 and 50~km~s$^{-1}$ 
and smoothed the derived spectrum to the same velocity resolution
obtained in the SMA observations of Henize~2-10 (5~km~s$^{-1}$).
The physical parameters of cloud~2 ``observed'' in this fashion are given in Table~\ref{param-13co63TOTspec}, where 
the FWHM line width is listed in col.~1. 
We list, for comparison, in Table~\ref{param-13co63TOTspec} also the parameters of one of the 
clouds resolved in the SMA emission from Henize~2-10.
\begin{table}
\centering
\caption{Summary of the parameters of the total integrated $^{13}$CO(1-0) emission of the GMC surrounding G19.61-0.23 and of a typical cloud in Henize~2-10, from SMA observations.}\label{param-13co63TOTspec}      
\smallskip 
\begin{threeparttable}
\renewcommand{\footnoterule}{}
\begin{tabular}{c c c c c}
\hline\hline       
FWHM & $M_{\rm VIR}$ & $M_{\rm LTE}$ & $\Sigma_{\rm VIR}$ & $\Sigma_{\rm LTE}$ \\
(km/s) & ($\times 10^5\, M_{\odot}$) & ($\times 10^5\, M_{\odot}$) & (g/cm$^{2}$) & (g/cm$^{2}$) \\ 
(1)&(2)&(3)&(4)&(5)\\
\hline                    
\multicolumn{5}{c}{\underline{G19.61-0.23}}  \\    
8 & 5 & 4 & 0.02 & 0.02               \\  
\hline
\multicolumn{5}{c}{\underline{Henize~2-10}}  \\    
 20 & 39 & 30 & 0.13 & 0.10 \\
\hline
\end{tabular}
\begin{tablenotes}
\item[] NOTES. -- Col.(1): FWHM line-width of the emission. Col.(2): Virial mass. Col.(3): LTE mass, assuming T$_{ex}$=20~K. Col.(4): Surface density from the virial mass. Col.(5): Surface density from the LTE mass.
\end{tablenotes}
\end{threeparttable}
\end{table}
We note that the line-widths (col.~1) are computed from the original total integrated $^{13}$CO(1-0) spectrum, before smoothing to 
the resolution of 5~km~s$^{-1}$, and that the total emission 
is computed by integrating the emission 
from the channels relative 
to the emission from cloud~2 (between 32 and 50~km~s$^{-1}$).
From the comparison with the typical line widths of the clouds identified in Henize~2-10 (see 
Table~\ref{param-13co63TOTspec}),
cloud~2 has a line width a factor of 2-3 smaller than the extragalactic clouds.

Following the same approach described in Sect.~\ref{parameters}, we computed the virial mass and the LTE mass of 
cloud~2, assuming that: (1) the $^{13}$CO(1-0) emission is optically thin, as for the extragalactic clouds; (2) the cloud 
is at a distance of 12.6~kpc; and (3) T$_{ex}=20$~K.
We also computed the surface density
of the cloud, using both the virial mass and the LTE mass. 
A summary of the derived masses and surface densities of cloud~2
is given in Table~\ref{param-13co63TOTspec}.
The derived masses are smaller than the masses derived for the CO clouds in Henize~2-10,
which are of the order of a few 10$^6\,M_{\odot}$.

\begin{figure}
\centering
\includegraphics[scale=0.44]{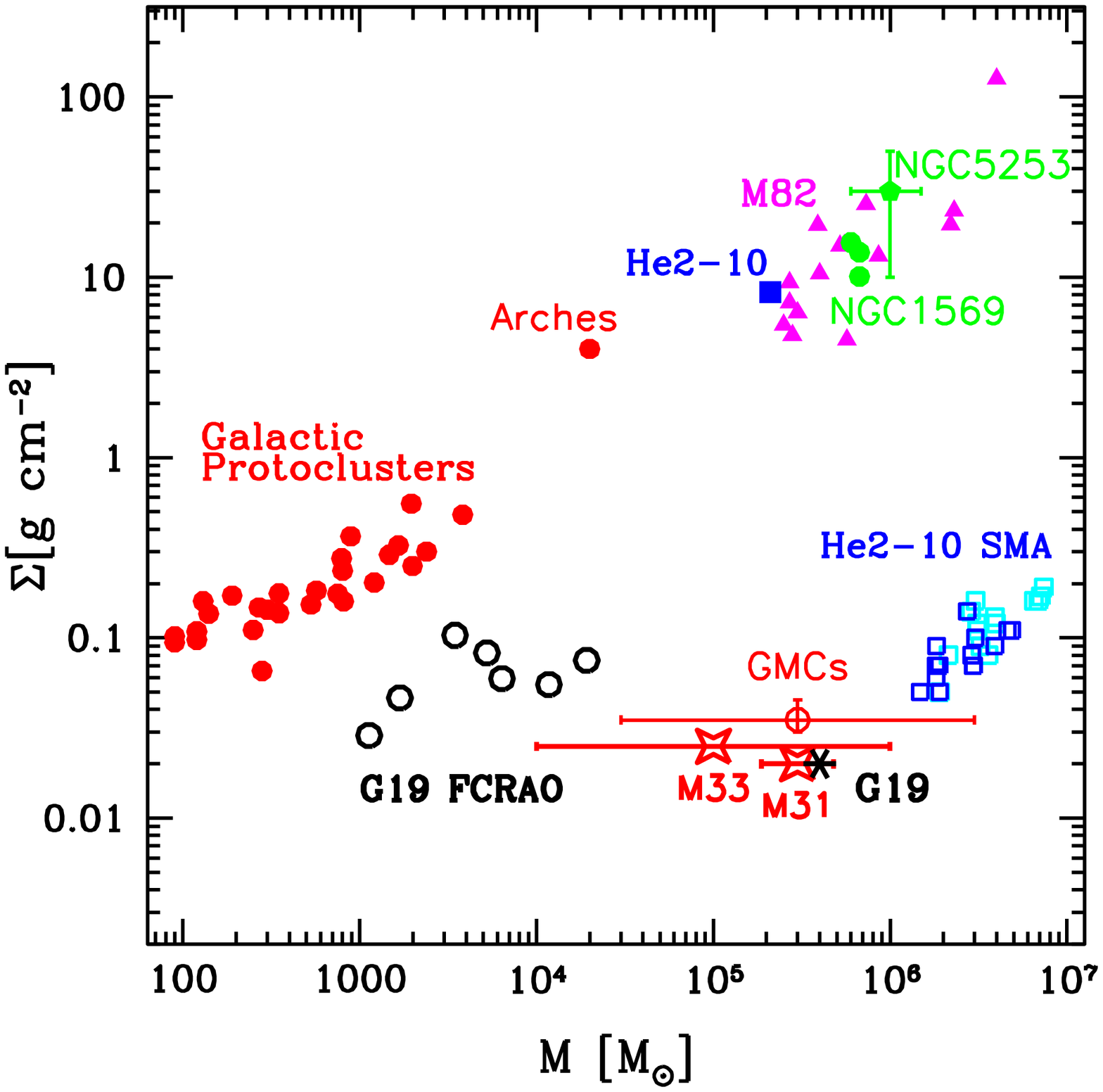}
\caption{Surface density, $\Sigma = M/(\pi R^2$), versus mass, $M$, 
for star clusters and interstellar clouds. 
Several massive clusters are indicated: the pink triangles represent the SSCs in M82
\citep{mccrady2007}; the green circles are the SSCs in NGC~1569 
\citep{demarchi1997,larsen2008,grocholski2008}; and 
the green pentagon represents the cluster in NGC~5253 \citep{mckee2003}.
Filled circles are star-forming 
clumps (\citealt{mueller2002}). The typical GMC parameters 
are shown (\citealt{solomon1987}).
The values for M31 (\citealt{sheth2008}) and M33 (\citealt{rosolowsky2003})
are shown in red asterisks in the bottom right part of the diagram.
The points labelled as \emph{He2-10 SMA} are the clouds in Henize~2-10 from 
\cite{santangelo2009}, where the cyan points are computed from the virial masses 
and the blue points are computed from the molecular gas masses. 
The points labelled as \emph{G19 FCRAO} are the clumps in G19.61-0.23 from this work, 
computed from the LTE masses. Finally, the point labelled as \emph{G19} 
represents the total emission from cloud~2 (see Table~\ref{param-13co63TOTspec}),
computed from the LTE mass.
}
\label{tan}
\end{figure}

In Fig.~\ref{tan} we show our results from Table~\ref{mass_nubi} and Table~\ref{param-13co63TOTspec}
on a mass-surface density
plot similar to that discussed by \cite{tan2007}.
The blue and cyan squares are the values for the clouds in Henize~2-10, 
respectively, from the molecular and virial masses, which are 
in the same general area of the diagram as Galactic GMCs, though with somewhat higher 
masses and surface densities. 
Moreover, due to our limited
angular resolution, the surface density may well be considerably 
underestimated. 
The black star represents the position of cloud~2
from the LTE mass (see Table~\ref{param-13co63TOTspec}). 
This cloud appears in the same area as the Galactic GMCs and the GMCs in M31 and M33, which
confirms that there are no significant differences in the cloud properties of the two 
nearby spiral galaxies and the Milky Way (e.g. \citealt{rosolowsky2003}; \citealt{sheth2008}).
However, cloud~2 shows lower surface density than the clouds in Henize~2-10, which
suggests that the clouds in Henize~2-10 might be different from the GMCs in the 
Galaxy as well as from those in M31 and M33.  
Finally, the black circles represent the position in the plot for the 
clumps resolved in cloud~2 (clumps~co1, co2, co3, co6, co8, co9 and co5), 
from the LTE masses.
It is evident from the plot that the clumps resolved in cloud~2 have higher pressure 
than the whole molecular cloud~2, seen at low angular resolution. 
In particular, from the whole mass and surface of cloud~2 and the masses and surface 
of the clumps resolved in the $^{13}$CO(1-0) emission from cloud~2, we can derive 
a filling factor in area and in mass for the dense molecular gas in the GMC of G19.61-0.23, 
computed respectively as: $\sum_i\,A_i/A_{TOT}$ and $\sum_i\,M_i/M_{TOT}$, where $A_i$ and $M_i$ 
are the area and mass of every clump resolved in cloud~2 (i.e. clumps~co1, co2, co3, co6, co8, co9 and co5; 
see Table~\ref{mass_nubi}) and $A_{TOT}$ and $M_{TOT}$ are the area and mass of cloud~2 (see 
Table~\ref{param-13co63TOTspec}).
We find that the fraction of dense 
molecular gas in cloud~2 is about 10-20\% (respectively from 
the LTE and the virial masses). 
Extrapolating those results to the extragalactic context, 
assuming that the fraction of the area and the mass of a 
GMC which is in the form of dense gas is the same as in the Galactic context, 
we would expect each cloud identified in Henize~2-10 
to be resolved in several molecular clumps which account for a total dense molecular 
gas mass up to 10$^6 \, M_{\odot}$, each having a surface density up to 1.5~g~cm$^{-2}$.
It seems thus reasonable to assume that 
the clusters in Henize~2-10 form from small parsec sized clumps at much higher 
surface density and pressure than 
the rest of the GMC and than the typical Galactic clumps.

Therefore, we suggest that finding molecular clouds in Henize~2-10 with lower 
surface densities than their end product, the SSCs,
might be due to the low linear resolution which is currently available.
High sensitivity observations at higher resolution are needed 
to confirm this scenario, as well as observations of higher density tracers.

\section{Summary and conclusions}\label{summary}

In this paper we have presented single dish observations of the molecular gas 
and the sub-mm continuum of the high-mass star-forming region G19.61-0.23.
Our observations, with a spatial resolution of about 2.8~pc at the adopted distance of 12.6~kpc, 
reveal a population of molecular clumps. 
The physical parameters of the identified clumps 
show a spread of virial parameters $M_{VIR}/M$, ranging from gravitationally unstable
to unbound. These results seem to be independent of the presence or otherwise of 
star-formation indicators in the cloud, which is in slight contrast to results 
for less distant GMCs (e.g. \citealt{williams1995}; \citealt{williams1998}). 
One needs however more precise mass determinations for the clumps in G19.61-0.23, 
using high density molecular tracers, such as CS. Analogously, it would be very useful 
to obtain molecular line data for the ATLASGAL continuum sources.

Towards the ATLASGAL continuum sources, we have studied the association with mid-infrared 
emission and found that most
ATLASGAL sources have counterparts in the mid-infrared emission 
at the peaks of the sub-millimeter continuum emission.
For those sources we derived crude spectral energy distributions and estimated the luminosities 
and dust temperatures. 
We find that the majority of the continuum sources are associated with star-forming 
regions of luminosity 1--5~$10^4 \, L_{\odot}$ and gas masses 400--6000~$M_{\odot}$. 
The luminosity to gas mass ratio, $L/M$, is in the range 10--100 $L_{\odot}/M_{\odot}$ 
with some evidence for a fall off for high mass clumps. The ATLASGAL sources are clearly good 
tracers of star formation.

We compared our results for the Galactic luminous high-mass star-forming region G19.61-0.23 with 
a previous extragalactic study of a dwarf starburst galaxy Henize~2-10. 
We find that the main cloud in G19.61-0.23 has physical properties comparable with the typical 
galactic GMCs and with the GMCs in M31 and M33. However this cloud shows less extreme 
properties than the clouds identified in Henize~2-10, in particular smaller surface densities 
and masses.

The new generation of telescopes is needed to address these questions. 
They will also allow to observe with higher sensitivity higher density tracers, 
to see the real sites of active star formation.

\begin{acknowledgements}
We thank the APEX support astronomers for performing excellent service mode
observations at APEX. The FCRAO observations were partially acquired by Carla Maxia.
This work was partially supported by ASI-INAF contracts.
Leonardo Bronfman acknowledges support from Center of Excellence in 
Astrophysics and Associated Technologies (PFB 06) and by FONDAP Center for 
Astrophysics 15010003.
\end{acknowledgements}

\Online

\appendix
\section{Online material}
\begin{figure*}
\centering
\includegraphics[angle=-90,scale=0.6]{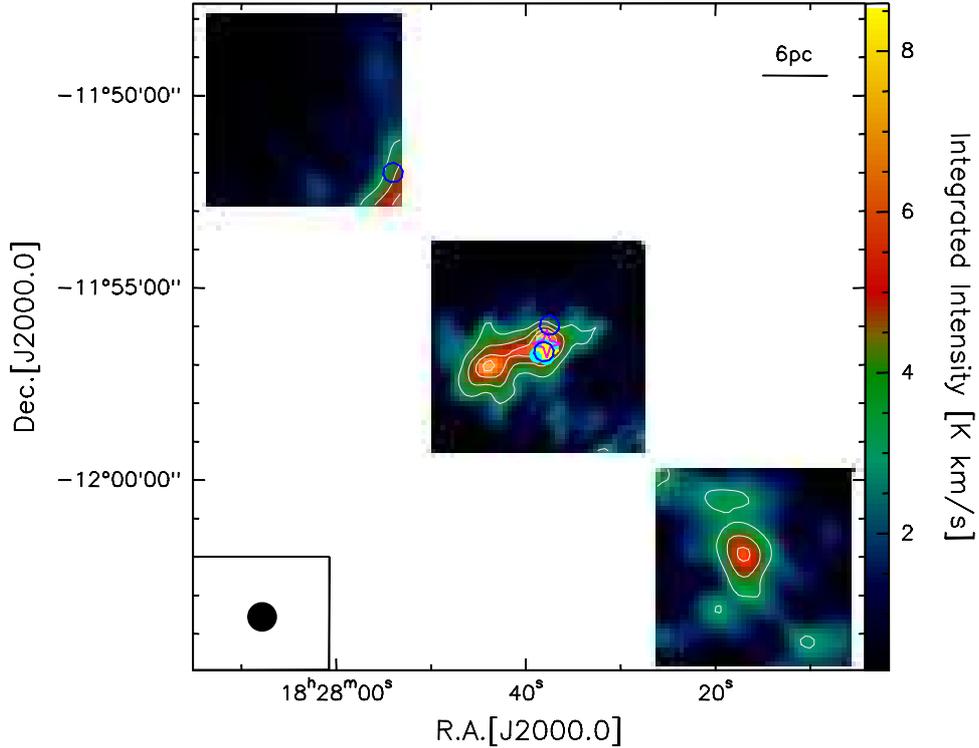}   
\caption{FCRAO intensity map of the C$^{18}$O(1-0) emission, integrated between 25 and 57~km~s$^{-1}$.
The contour levels are from 10~$\sigma$ in steps of 5~$\sigma$ ($\sigma$= 0.3 K km s$^{-1}$).
Symbols are as in Fig.~\ref{integr_tot} (see Table~\ref{SIMBAD}).
}
\label{integr_tot2}
\end{figure*}

\begin{figure*}
\centering
\includegraphics[width=\textwidth]{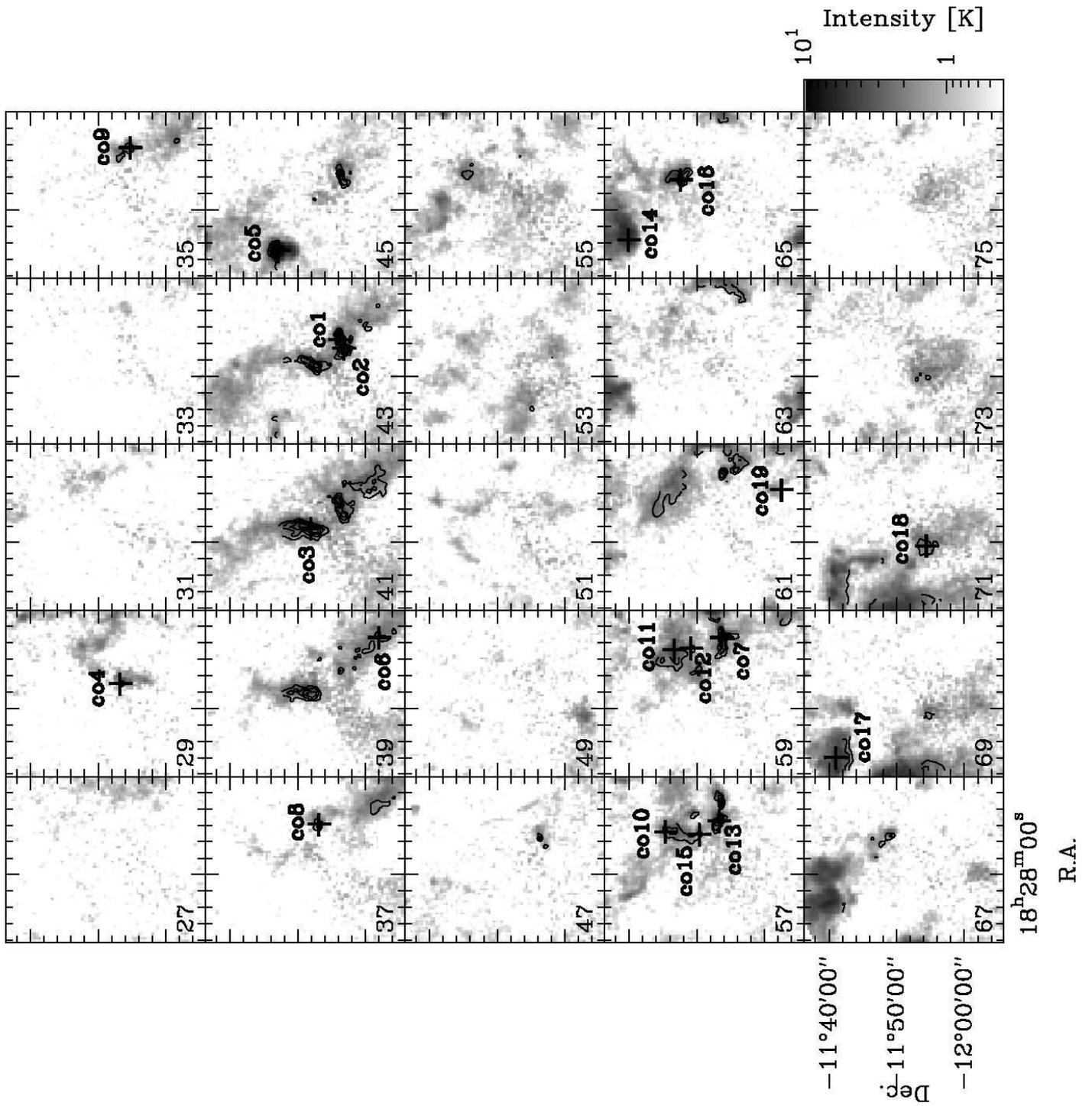}
\caption{Channel map of the BU-FCRAO GRS $^{13}$CO(1-0) data in grey scale and 
our FCRAO $^{13}$CO(1-0) emission in black contours.
Center velocities of the channels are indicated in the bottom left 
corner of each panel and increase in steps of 2~km s$^{-1}$. Contour levels are 
drawn from 10~$\sigma$ ($\sigma$ = 0.2~K) in steps of 5~$\sigma$. 
The black numbered crosses indicate the peak positions (in the channel 
corresponding to the peak velocity) of the emission of the 19 molecular clumps
identified in the FCRAO $^{13}$CO(1-0) emission.}
\label{13cototcanali}
\end{figure*}

\begin{figure*}
\hspace{0.4cm} {\bfseries\Large FCRAO $^{13}$CO(1-0)} \hspace{4.85cm} {\bfseries\Large APEX 870~$\mu$m}
\centering
 \includegraphics[height=0.27\textheight]{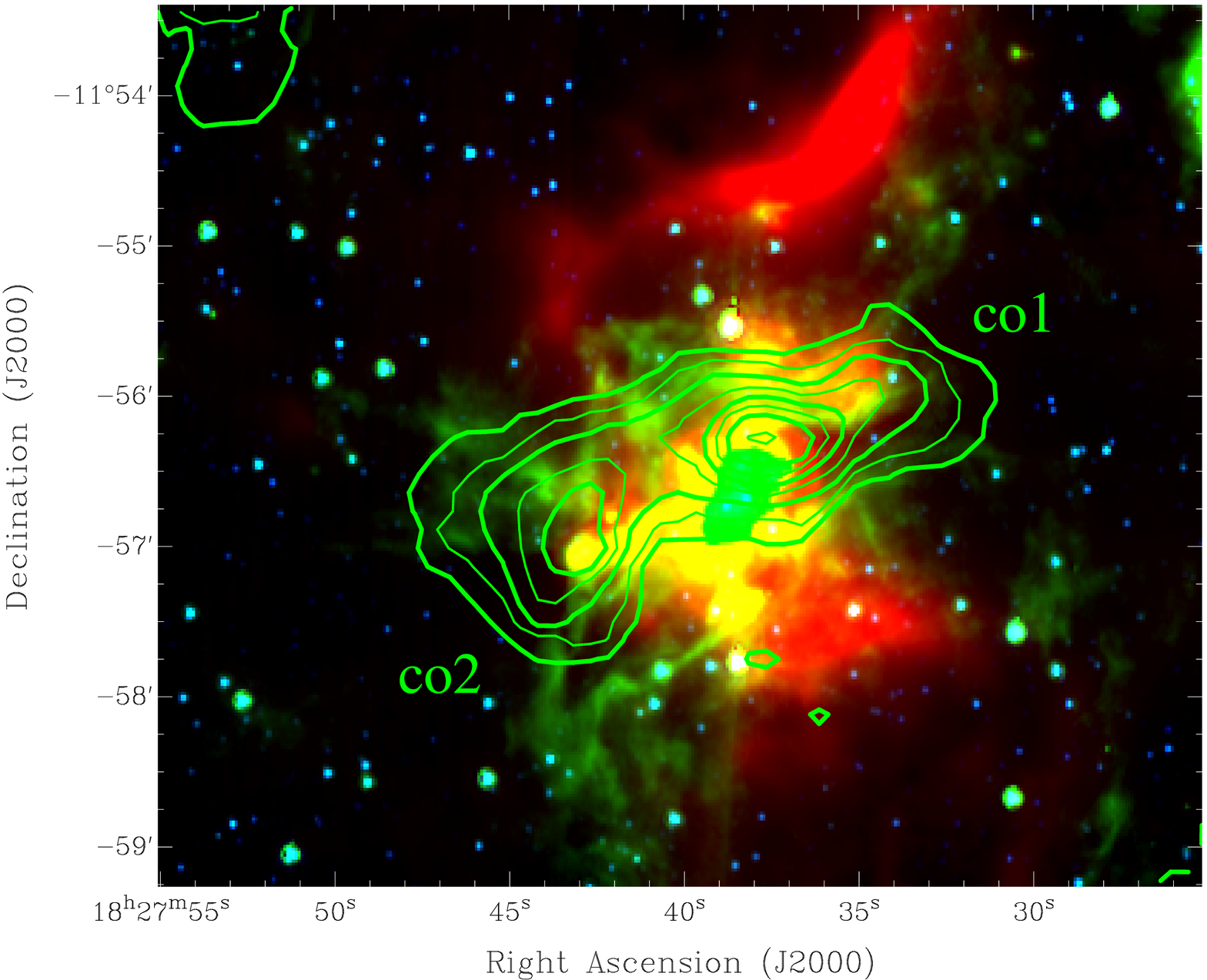}        
 \includegraphics[height=0.27\textheight]{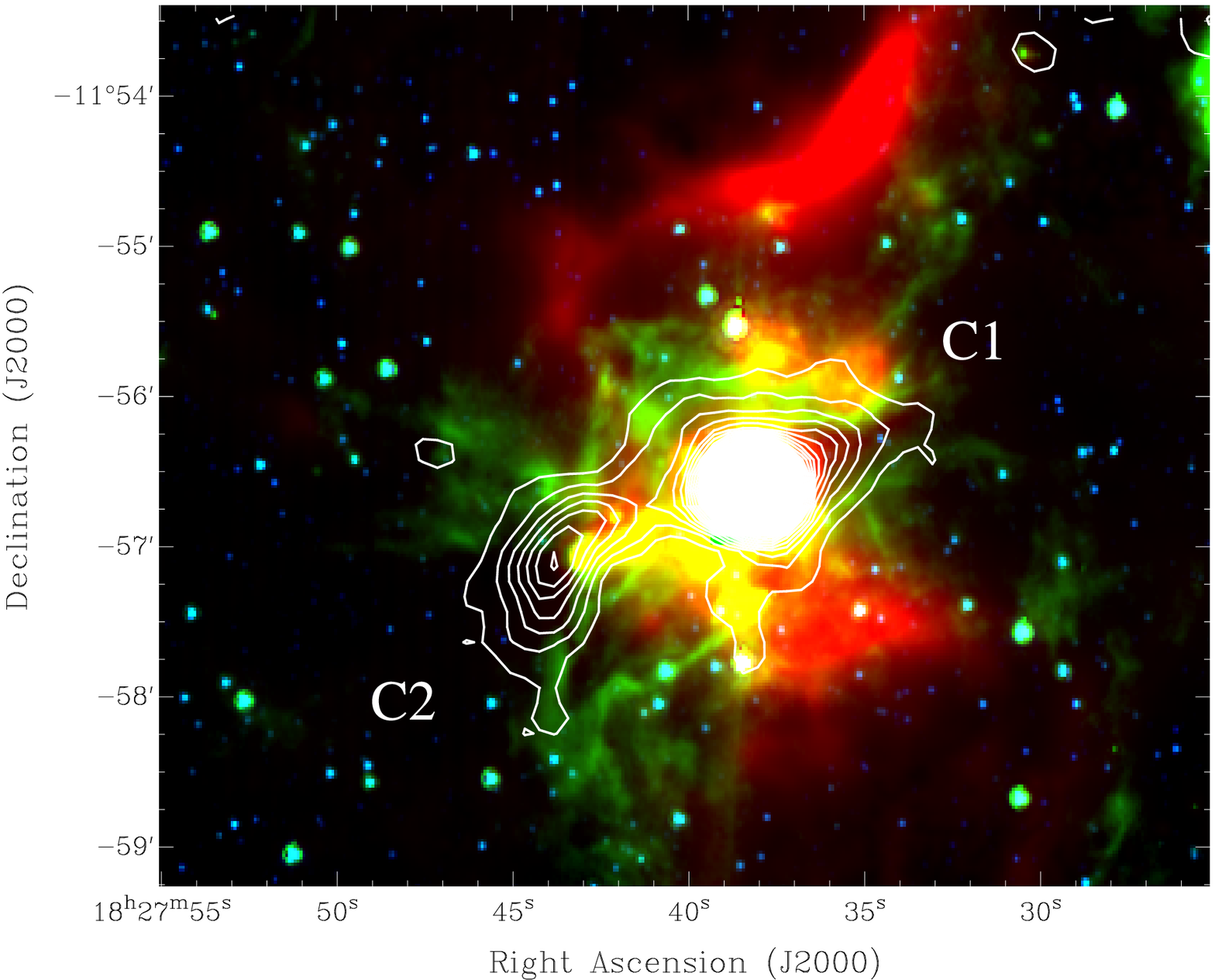}       
 \includegraphics[height=0.25\textheight]{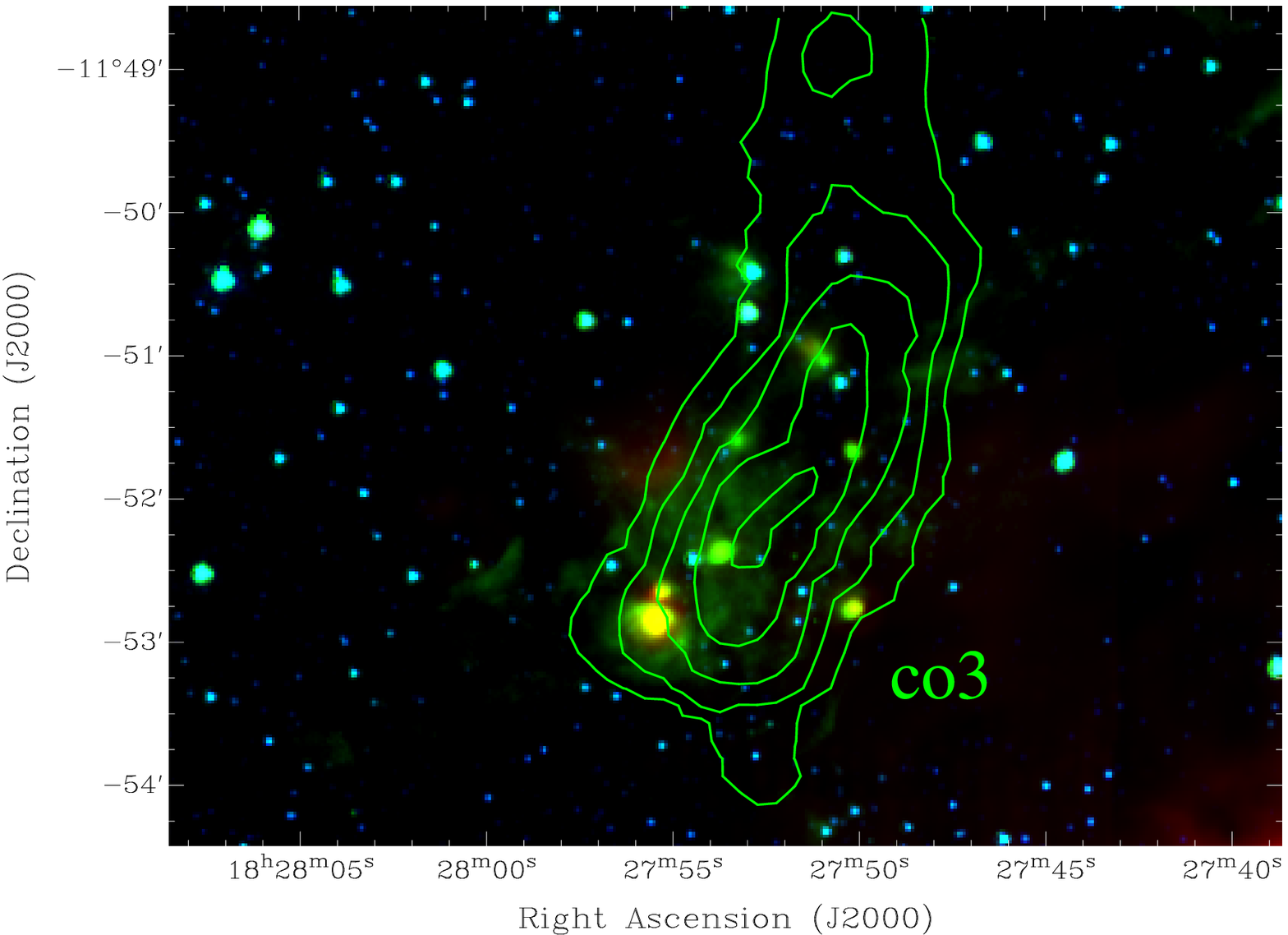}          
 \includegraphics[height=0.25\textheight]{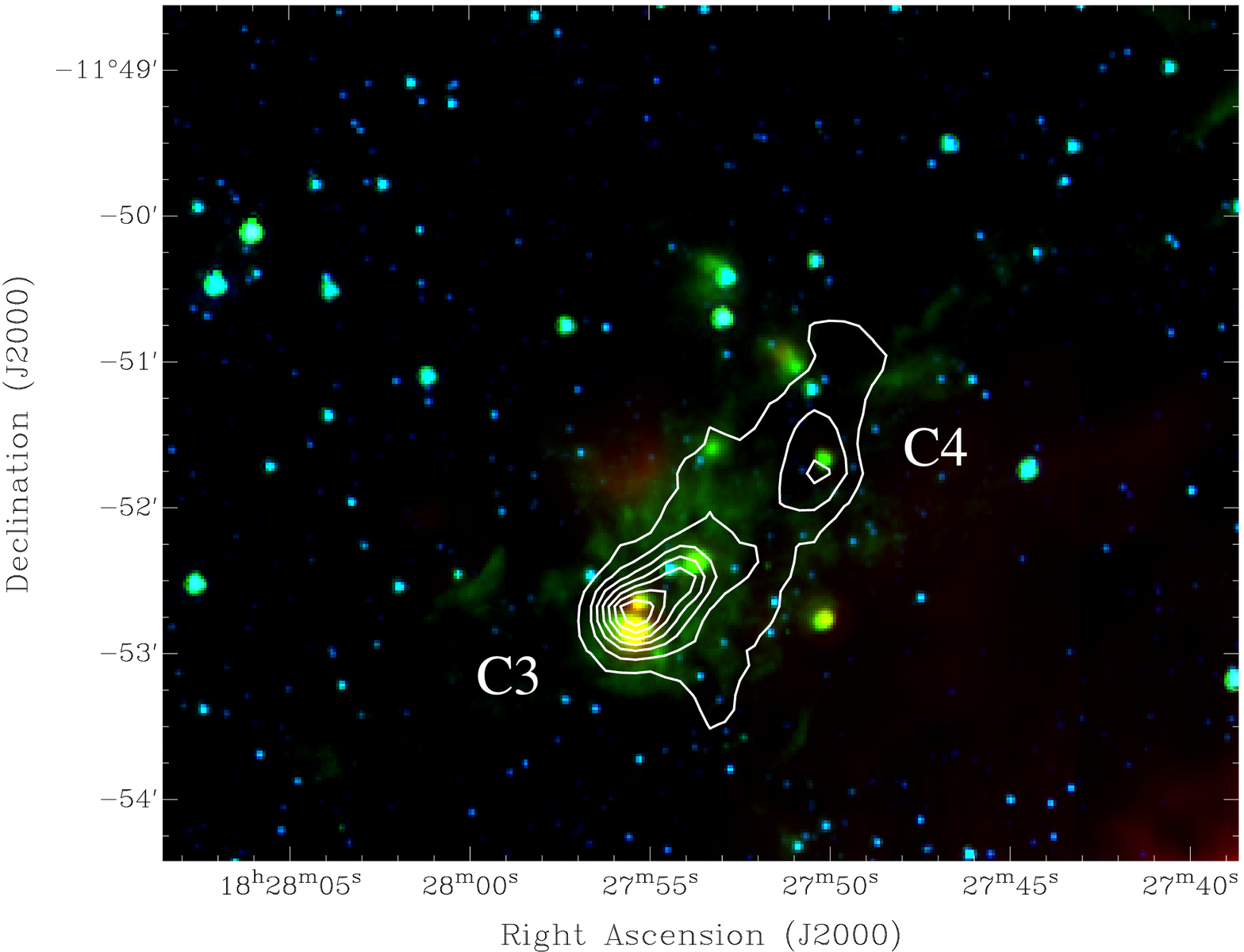}        
 \includegraphics[height=0.27\textheight]{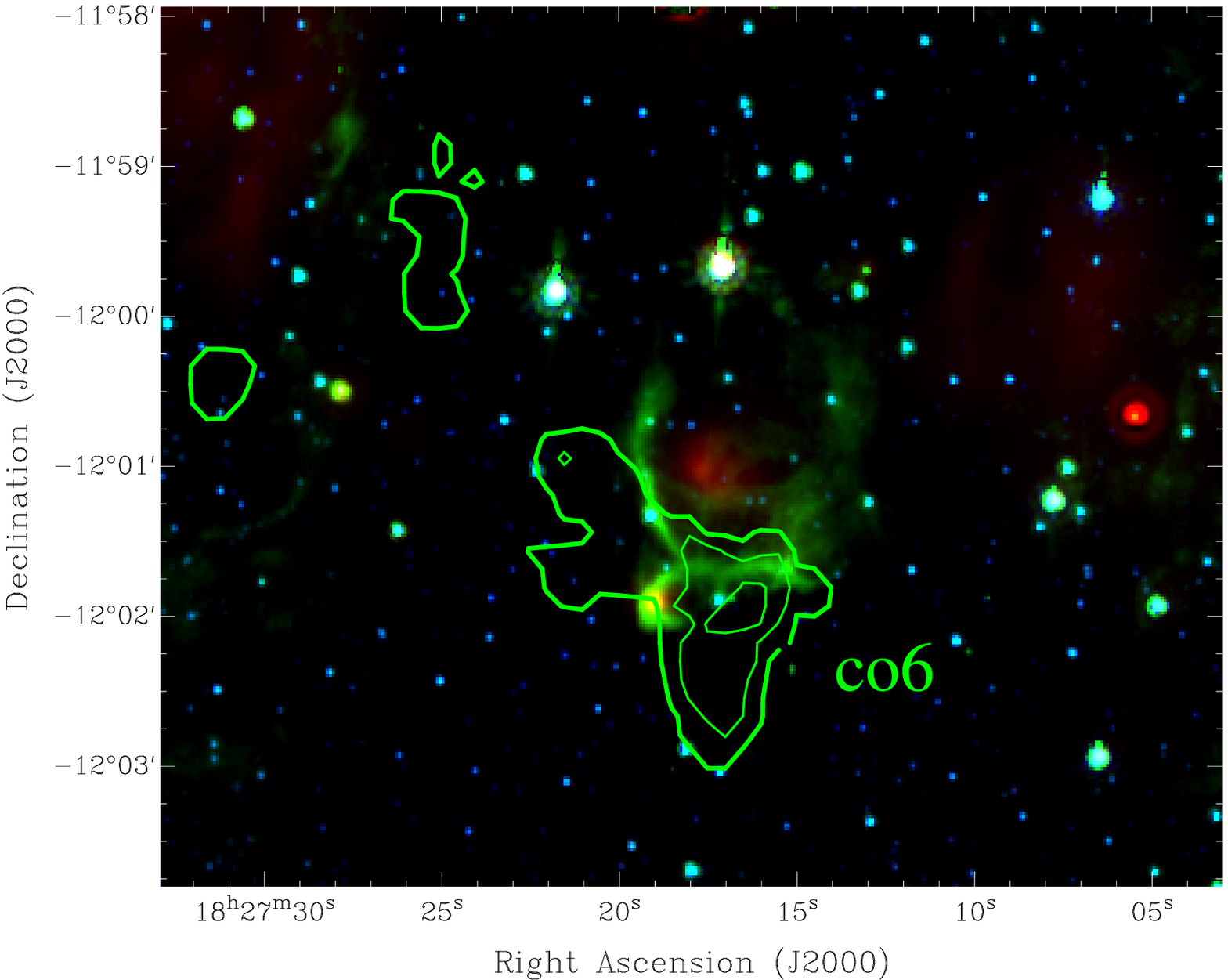}          
 \includegraphics[height=0.27\textheight]{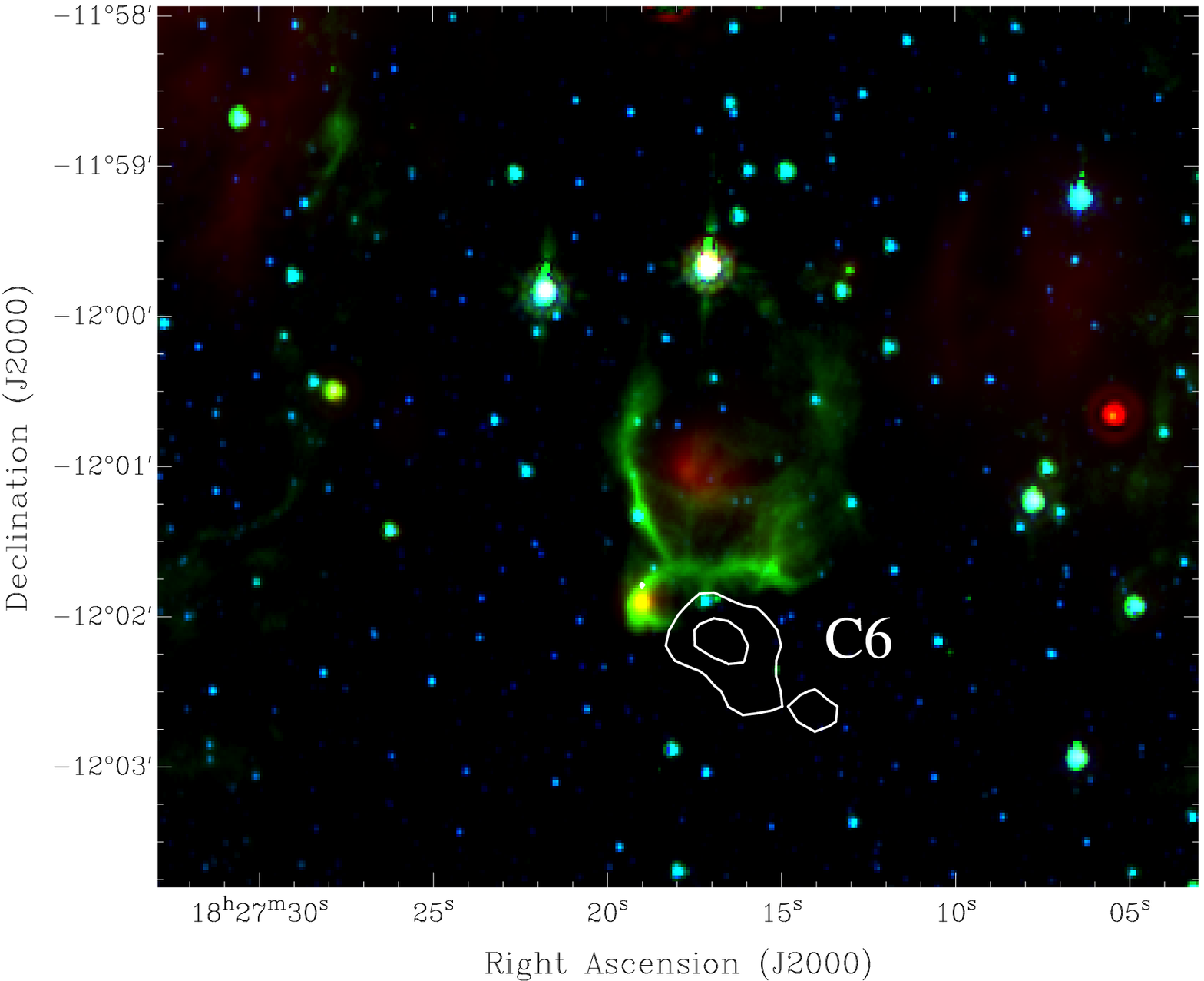}        
\caption{Composite image of the 24~$\mu$m (red), the 8~$\mu$m (green) 
and 3.6~$\mu$m (blue) emissions from Spitzer, covering several regions 
of the large-scale map, corresponding to the single clumps associated with the APEX continuum emission at 870~$\mu$m.
The green contours are 
the $^{13}$CO(1-0) emission of the single clumps, integrated over the velocity range of every clump
(see col.~4 of Table~\ref{param_nubi}),
and the white contours are the APEX continuum emission
at 870~$\mu$m (from 5~$\sigma$ in steps of 5~$\sigma$, where $\sigma = 40$~mJy~beam$^{-1}$).
The $^{13}$CO clumps and the correspondent continuum sources are labelled in the single diagrams.
Note that some of the panels are similar because for each clump we integrated on the velocity 
channels of emission of the clump (see col.~4 of Table~\ref{param_nubi}).
}
\label{CO-apex-spitzer}
\end{figure*}

In this appendix we present additional material, which may be useful.
In particular, we show the list of all the sources in the sampled field, 
from the SIMBAD Astronomical Database (Table~\ref{SIMBAD}); 
the integrated intensity map of the C$^{18}$O(1-0) emission from our FCRAO 
observations (Fig.~\ref{integr_tot2}); 
the channel map of our FCRAO $^{13}$CO(1-0) data (Fig.~\ref{13cototcanali}); 
and finally the overlay between the $^{13}$CO(1-0) clumps 
and the APEX continuum clumps with the Spitzer emission at 3.6, 8 and 24~$\mu$m 
(Fig.~\ref{CO-apex-spitzer}).

\begin{table*}
\caption{List of the sources in the sampled field, from the SIMBAD Astronomical Database. Col.(1)-(2): Coordinates (J2000.0). Col(3): Distance from the peak emission of the associated clump (see Sect.~\ref{identif}). Col.(4): References.}
\label{SIMBAD}
\centering      
\smallskip 
\begin{threeparttable}
\renewcommand{\footnoterule}{} 
\begin{tabular}{c c c c | c c c c}
\hline\hline
\multicolumn{2}{c}{Position} & Distance & Ref.   &     \multicolumn{2}{c}{Position} & Distance & Ref.  \\
R.A[J2000.0] & Dec.[J2000.0] & (arcsec) &        &     R.A[J2000.0] & Dec.[J2000.0] & (arcsec) & \\
(1)&(2)&(3)&(4)&(1)&(2)&(3)&(4)\\
\hline
\multicolumn{4}{c|}{\underline{HII Regions}}        &                \multicolumn{4}{c}{\underline{H$_2$O Masers}}   \\ 
18:27:38.29  &  -11:56:29.4   &    16.8 &  1      &           	     18:27:38.00  &  -11:56:36.0  &  21.0 &  13     \\ 
18:27:37.34  &  -11:56:32.4   &    17.2 &  2,3,7  &	       	     18:27:37.80  &  -11:56:37.0  &  21.5 &  13     \\ 
18:27:37.30  &  -11:56:32.3   &    17.3 &  1  	   &	       	     18:27:38.00  &  -11:56:37.0  &  22.0 &  13     \\ 
18:27:38.20  &  -11:56:31.4   &    17.8 &  2,3    &	       	     18:27:37.40  &  -11:56:38.0  &  22.6 &  13     \\ 
18:27:38.02  &  -11:56:32.9   &    18.1 &  1  	   &	       	     18:27:38.20  &  -11:56:37.0  &  22.9 &  13     \\ 
18:27:38.01  &  -11:56:34.9   &    20.0 &  3  	   &	       	     18:27:37.60  &  -11:56:41.0  &  25.3 &  13     \\ 
18:27:38.08  &  -11:56:35.5   &    20.9 &  4  	   &	       	     \multicolumn{4}{c}{\underline{OH Masers}}       \\ 
18:27:37.70  &  -11:56:37.0   &    21.3 &  10     &                 18:27:37.80  &  -11:56:27.0  &  11.6 &  16     \\ 
18:27:38.18  &  -11:56:36.0   &    21.9 &  5  	   &	       	     18:27:16.20  &  -11:53:36.0  &  13.3 &  17     \\ 
18:27:38.30  &  -11:56:35.4   &    22.0 &  6  	   &	       	     \multicolumn{4}{c}{\underline{Methanol Masers}} \\ 
18:27:38.09  &  -11:56:39.6   &    24.8 &  1  	   &	       	     18:27:37.50  &  -11:55:58.0  &  17.8 &  12     \\ 
18:27:38.14  &  -11:56:39.7   &    25.1 &  2,3    &	             18:27:38.09  &  -11:56:39.5  &  24.7 &  14     \\ 
18:27:38.16  &  -11:56:40.2   &    25.7 &  7      &	       	     18:27:54.00  &  -11:52:00.0  &  33.4 &  15     \\ 
18:27:38.00  &  -11:56:42.0   &    26.9 &  8,10,11  &         	     18:27:17.30  &  -11:53:46.0  &  6.1  &  12     \\  
18:27:37.80  &  -11:56:48.0   &    32.4 &  9  	   &                 18:27:14.40  &  -11:53:25.0  &  41.9 &  12,15  \\
\cline{5-8}							     
18:27:38.71  &  -11:56:44.0   &    32.4 &  1      &&&&\\   
18:27:38.79  &  -11:56:44.7   &    33.6 &  2,3    &&&&\\   
18:27:17.20  &  -11:53:45.0   &     4.4 &  10     &&&&\\     
18:27:16.30  &  -11:53:30.0   &    15.7 &  8      &&&&\\     
\cline{1-4}							     
\end{tabular}
\begin{tablenotes}
\item[] REFERENCES. -- (1) \cite{ho1981}; (2) \cite{debuizer2003}; (3) \cite{furuya05}; (4) \cite{kurtz2004}; (5) \cite{becker1994}; (6) \cite{kolpak2003}; (7) \cite{forster&caswell2000}; (8) \cite{lockman1989}; (9) \cite{chini1987}; (10) \cite{walsh1998}; (11) \cite{testi1998}; (12) \cite{caswell1995}; (13) \cite{hofner1996}; (14) \cite{valtts2000}; (15) \cite{szymczak2000}; (16) \cite{caswell1983}; (17) \cite{forster1989}. 
\end{tablenotes}
\end{threeparttable}
\end{table*}

\begin{figure*}
\hspace{0.4cm} {\bfseries\Large FCRAO $^{13}$CO(1-0)} \hspace{4.85cm} {\bfseries\Large APEX 870~$\mu$m}
\centering
 \includegraphics[height=0.27\textheight]{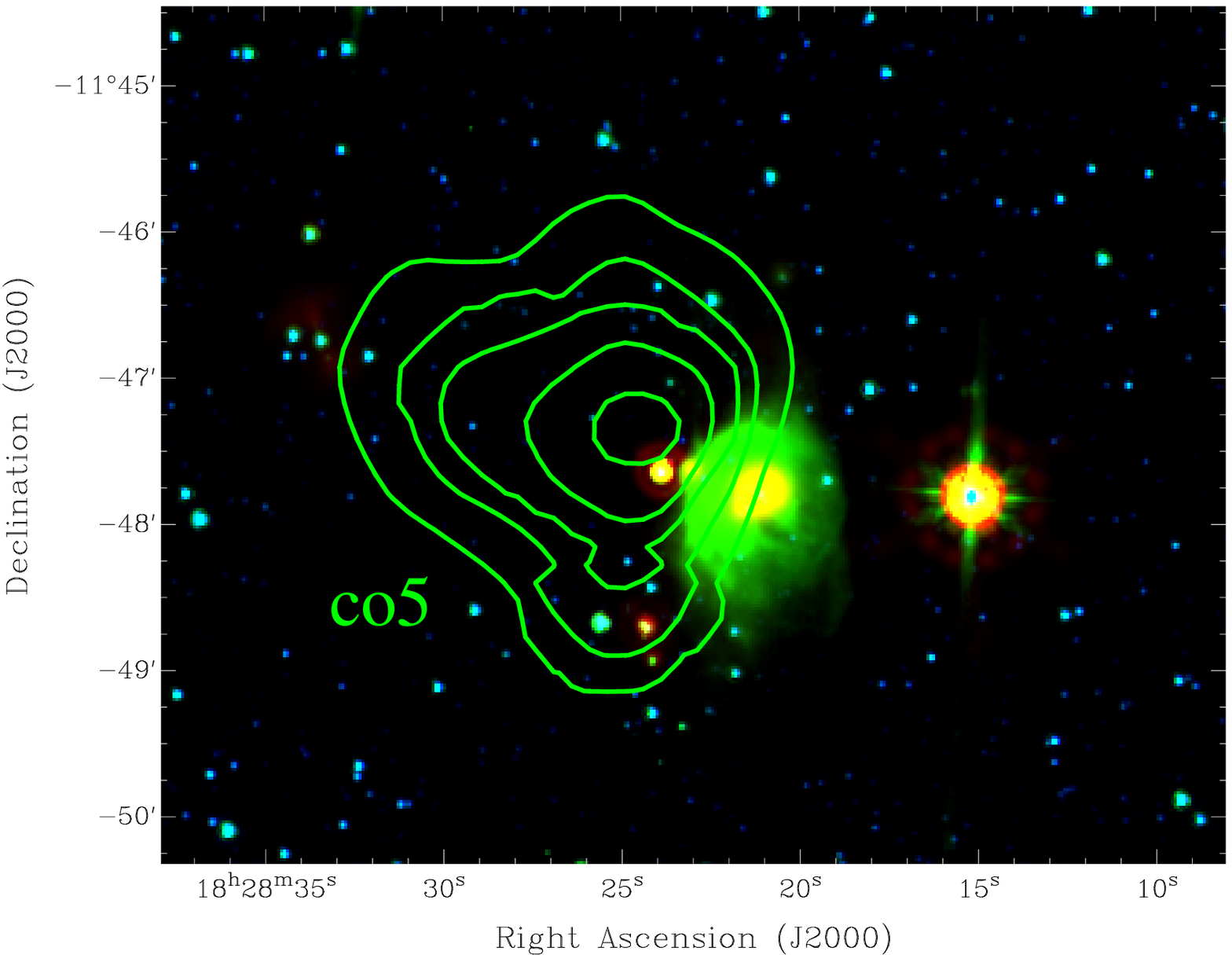}       
 \includegraphics[height=0.27\textheight]{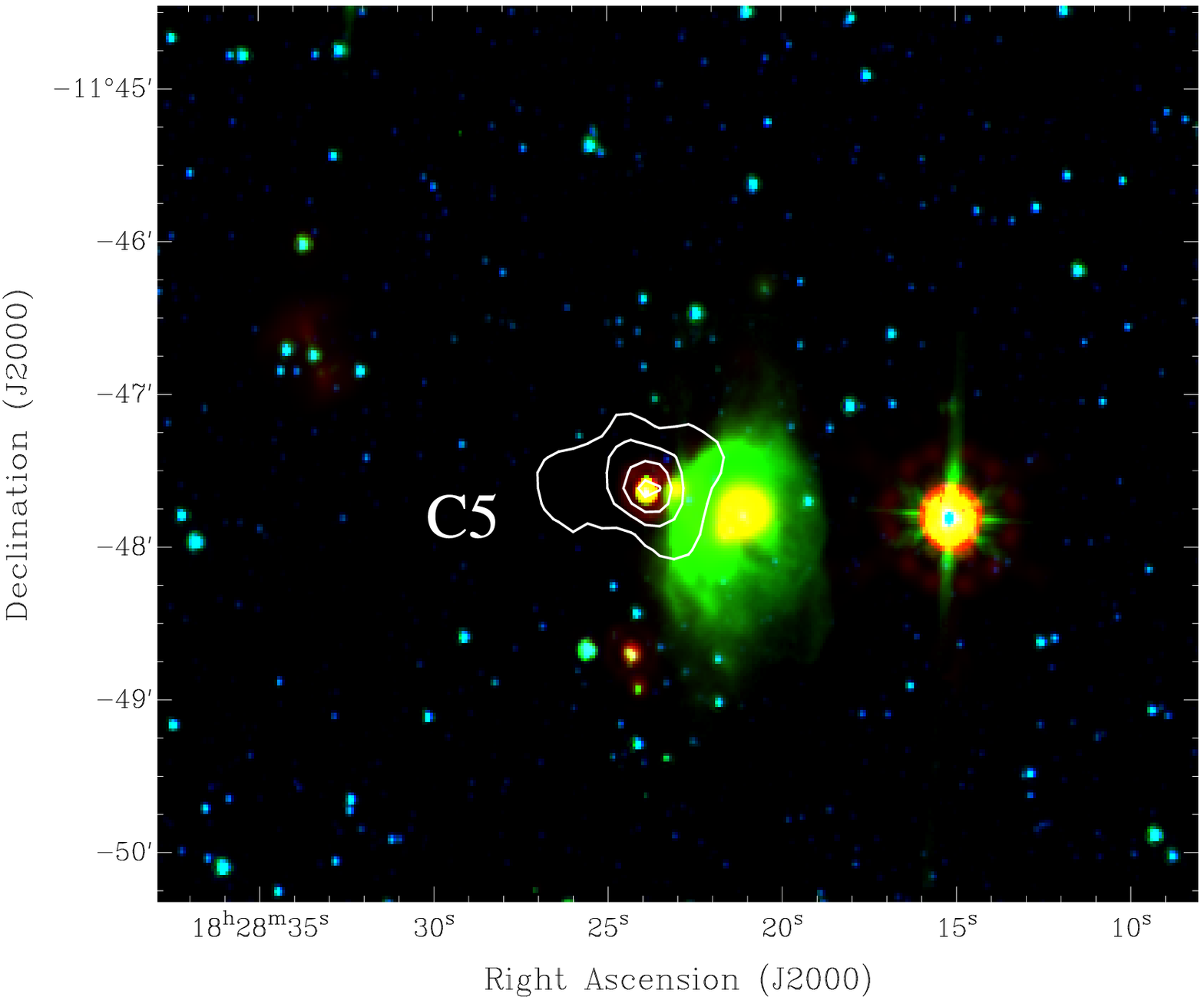}     
 \includegraphics[height=0.27\textheight]{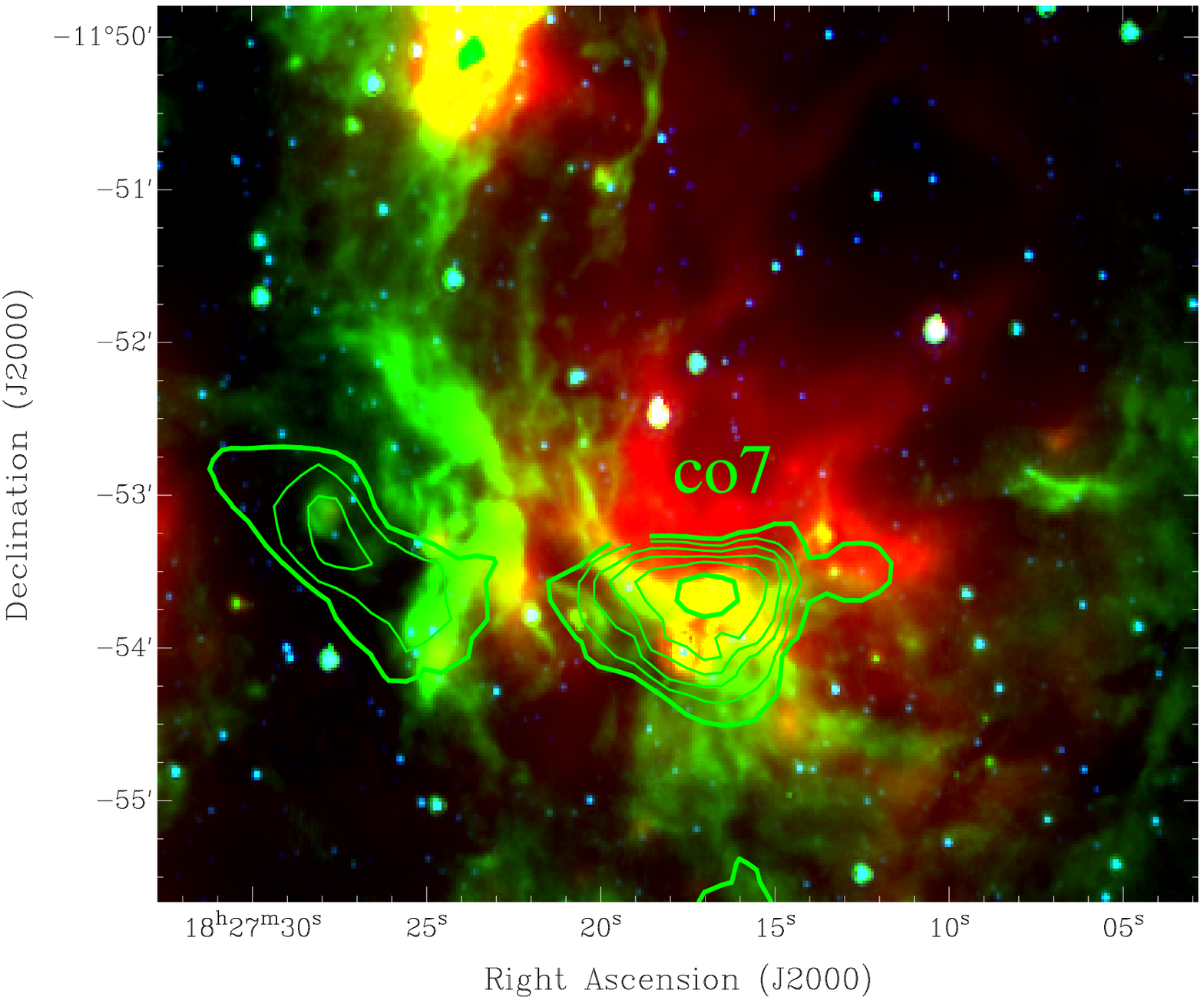}       
 \includegraphics[height=0.27\textheight]{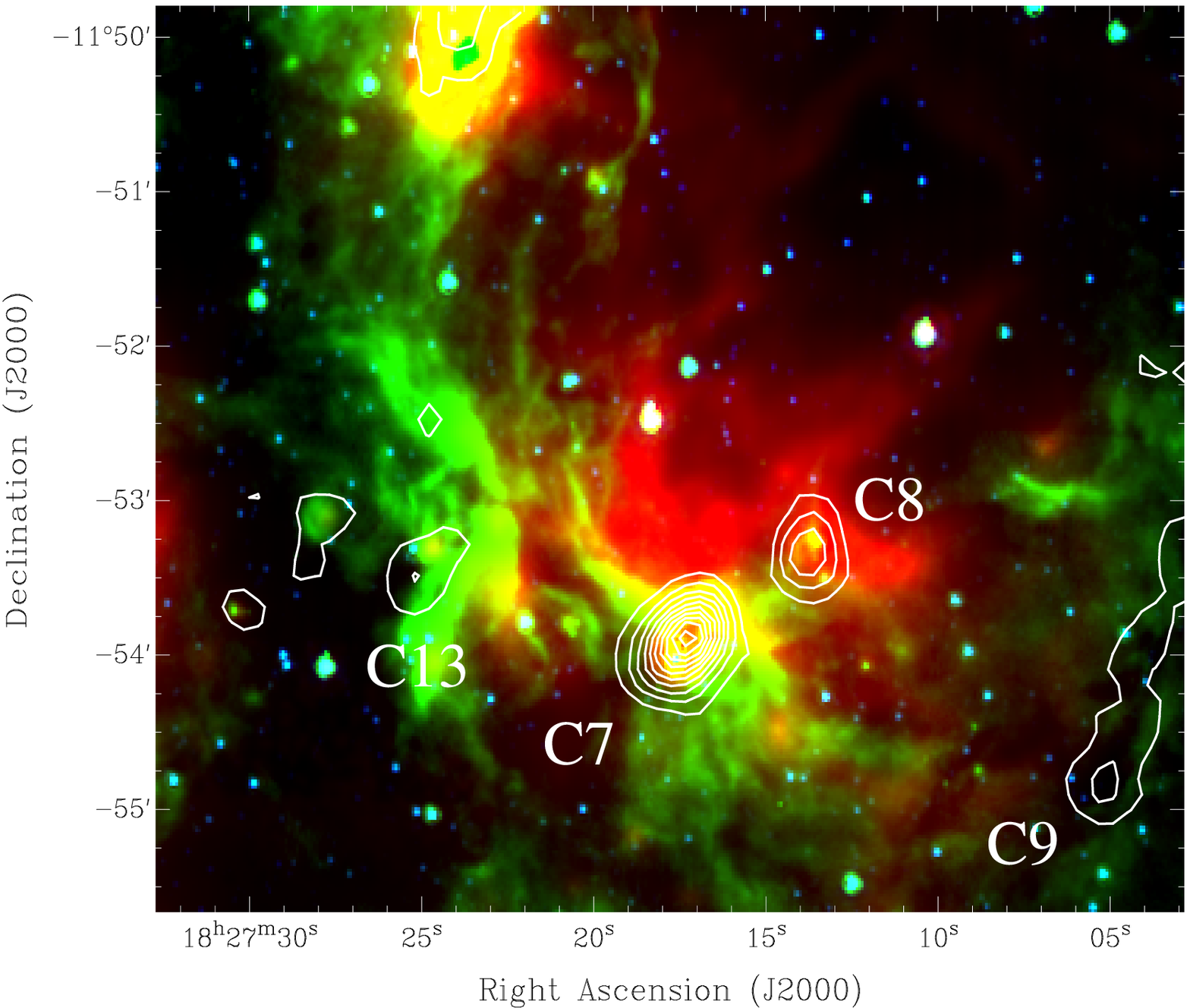}     
 \includegraphics[height=0.27\textheight]{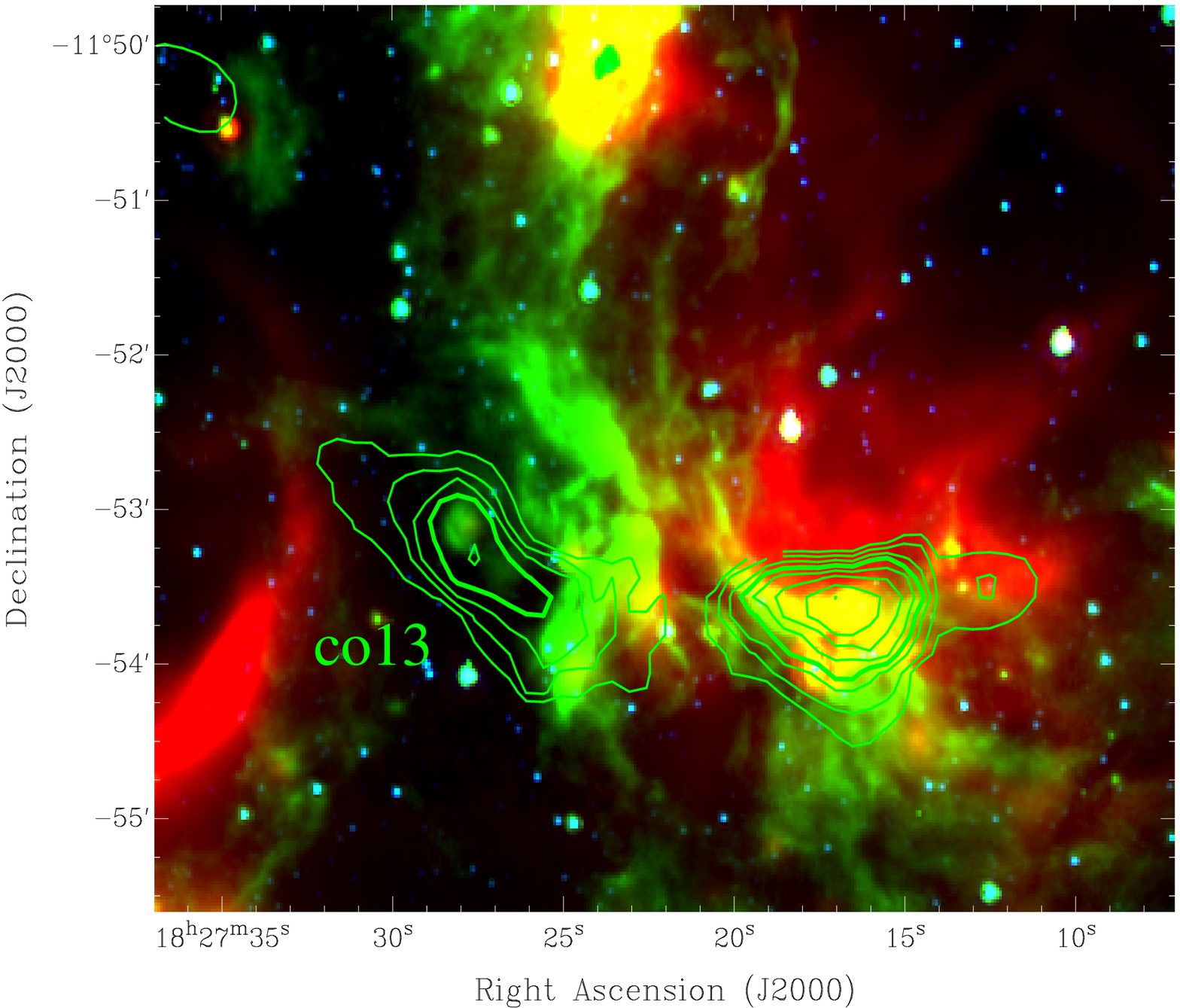}       
 \includegraphics[height=0.27\textheight]{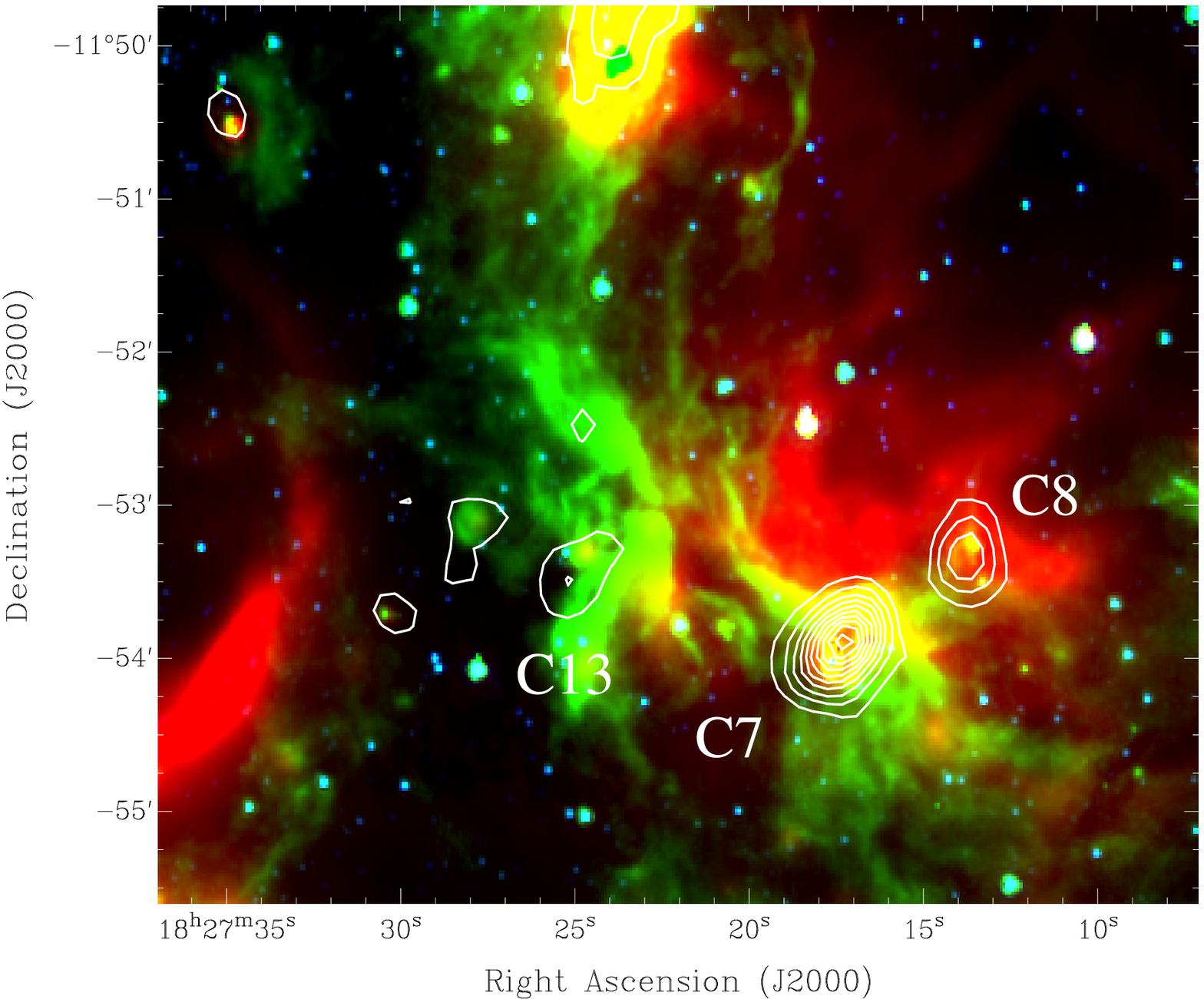}     
\begin{center}
Fig.~\ref{CO-apex-spitzer} -- Continued.
\end{center}
\end{figure*}
\begin{figure*}
\hspace{0.1cm} {\bfseries\Large FCRAO $^{13}$CO(1-0)} \hspace{3cm} {\bfseries\Large APEX 870~$\mu$m} \hspace{2cm} 
\centering
 \includegraphics[height=0.235\textheight]{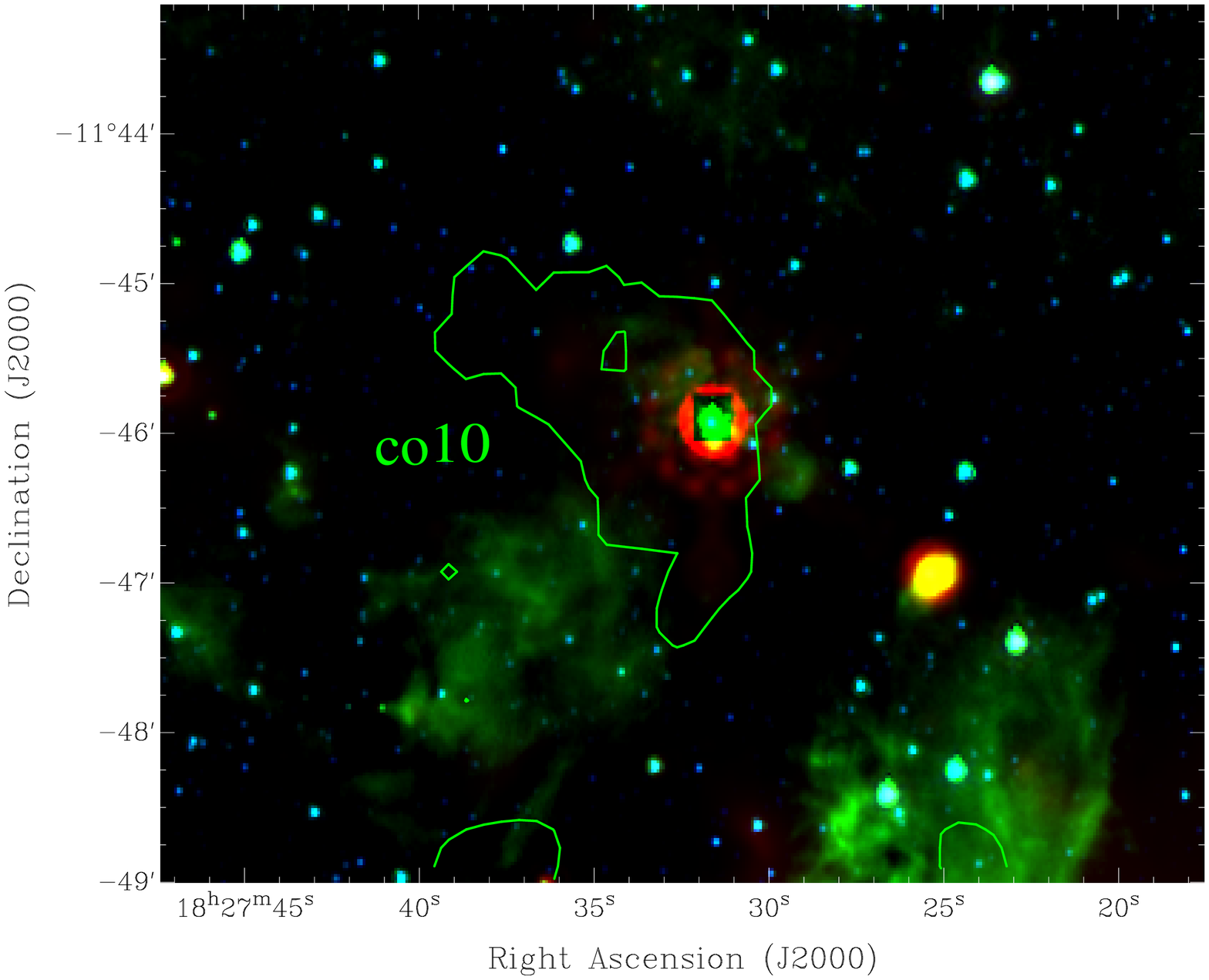}         
 \includegraphics[height=0.235\textheight]{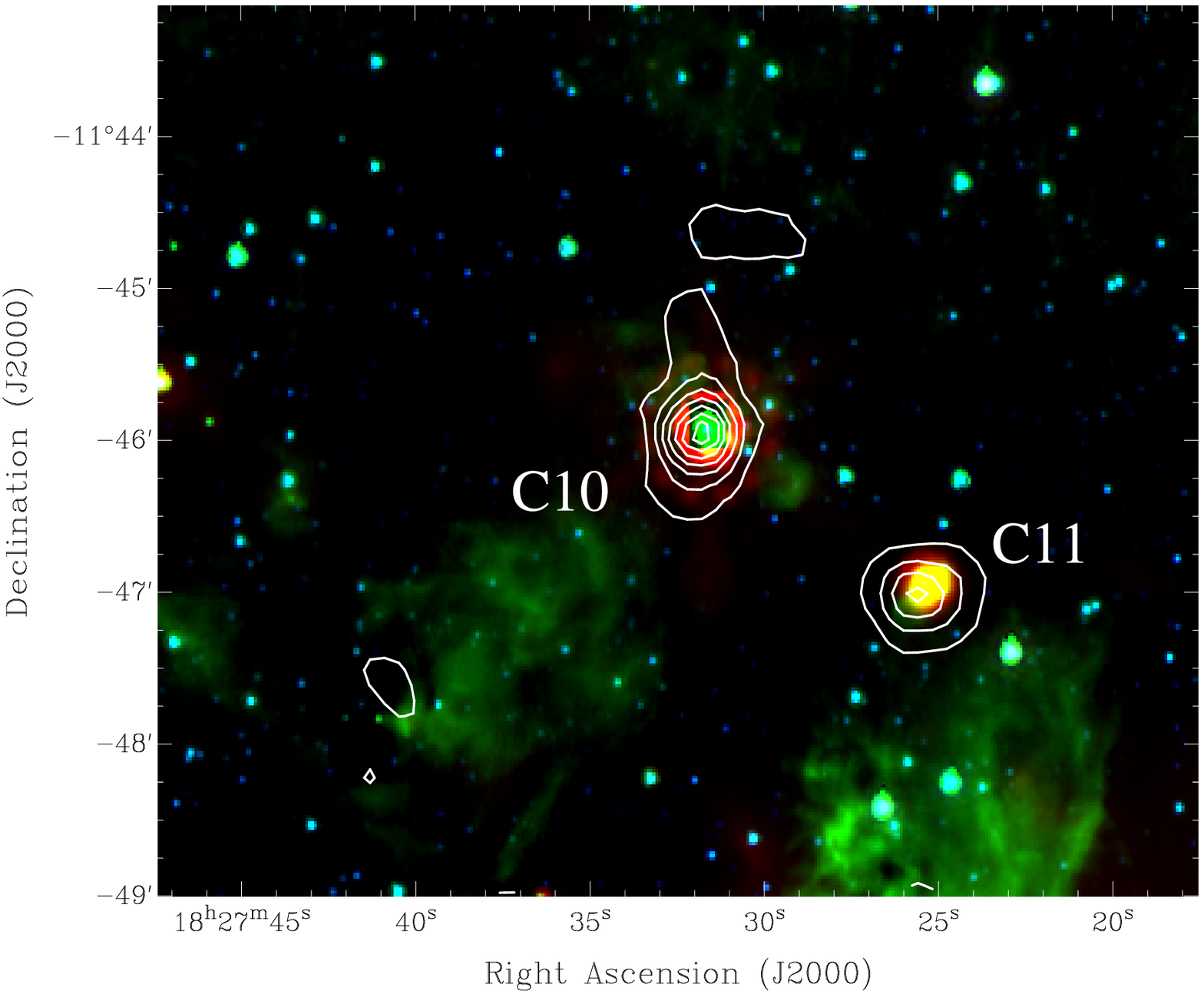}	   
 \includegraphics[height=0.23\textheight]{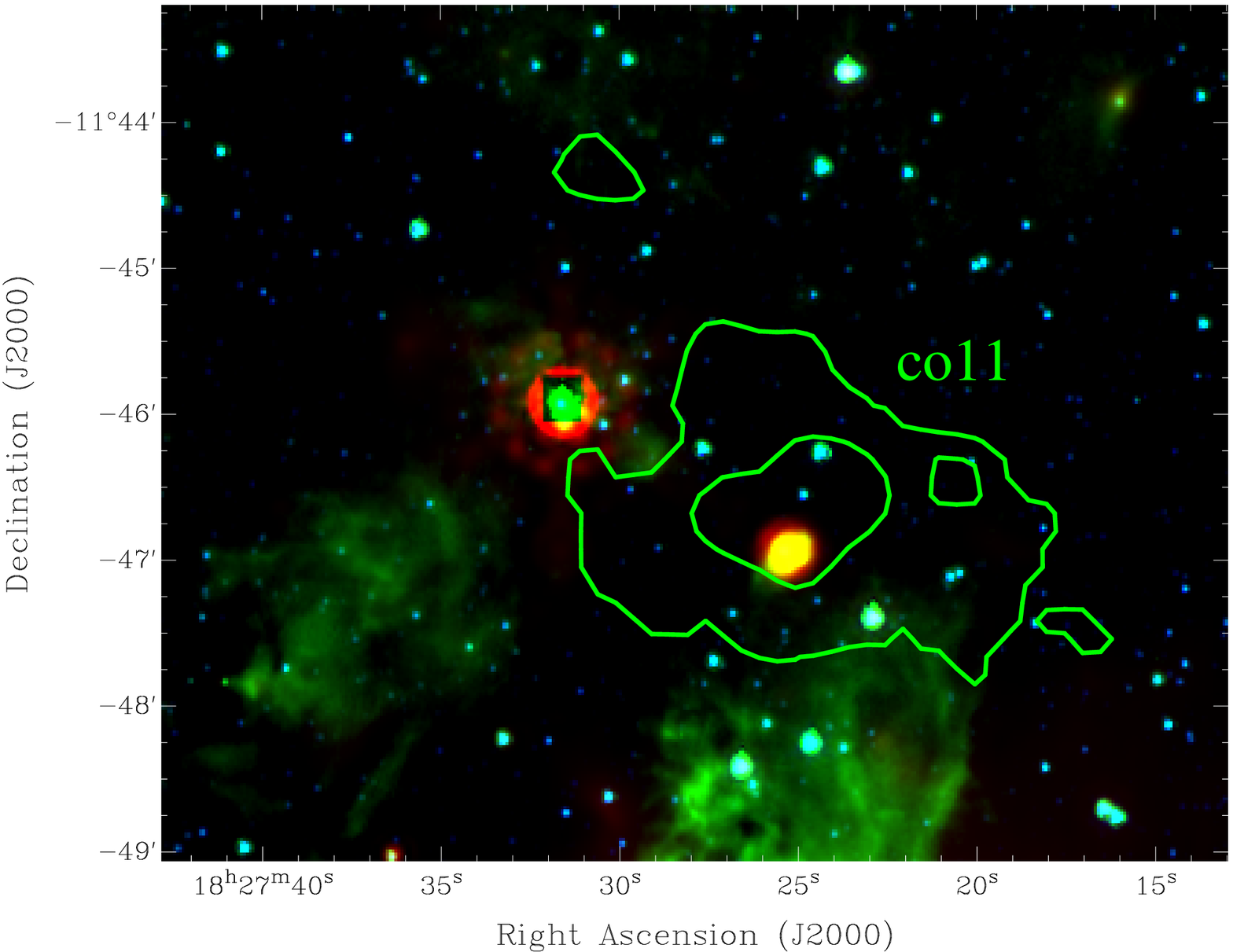}	   
 \includegraphics[height=0.23\textheight]{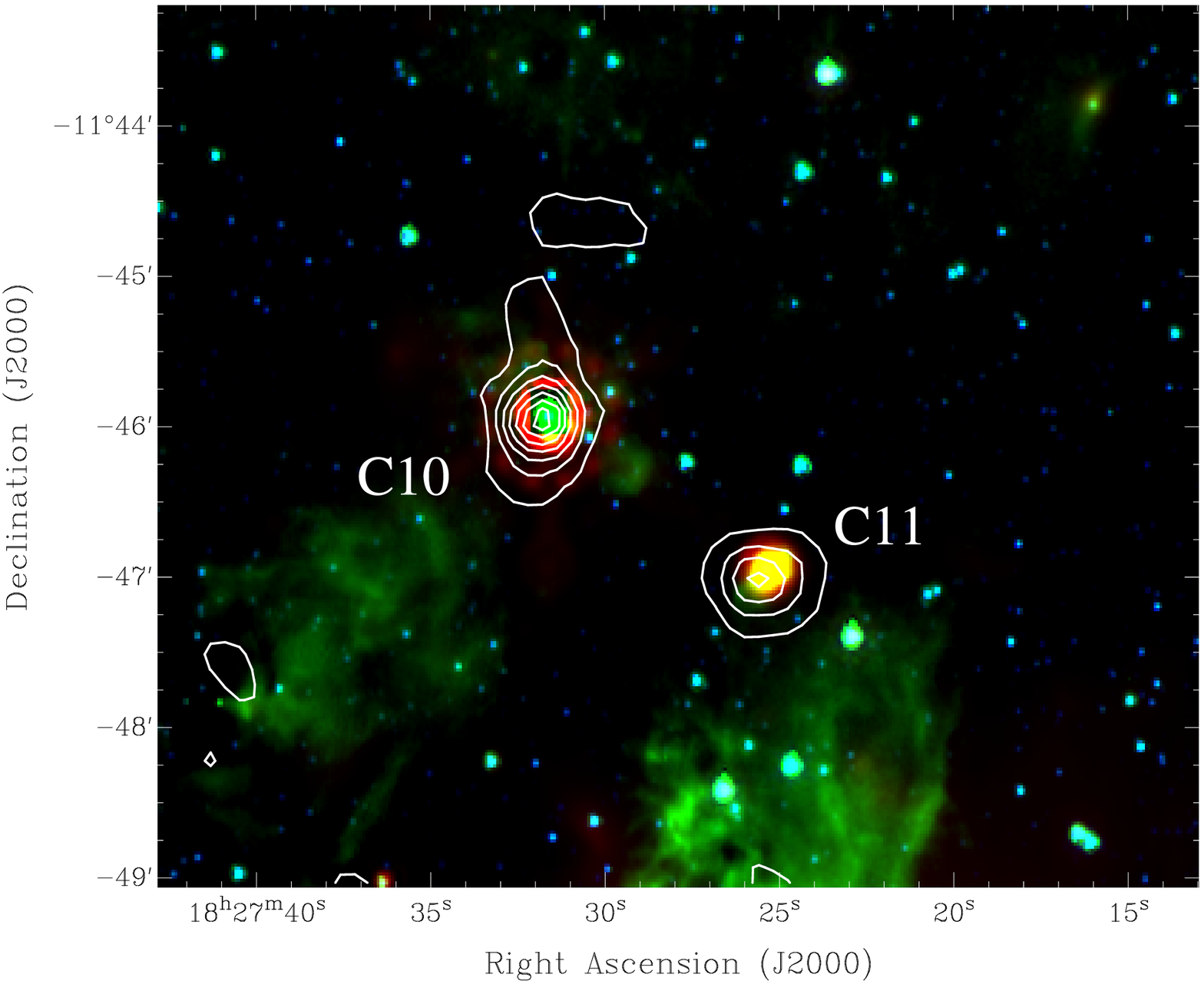}	   
 \includegraphics[height=0.24\textheight]{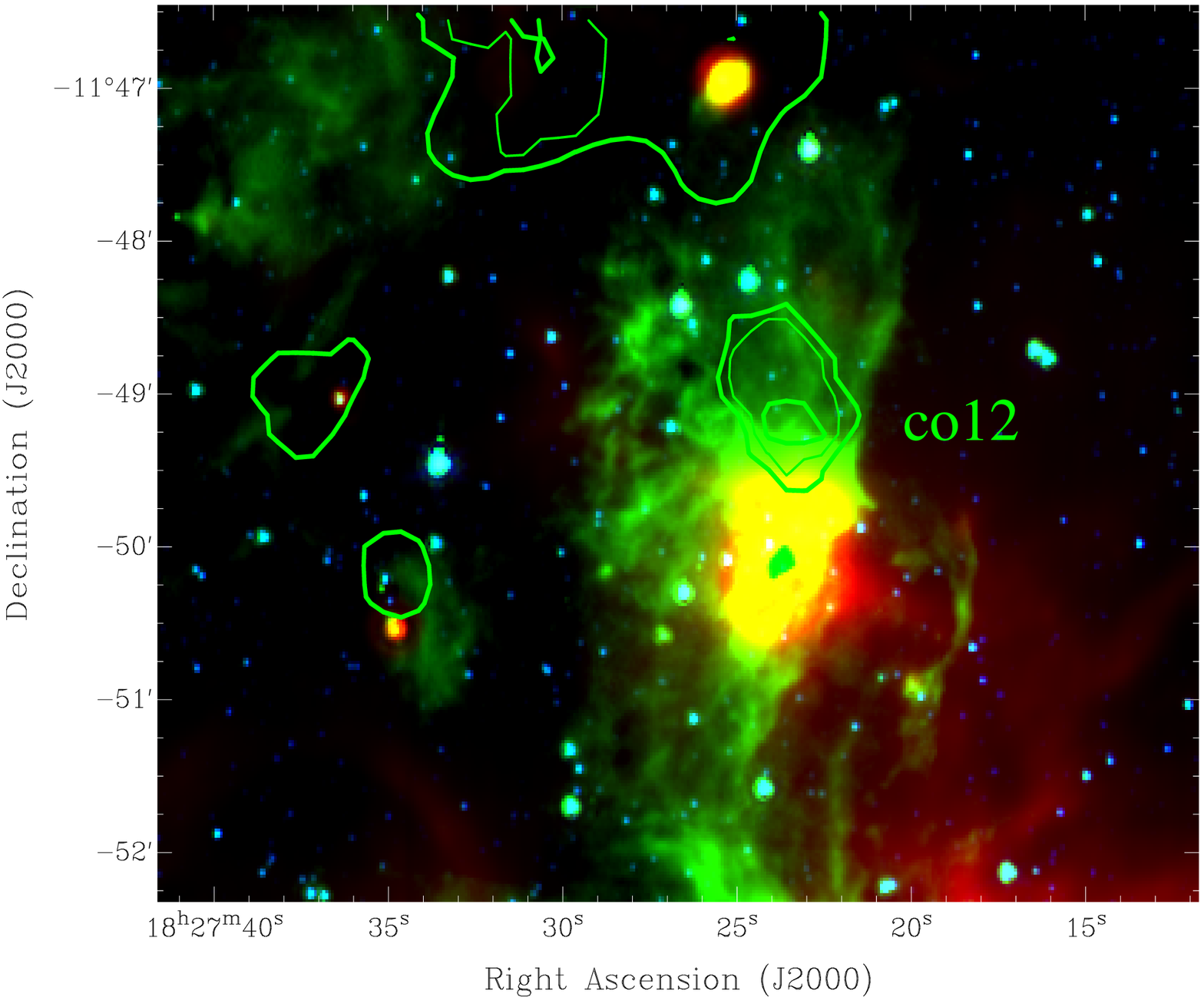}         
 \includegraphics[height=0.24\textheight]{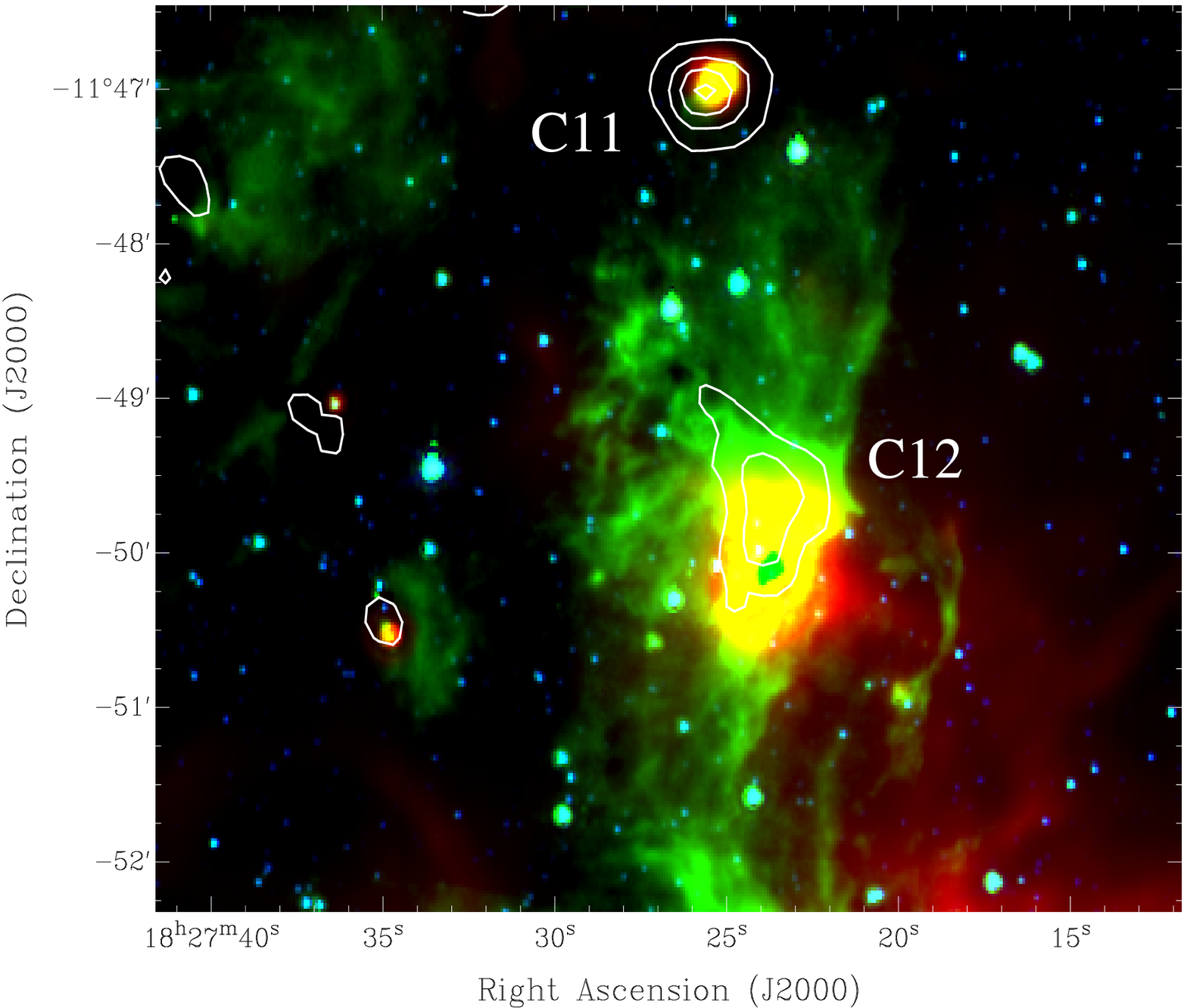}	   
 \includegraphics[height=0.22\textheight]{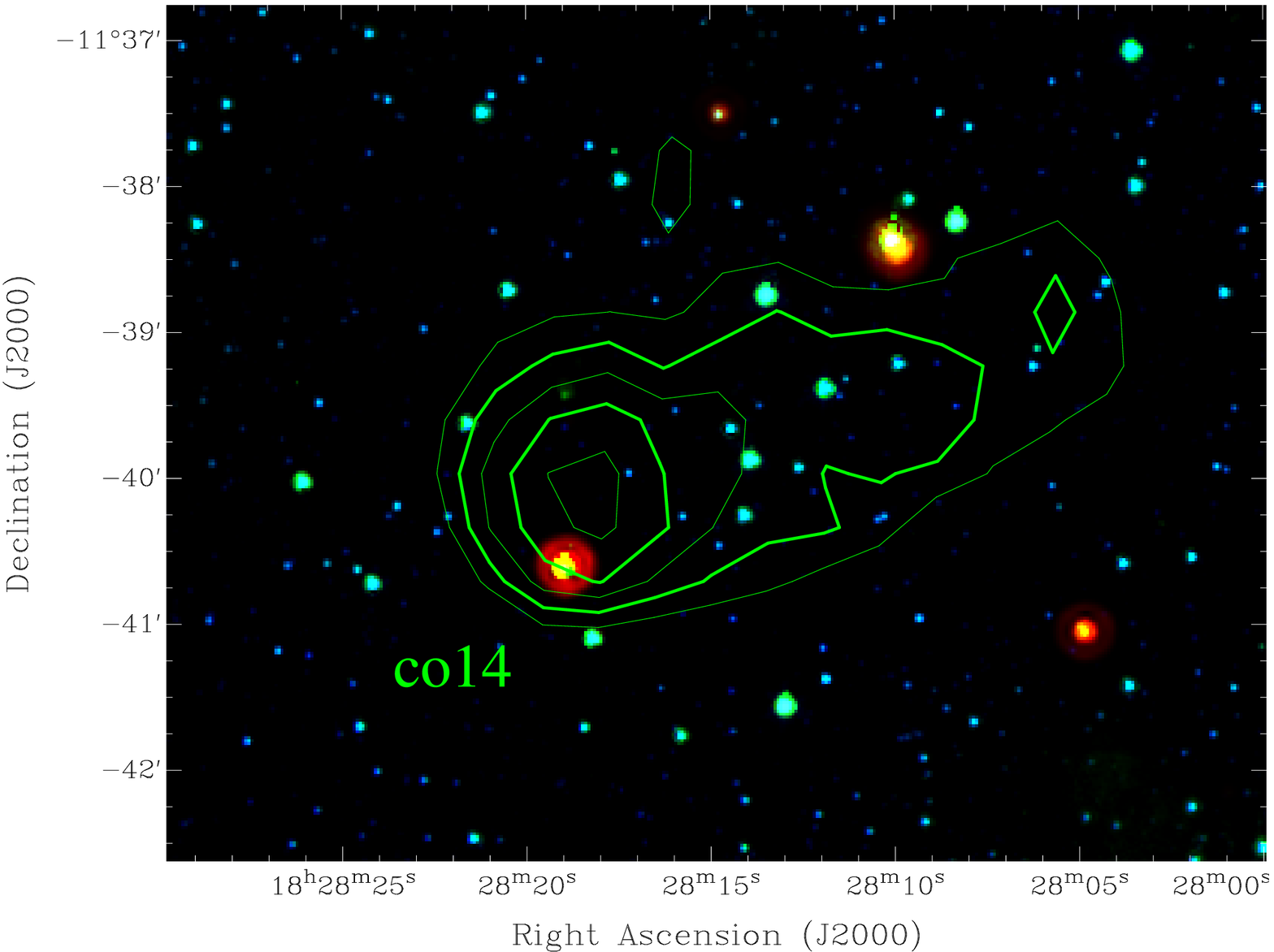}	   
 \includegraphics[height=0.22\textheight]{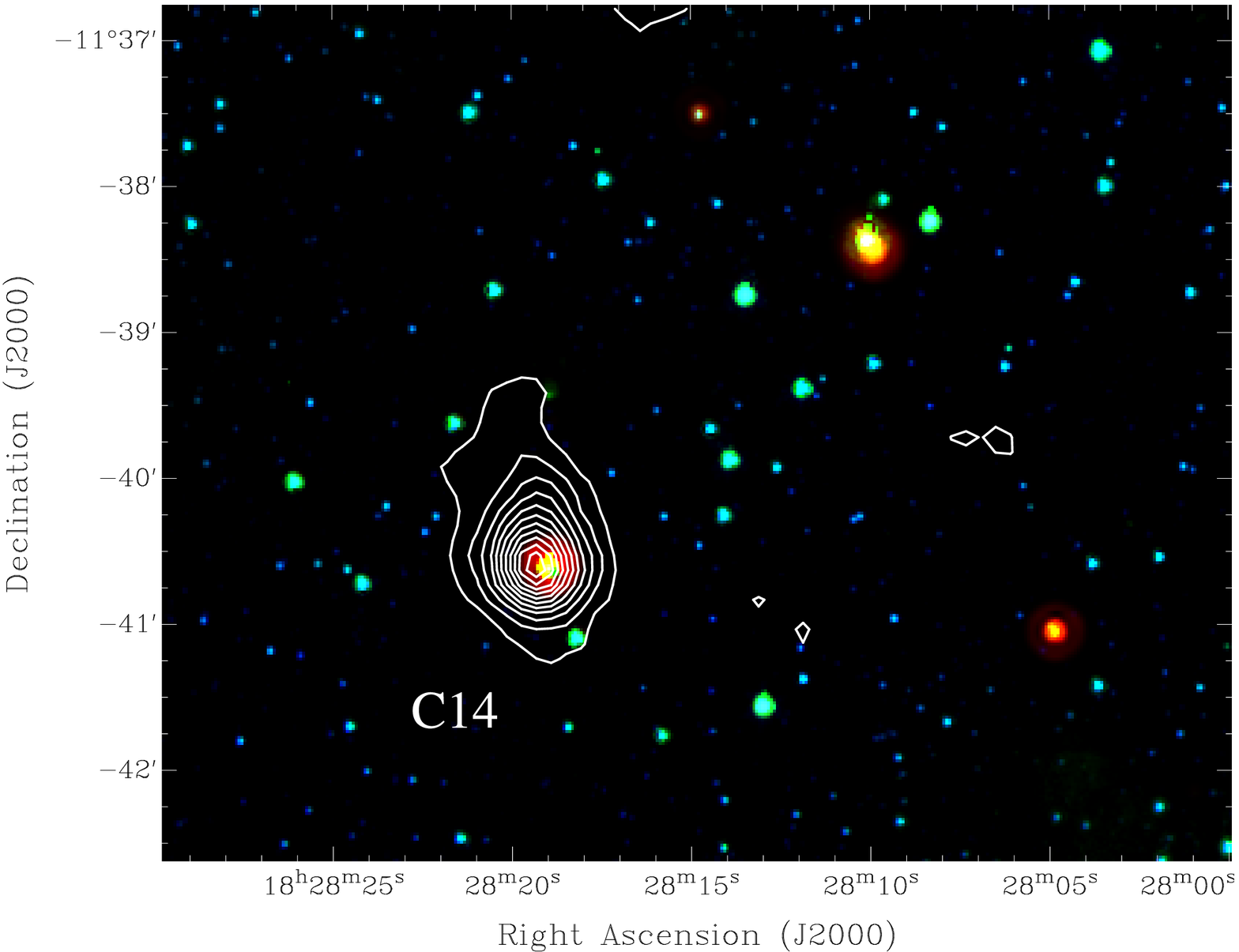}	   
\begin{center}
Fig.~\ref{CO-apex-spitzer} -- Continued.
\end{center}
\end{figure*}

\end{document}